\documentclass[reprint,
superscriptaddress,
showpacs,preprintnumbers,
notitlepage,amsmath,amssymb,
prb,
]{revtex4-1}

\usepackage{comment}
\usepackage{graphicx}
\usepackage{multirow}

\usepackage[usenames,dvipsnames]{xcolor}
\usepackage[colorlinks, breaklinks=true, linkcolor=red, citecolor=blue, linktocpage=true]{hyperref}

\newcommand{\chenjie}[1]{{\color{red}{\small Chenjie's notes: #1}}}

\newcommand{\dd}{\mathrm{d}}
\newcommand{\be}{\beta}
\newcommand{\om}{\omega}
\newcommand{\Z}{\mathbb{Z}}
\newcommand{\Zf}{\mathbb{Z}_2^f}
\newcommand{\timesw}{\times_{\om_2}}
\newcommand{\floor}[1]{\left\lfloor #1 \right\rfloor}
\newcommand{\eq}[1]{Eq.~(\ref{#1})}
\newcommand{\eqs}[2]{Eqs.~(\ref{#1}) and (\ref{#2})}

\begin{document}

\title{Non-Abelian Three-Loop Braiding Statistics for 3D Fermionic
Topological Phases}
\author{Jing-Ren Zhou}
\thanks{These two authors contributed equally to this work.}
\affiliation{Department of Physics, The Chinese University of Hong Kong, Shatin, New Territories, Hong Kong}

\author{Qing-Rui Wang}
\thanks{These two authors contributed equally to this work.}
\affiliation{Department of Physics, The Chinese University of Hong Kong, Shatin, New Territories, Hong Kong}

\author{Chenjie Wang}
\email{cjwang@hku.hk}
\affiliation{Department of Physics and HKU-UCAS Joint Institute for Theoretical and Computational Physics, The University of Hong Kong, Pokfulam Road, Hong Kong, China}

\author{Zheng-Cheng Gu}
\email{zcgu@phy.cuhk.edu.hk}
\affiliation{Department of Physics, The Chinese University of Hong Kong, Shatin, New Territories, Hong Kong}

\begin{abstract}
Fractional statistics is one of the most intriguing features of topological phases in 2D. In particular, the so-called non-Abelian statistics plays a crucial role towards realizing topological quantum computation. Recently, the study of topological phases has been extended to 3D and it has been proposed that loop-like extensive objects can also carry fractional statistics.
In this work, we systematically study the so-called three-loop braiding statistics for 3D interacting fermion systems. Most surprisingly, we discover new types of non-Abelian three-loop braiding statistics that can only be realized in fermionic systems (or equivalently bosonic systems with emergent fermionic particles). 
On the other hand, due to the correspondence between gauge theories with fermionic particles and classifying fermionic symmetry-protected topological (FSPT) phases with unitary symmetries, our study also gives rise to an alternative way to classify FSPT phases. We further compare the classification results for FSPT phases with arbitrary Abelian unitary total symmetry $G^f$ and find systematical agreement with previous studies.
\end{abstract}

\maketitle

\tableofcontents

\section{Introduction}
\begin{figure}
\centering
\includegraphics[
width=0.3\textwidth]{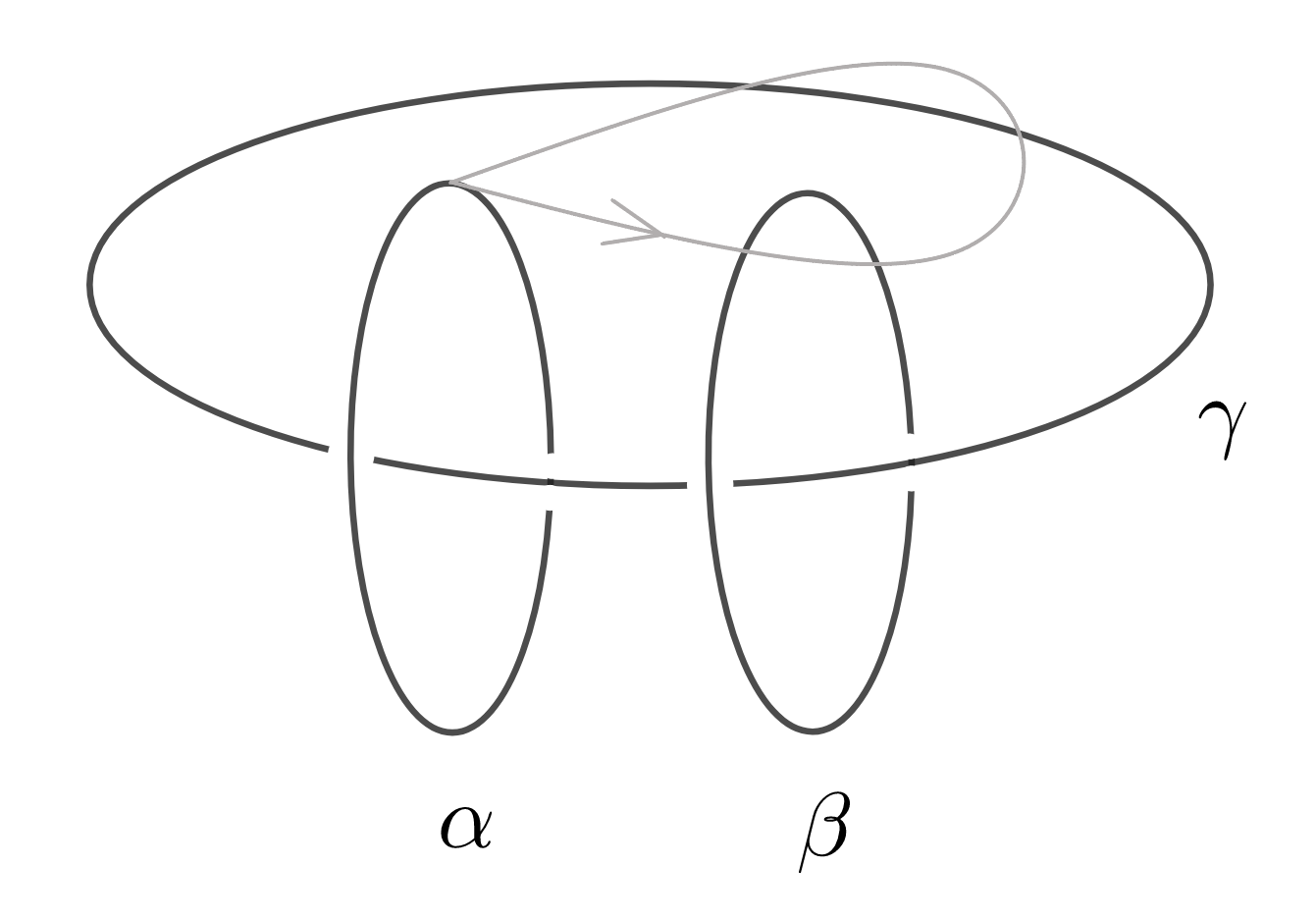}

\caption{The so-called three-loop braiding process, that is, braiding one loop $\alpha$ around another loop $\beta$, while both of them are linked to a third loop $\gamma$}
\label{Fig1}
\end{figure}

Topological phases of quantum matter are a new kind of quantum phases beyond Landau's paradigm. Since the discovery of fractional quantum Hall effect (FQHE), fractionalized statistics of point-like excitations in topological phases has been intensively studied in 2D strongly correlated electron systems.
In the past decade, the theoretical prediction and experimental discovery of topological insulator and topological superconductor in 3D systems 
have further extended our knowledge of topological phases into higher dimensions. As a unique feature, the excitations of 3D topological phases not only contain point-like excitations, but also contain loop-like excitations. Therefore, the fundamental braiding process is not only limited to particle-particle braiding, but is also extended to particle-loop braiding and loop-loop braiding. It is well known that due to topological reasons, point-like excitations in 3D can only be bosons or fermions. In addition, particle-loop braiding can be understood in terms of Aharonov-Bohm effect and loop-loop braiding is equivalent to particle-loop braiding(one can always shrink one of the loops into a point-like excitation). As a result, for long time people thought there was no interesting fractional statistics in 3D beyond the Aharonov-Bohm effect. Surprisingly, a recent breakthrough pointed out that loop-like excitations can indeed carry interesting fractional statistics via the so-called three-loop braiding process\cite{threeloop,WCWG2018,lin15}: braiding a loop $\alpha$ around another loop $\beta$, while both are linked to a third loop $\gamma$, as shown in Fig. \ref{Fig1}. Apparently, such kind of braiding process can not be reduced to the particle-loop braiding due to the linking with a third loop. So far, it has been believed that the three-loop braiding process is the most elementary loop braiding process in 3D.

Another natural question would be: Whether we can use three-loop braiding process to characterize and classify all possible topological phases for  interacting fermion systems in 3D? Recent studies on the classification of topological phases for interacting bosonic and fermionic systems in 3D suggest a positive answer to the above question\cite{Tian1,Tian2}. Basically, it has been conjectured that all topological phases in 3D can be realized by ``gauging" certain underlying symmetry-protected topological (SPT) phases \cite{braiding12,gu14b}.  For bosonic systems, the ``gauged" SPT states are known as Dijkgraaf-Witten gauge theory, and it has been shown (at least for Abelian gauge groups) that three-loop braiding process of their corresponding flux lines can uniquely characterize and exhaust all Dijkgraaf-Witten gauge theories\cite{topoinvar15}. For fermionic systems, some particular examples with Abelian three-loop braiding process are also studied recently\cite{threeloop2018}. However, it is still unclear how to understand general cases. 
On the other hand, it is well known that in low dimensions (up to 3D), the group cohomology theory\cite{BSPT2012,BSPT13,gu09} gives rise to a complete classification of bosonic symmetry-protected topological (BSPT) phases for arbitrary finite unitary symmetry groups.  
The classification can be generalized to fermionic symmetry-protected topological (FSPT)\ phases by more advanced constructions\cite{supercohomology,kapustin14,freed14,cheng15,Gaiotto2016,freed16,Kapustin2017,WangGu2017,WangGu2018,Meng2018}. 

\begin{figure}
\centering
\includegraphics[
width=0.3\textwidth]{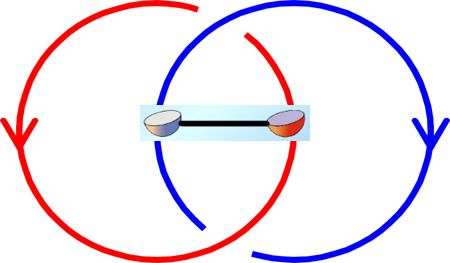}
\caption{Attaching an open Majorana chain onto a pair of linked loops realizes the so-called Ising non-Abelian three-loop braiding process.}
\label{Fig2}
\end{figure}

In this work, we attempt to systematically understand the three-loop braiding statistics for gauged interacting FSPT systems with general Abelian unitary symmetries. In particular, we discover new types of non-Abelian three-loop braiding statistics that can be only realized in the presence of fermionic particles (accordingly beyond Dijkgraaf-Witten theories). The simplest symmetry group supporting such kind of non-Abelian three-loop braiding process is $\mathbb{Z}_2 \times \mathbb{Z}_8$ or $\mathbb{Z}_4 \times \mathbb{Z}_4$. (More precisely, the corresponding total groups are $G^f=\mathbb{Z}_2^f\times\mathbb{Z}_2 \times \mathbb{Z}_8$ or $\mathbb{Z}_2^f\times\mathbb{Z}_4 \times \mathbb{Z}_4$ if we include the total fermion parity symmetry $\mathbb{Z}_2^f$.)  A simple physical picture describing the corresponding non-Abelian statistics can be viewed as attaching an open Kitaev's Majorana chain onto a pair of linked flux lines ($\mathbb{Z}_2$ and $\mathbb{Z}_8$ unit flux lines for the former case and two different $\mathbb{Z}_4$ unit flux lines for the latter case). 

In 1D, it has been shown that a Majorana chain will carry two protected Majorana zero modes on its open ends \cite{Kitaev2001}. In 2D, it is also well known the vortex(anti-vortex) of a p+ip topological superconductor can carry a single topological Majorana zero mode. 
Thus, it is very natural to ask if flux lines in 3D can also carry topological Majorana zero mode or not. Surprisingly, we find that flux lines carrying topological Majorana zero modes must be linked to each other, as shown in Fig. \ref{Fig2}. In contrast, if the loops are unlinked, they can never carry Majorana zero modes, and this is because one can always smoothly shrink the flux loops into a point like excitation with a single Majorana zero mode on it. However, as a point like object in 3D can only be boson or fermion, and it is impossible to be an anyon with Ising non-Abelian statistics.   

The non-Abelian nature of the new type three-loop braiding process we discovered can be understood as the two-fold degeneracy carried by a pair of linked flux lines, and the braiding statistics between two loops that linked with a third loop should be characterized by a unitary 2 by 2 matrix instead of a simple $U(1)$ phase factor. An alternative way to understand the non-Abelian nature of the three-loop braiding statistics is to use the standard dimension reduction method to deform the 3D lattice model into a 2D lattice model\cite{Tantivasadakarn}, i.e., by shrinking the z-direction to single lattice spacing such that the flux line along the z-direction can be regarded as a 2D particle which is exactly the Ising non-Ableian anyon\cite{Kitaev2006} with quantum dimension $\sqrt{2}$. Finally, by explicitly working out all the algebraic constraints of three-loop braiding process for fermionic systems(or equivalently, bosonic systems with emergent fermionic particles), we not only uncover new types of Ising non-Abelian three-loop braiding, but also derive a complete classification of 3D FSPT phases with Abelian unitary symmetry.

\section{Results}

\subsection{Preliminaries}

We begin with the basics of symmetries in fermionic systems and loop braiding statistics in gauged 3D FSPT phase. 

\subsubsection{Symmetries in interacting fermion systems}
\label{sec:symmetry}

Fermionic systems have a fundamental symmetry---the conservation of fermion
parity: $P_{f}=(-1)^{N_{f}}$, where $N_{f}$ is the total number of fermions. The corresponding symmetry group is denoted as $%
\mathbb{Z}
_{2}^{f}=\{1,P_{f}\}$. In the presence of other global on-site
symmetries, the total symmetry group $G_f$ is the central extension of the bosonic symmetry group $G_b$ by the fermion parity $\mathbb{Z}_2^f$, determined 
by the 2-cocycle $\omega_2 \in H^2(G_b, \mathbb{Z}_2)$.  In this work, we consider a general Abelian unitary symmetry group of the following form:
\begin{equation}
\label{symmetry}
G_{f}=%
\mathbb{Z}
_{N_{0}}^{f}\times \overset{K}{\underset{i=1}{\prod }}%
\mathbb{Z}
_{N_{i}}
\end{equation}
where $N_{0}=2m$ is an even integer. One can show that any finite Abelian symmetry group in fermionic systems can be written in this form, after a proper isomorphic transformation. 


The bosonic symmetry group is expressed as 
\begin{equation}
G_{b}=G_{f}/%
\mathbb{Z}
_{2}^{f}=%
\mathbb{Z}
_{N_{0}/2}\times \prod_{i=1}^{K}%
\mathbb{Z}
_{N_{i}}
\end{equation}
For simplicity, we will mainly consider the case that $m$ and all $N_{i}$ are powers of $2 $, i.e.,
\begin{equation}
N_\mu = 2^{n_\mu}, \quad m= 2^{n_0-1}
\label{eq:specialN}
\end{equation}
where $n_0 \ge 1$. When $n_0=1$ (i.e. $m=1$), the central extension of $G_b$ is trivial; when $n_0\ge 2$, the central extension of $G_b$ is nontrivial. This simplification does not exclude any interesting FSPT phases because odd factors of each $N_\mu$ can be factored out. Moreover, $n_2$ and $n_3$ are always trivial if all $N_i$'s are odd integers. Accordingly, neglecting the odd factors, we only lose some BSPT phases, whose classification and characterization are well studied\cite{BSPT2012}. 

\subsubsection{Topological excitations and three-Loop braiding in 3D}

To study FSPT phases with symmetry group $G_f$, we will gauge the full symmetry. That is, we introduce a gauge field of gauge group $G_f$  and couple it to the FSPT system through the minimal coupling procedure (see Refs.~\onlinecite{topoinvar15,braiding12} for details of the procedure).  The resulting gauged system is guaranteed to be gapped through that procedure, which is actually topologically ordered. It contains two types of topological excitations:

(i) Point-like excitations that carry gauge charge. We label them by  a vector $q=(q_{_{0}},...,q_{K})$, where $q_{\mu }$ is an
integer defined modulo $N_{\mu }$. We will use $q$ to denote both the excitation and its gauge charge. This is legitimate because gauge charge uniquely determines charge excitations.  Charge excitations are Abelian anyons.  Fusing two charge excitations $q_1$ and $q_2$, we obtain a unique charge excitation $q = q_1 + q_2$.


(ii) Loop-like excitations that carry gauge flux. We call them vortices, vortex loops or simply loops, and label them by $\alpha, \beta, \dots$. The gauge flux carried by loop $\alpha$ is denoted by $
\phi _{\alpha }=(\frac{2\pi }{N_{0}}a_{_{0}},...,\frac{2\pi }{N_{K}}a_{K})$, where $a_{\mu
}$ is an integer defined modulo $N_{\mu }$. There exist many loops that carry the same gauge flux, which differ from each other by attaching charges. Unlinked loops are Abelian, however, they may become non-Abelian when they are linked with other loops. Hence, fusion of vortex loops depend on whether they are linked or not. Nevertheless, regardless Abelian or non-Abelian, gauge flux always adds up. General vortex excitations are not limited to simple loops. For example, they may be knots or even more complicated structure. In this work, we only consider simple loops and links of them. So far, properties of loops are enough to characterize gauged FSPT systems.


We need to consider three types of braiding statistics between the loops and charges\cite{threeloop}:

First, charge-charge exchange statistics.  A charge is either a boson or
fermion, depending on the gauge charge it carries. More explicitly, the exchange statistics of charge $q$ is given by
\begin{equation}
\theta_q = \pi q_0
\end{equation}
That is, when $q_0$ is odd, it is a fermion. Otherwise, it is a boson. Mutual statistics between charges are always trivial.

Second, charge-loop braiding statistics, which is the Aharonov-Bohm phase given by
\begin{equation}
\theta_{q, \alpha} =q\cdot \phi_\alpha\label{theta1}
\end{equation}%
where ``$\cdot $'' is the vector dot product. We single out a special class of vortex loops, those carrying the fermion parity gauge flux $\phi=(\pi, 0,\dots, 0)$. We denote these \textit{fermion-parity loops}
as $\xi _{f}$. The mutual statistics between charges and fermion-parity loops are simply given by\cite{2DFSPTbraiding}	:
\begin{equation}
\theta _{q,\xi_f}=q\cdot \phi _{\xi _{f}}=\pi q_{0}\label{theta2}
\end{equation}%
We notice that the self-exchange statistics of a charge $q$ is equal to Aharonov-Bohm phase $\theta_{q, \xi_f}$, which is required by the very definition of fermion parity symmetry.

Third, loop-loop braiding statistics. It was shown in Ref.~\onlinecite{threeloop} that the fundamental braiding process between loops is the so-called three-loop braiding statistics (Fig. \ref{Fig1}):

\textit{Let }$\alpha ,\beta ,\gamma $\textit{\ be three loop-like
excitations. A three-loop braiding is a process that a loop }$\alpha $%
\textit{\ braids around loop }$\beta $\textit{\ while both linked to a base
loop }$\gamma $.

On the other hand, if there is no base loop, the two-loop braiding process can always be reduced to charge-loop braiding statistics:\cite{threeloop}
\begin{equation}
\theta _{\alpha \beta }=q_{\alpha }\cdot \phi _{\beta }+q_{\beta }\cdot \phi
_{\alpha }\label{theta3}
\end{equation}%
Here $q_\alpha$ is the absolute charge carried by loop $\alpha$, which can be obtained by shrinking the loop to a point. Since charge-loop braiding statistics is universal for all FSPT phases
with the same symmetry group $G_{f}$, two-loop braiding is not able to
distinguish different FSPT phases. In the presence of a base loop $\gamma$, the notion of absolute charge is not well defined as shrinking loop $\alpha$ to a point will inevitably touch the base loop.  Accordingly, three-loop braiding statistics can go beyond Aharonov-Bohm phases, as already demonstrated in many previous works.\cite{threeloop,topoinvar15,threeloop2018}

While the gauge group $G_f$ is Abelian, three-loop braiding process is not limited to be Abelian. As mentioned above, linked loops can be non-Abelian in general, and three-loop process involves linked loops. Let us consider loops $\alpha,\beta$, which are linked to the base loop $\gamma$. The base loop $\gamma$ carries gauge flux $\phi _{\gamma }=(\frac{%
2\pi }{N_{0}},...,\frac{2\pi }{N_{K}})\cdot c$, where $c$ is an integer vector. Generally speaking,  the fusion space  between $\alpha$ and $\beta$, denoted as $V_{\alpha\beta,c}$, is multi-dimensional (we use this notation because
the fusion and braiding process only depend on the gauge flux of the base loop). More explicitly, 
\begin{equation}
V_{\alpha \beta ,c}=%
\underset{\delta }{\bigoplus }V_{\alpha \beta ,c}^{\delta }
\end{equation}
where loop $\delta$ are the possible fusion channels of $\alpha$ and $\beta$. Braiding between $\alpha$ and $\beta$ is a unitary transformation in the fusion space, which in general is not just a phase, but a matrix, leading to non-Abelian three-loop braiding statistics. Similarly to anyons in 2D, one can define fusion multiplicities $N_{\alpha\beta,c}^\delta$, $F$- and $R$-matrices to describe the loop fusion and braiding properties.\cite{topoinvar15} We give more detailed descriptions in Appendix.


\begin{widetext}

\begin{tabular}{|l|l|l|}
\hline 
Stacking Group & Cases & Classification \\ \hline

\begin{tabular}{l}
$A$ \\
\end{tabular}
&
\begin{tabular}{l}
If $m$ is odd \\ 
\end{tabular}
&
\begin{tabular}{l}
$\mathbb{Z}_{1}$  \\
\end{tabular}
\cr

\begin{tabular}{l}
\\
\end{tabular}
&
\begin{tabular}{l}
If $m$ is even \\ 
\end{tabular}
&
\begin{tabular}{l}
$\mathbb{Z}_{1}$ \\ 
\end{tabular}
\cr

\hline

\begin{tabular}{l}
$B_{i}$ \\
\end{tabular}
&
\begin{tabular}{l}
If $m$ is odd \\ 
\end{tabular}
&
\begin{tabular}{l}
$\mathbb{Z}_{1}$ \\ 
\end{tabular}
\cr

\begin{tabular}{l}
  \\
\end{tabular}
&
\begin{tabular}{l}
If $m$ is even
\\ 
\end{tabular}
&
\begin{tabular}{l}
$
\mathbb{Z}
_{\gcd \{N_{0}/2,2N_{i}\}}\times 
\mathbb{Z}
_{\gcd \{N_{0}/2,N_{i}\}/2}$ \\
\end{tabular}
\cr
\hline

\begin{tabular}{l}
$C_{ij}$ \\
\end{tabular}
&
\begin{tabular}{l}
If $m$ is odd and $N_{i}=N_{j}=2$
\\ 
\end{tabular}
&
\begin{tabular}{l}
$
\mathbb{Z}
_{2}\times 
\mathbb{Z}
_{2}$  \\ 
\end{tabular}
\cr
\begin{tabular}{l}
  \\
\end{tabular}
&
\begin{tabular}{l}
If $m$ is odd and  $N_{i}=2$, $N_{j}=4$
\\ 
\end{tabular}
&
\begin{tabular}{l}
$
\mathbb{Z}
_{4}\times 
\mathbb{Z}
_{2}$ \\
\end{tabular}
\cr

\begin{tabular}{l}
  \\
\end{tabular}
&
\begin{tabular}{l}
If $m$ is odd and $N_{i}=2$, $N_{j}\geq 8$
\\ 
\end{tabular}
&
\begin{tabular}{l}
$
\mathbb{Z}
_{8}\times 
\mathbb{Z}
_{2}$ \\
\end{tabular}
\cr

\begin{tabular}{l}
  \\
\end{tabular}
&
\begin{tabular}{l}
If $m$ is odd and $4\leq N_{i}\leq N_{j}$
\\ 
\end{tabular}
&
\begin{tabular}{l}
$
\mathbb{Z}
_{\gcd \{2N_{i},N_{j}\}}\times 
\mathbb{Z}
_{\gcd \{2N_{j},N_{i}\}}\times 
\mathbb{Z}
_{2}$\\
\end{tabular}
\cr

\begin{tabular}{l}
  \\
\end{tabular}
&
\begin{tabular}{l}
If $m$ is even
\\ 
\end{tabular}
&
\begin{tabular}{l}
$
\mathbb{Z}
_{\gcd \{2N_{i},N_{j}\}}\times 
\mathbb{Z}
_{\gcd \{2N_{j},N_{i}\}}\times 
\mathbb{Z}
_{\gcd \{N_{0}/2,N_{ij}\}}\times 
\mathbb{Z}
_{N_{0ij}/2}$ \\ 
\end{tabular}
\cr

\hline

\begin{tabular}{l}
$D_{ijk}$ \\
\end{tabular}
&
\begin{tabular}{l}
If $m$ is odd and $N_{i}=N_{j}=N_{k}=2$
\\ 
\end{tabular}
&
\begin{tabular}{l}
$
\mathbb{Z}
_{2}\times 
\mathbb{Z}
_{2}$  \\ 
\end{tabular}
\cr

\begin{tabular}{l}
  \\
\end{tabular}
&
\begin{tabular}{l}
If $m$ is odd, $N_{i}=N_{j}=2$ and $N_{k}\geq 4$
\\ 
\end{tabular}
&
\begin{tabular}{l}
$
\mathbb{Z}
_{4}\times 
\mathbb{Z}
_{2}$ \\
\end{tabular}
\cr

\begin{tabular}{l}
  \\
\end{tabular}
&
\begin{tabular}{l}
If $m$ is odd and otherwise
\\ 
\end{tabular}
&
\begin{tabular}{l}
$
\mathbb{Z}
_{N_{ijk}}\times 
\mathbb{Z}
_{N_{ijk}}\times 
\mathbb{Z}
_{2}$ \\
\end{tabular}
\cr

\begin{tabular}{l}
  \\
\end{tabular}
&
\begin{tabular}{l}
If $m$ is even
\\ 
\end{tabular}
&
\begin{tabular}{l}
$
\mathbb{Z}
_{N_{ijk}}\times 
\mathbb{Z}
_{N_{ijk}}\times 
\mathbb{Z}
_{N_{0ijk}}$\\
\end{tabular}
\cr

 \hline

\begin{tabular}{l}
$E_{ijkl}$ \\
\end{tabular}
&
\begin{tabular}{l}
If $m$ is odd
\\ 
\end{tabular}
&
\begin{tabular}{l}
$\mathbb{Z}
_{N_{ijkl}}$ \\ 
\end{tabular}
\cr

\begin{tabular}{l}
\\
\end{tabular}
&
\begin{tabular}{l}
If $m$ is even
\\ 
\end{tabular}
&
\begin{tabular}{l}
$\mathbb{Z}
_{N_{ijkl}}$ \\ 
\end{tabular}
\cr

\hline
\end{tabular}

\begin{center}
Table I. Classification of 3D FSPT phases with finite unitary Abelian
symmetry groups (For simplicity, we only consider symmetry groups $%
\mathbb{Z}
_{N_{\mu }}$ with $N_{\mu }$ being power of 2, and we assume $N_{i}\leq
N_{j}\leq N_{k}\leq N_{l}$ without loss of generality), where $m=N_{0}/2$ and "gcd" means
the greatest common divisor. $N_{ij}$ denotes for the greatest common divisor of $N_{i}$ and $N_{j}$, similarly for $N_{0ij}$, $N_{ijk}$, $N_{0ijk}$ and $N_{ijkl}$.
\end{center}

\end{widetext}

\subsection{Classification of FSPT phases via three-loop braiding statistics}

The main purpose of this work is to obtain a classification of 3D FSPT phases via three-loop braiding statistics, and to study non-Abelian three-loop braiding statistics of gauged FSPT phases. We focus on finite Abelian groups of unitary symmetries, which can generally be written as Eq. (\ref{symmetry}).

We start by defining a set of 3D
topological invariants $\{\Theta _{\mu ,\sigma },\Theta _{\mu \nu ,\sigma
},\Theta _{\mu \nu \lambda ,\sigma }\}$ through the three-loop braiding processes (see method section for full details).  Our definitions are very similar to those for 2D FSPTs given in Ref.~\onlinecite{2DFSPTbraiding}, which actually can be related by dimension reduction\cite{topoinvar15}. 
Next, we  find 14 constraints on $\{\Theta _{\mu ,\sigma },\Theta _{\mu \nu ,\sigma
},\Theta _{\mu \nu \lambda ,\sigma }\}$, listed in method section.  Out of these constraints, 7 follow directly from 2D constraints\cite{2DFSPTbraiding}, while the
other 7 are intrinsically 3D. All intrinsically 3D constraints can be traced back to either the 3D Abelian case\cite{threeloop2018} or 3D
non-Abelian bosonic case\cite{topoinvar15}. Unfortunately, we are not able to prove all the constraints; those we can prove are discussed in Appendix. Finally, by solving the constraints, we obtain a classification of 3D FSPT phases in Table I.  
The classification group $H_{stack}$ under the stacking operation has the following general form:
\begin{equation}
H_{stack}=A\times \underset{i}{\prod }B_{i}\times \underset{i<j}{\prod }
C_{ij}\times \!\! \underset{i<j<k}{\prod } \!\!D_{ijk}\times \!\!\!\!\underset{i<j<k<l}{%
\prod }\!\!E_{ijkl} \label{H1}
\end{equation}
where $i,j,k,l$ take values in $1,2,...,K$, and $%
A,B_{i},C_{ij},D_{ijk},E_{ijkl}$ are finite Abelian groups. 
This classification is one of the main results. While it is obtained from a set of partially conjectured constraints, it agrees with all previously known examples. This justifies the validity of the classification. Below we discuss more details for the stacking group structure of the classification results.

According to the stacking group Eq.(\ref{H1}) for classifying 3D FSPT phases with Abelian total symmetry $G^f$,
we can divide the corresponding topological invariants into five categories, such that the
topological invariants in each category are independent of those in other
categories, i.e. the constraints only relate topological invariants inside
each category. The five categories are:
\\

(A) $\Theta _{0,0}$, $\Theta _{00,0}$, $\Theta _{000,0}$

(B) \ (B1) $\Theta _{0,i}$, $\Theta _{00,i}$, $\Theta _{000,i}$

\qquad (B2) $\Theta _{i,0}$, $\Theta _{0i,0}$, $\Theta _{ii,0}$, $\Theta
_{00i,0}$, $\Theta _{0ii,0}$, $\Theta _{iii,0}$

\qquad (B3) $\Theta _{i,i}$, $\Theta _{0i,i}$, $\Theta _{ii,i}$, $\Theta
_{00i,i}$, $\Theta _{0ii,i}$, $\Theta _{iii,i}$

(C) \ (C1) $\Theta _{ij,0}$, $\Theta _{0ij,0}$, $\Theta _{iij,0}$, $\Theta
_{jji,0}$

\qquad (C2) $\Theta _{ij,i}$, $\Theta _{0ij,i}$, $\Theta _{iij,i}$, $\Theta
_{jji,i}$

\qquad (C3) $\Theta _{ij,j}$, $\Theta _{0ij,j}$, $\Theta _{iij,j}$, $\Theta
_{jji,j}$

\qquad (C4) $\Theta _{i,j}$, $\Theta _{0i,j}$, $\Theta _{ii,j}$, $\Theta
_{00i,j}$, $\Theta _{0ii,j}$, $\Theta _{iii,j}$

\qquad (C5) $\Theta _{j,i}$, $\Theta _{0j,i}$, $\Theta _{jj,i}$, $\Theta
_{00j,i}$, $\Theta _{0jj,i}$, $\Theta _{jjj,i}$

(D) (D1) $\Theta _{ij,k}$, $\Theta _{0ij,k}$, $\Theta _{iij,k}$, $\Theta
_{jji,k}$

\qquad (D2) $\Theta _{jk,i}$, $\Theta _{0jk,i}$, $\Theta _{jjk,i}$, $\Theta
_{kkj,i}$

\qquad (D3) $\Theta _{ki,j}$, $\Theta _{0ki,j}$, $\Theta _{kki,j}$, $\Theta
_{iik,j}$

\qquad (D4) $\Theta _{ijk,0}$, $\Theta _{ijk,i,}$, $\Theta _{ijk,j}$, $%
\Theta _{ijk,k}$

(E) $\Theta _{ijk,l}$, $\Theta _{jkl,i}$, $\Theta _{kli,j}$, $\Theta
_{lij,k} $
\\

where $A$ is the classification group protected by the symmetry group $%
\mathbb{Z}
_{N_{0}}^{f}$, $B_{i}$ is protected by $%
\mathbb{Z}
_{N_{0}}^{f}$ and $%
\mathbb{Z}
_{N_{i}}$, $C_{ij}$ is protected by $%
\mathbb{Z}
_{N_{0}}^{f},%
\mathbb{Z}
_{N_{i}},%
\mathbb{Z}
_{N_{j}}$, $D_{ijk}$ is protected by $%
\mathbb{Z}
_{N_{0}}^{f},%
\mathbb{Z}
_{N_{i}},%
\mathbb{Z}
_{N_{j}},%
\mathbb{Z}
_{N_{k}}$, and $E_{ijkl}$ is protected by $%
\mathbb{Z}
_{N_{i}},%
\mathbb{Z}
_{N_{j}},%
\mathbb{Z}
_{N_{k}},%
\mathbb{Z}
_{N_{l}}$. {We note that in the classification of 3D FSPT phases, the classification group $A$ is always trivial,. However, $A$ is nontrivial for 2D FSPT phases. A newly involved 3D constraint Eq. (\ref{c11})(see method section for more details) trivializes it in 3D.}

{
$H_{stack}$ is exactly our classification group of 3D FSPT phases.  The total number of FSPT phases is given by the order of the group, $\lvert H_{stack} \rvert$.  And it is called "stacking" group because the topological invariants are additive under stacking operations: if we consider two FSPT phases with the values of the topological invariants being $\{\Theta _{\mu ,\sigma }^a,\Theta
_{\mu \nu ,\sigma }^a,\Theta _{\mu \nu \lambda ,\sigma }^a\}$ and
$\{\Theta _{\mu ,\sigma }^b,\Theta
_{\mu \nu ,\sigma }^b,\Theta _{\mu \nu \lambda ,\sigma }^b\}$ respectively, the values of the topological invariants for the new phase obtained by stacking them are given by
}${
\{\Theta _{\mu ,\sigma }^a+\Theta _{\mu ,\sigma }^b,
\Theta_{\mu \nu ,\sigma }^a+\Theta_{\mu \nu ,\sigma }^b,
\Theta _{\mu \nu \lambda ,\sigma }^a+\Theta _{\mu \nu \lambda ,\sigma }^b\}.}$

{
Clearly, $H_{stack}$ is an Abelian group. It satisfies the following group properties: (1) "Identity": there exists a trivial phase, the conventional atomic insulators; (2) "group multiplication": stacking two FSPT phases, we obtain a new phase; (3) "Inverse": given an FSPT phase, there exists an inverse phase, such that stacking the two produces the trivial phase. }

We believe the the topological invariants are complete for characterizing FSPT phases with Abelian symmetry group $G_f$, and the constraints are complete so that all solutions are physical. Both completenesses are justified by a comparison with the general group super-cohomology method in Appendix. 

\subsection{Statistics-type indicators}
Our exploration of the classification scheme also uncovers several new kinds of non-Abelian loop braiding statistics, in particular the new kind that involves Majorana zero modes (Fig. \ref{Fig2}), which we have briefly mentioned in the introduction. In fact, the correspondence between the layer construction in
Refs.~\onlinecite{WangGu2017,WangGu2018} and the three-loop braiding statistics data can be extracted. More explicitly, we pick out several special topological invariants, named \textit{statistics-type indicators}, to indicate non-Abelian loop braiding statistics with different origins: 




(1) $\Theta _{00i,j}=\pi $ ($m$ is odd) is the indicator of the non-Abelian
statistics in the Majorana-chain layer, which is generated by the loops
carrying unpaired Majorana modes, and a loop carrying one Majorana mode is
characterized by its quantum dimension $\sqrt{2}$.

(2) $\Theta _{fi,j}=\frac{N_{0i}}{\gcd (2,N_{i})}\Theta _{0i,j}\neq 0$ ($%
i\neq j$) is the indicator of the complex fermion layer, where "$f$" stands
for the fermion-parity loop $\xi _{f}$\ with gauge flux $\phi _{\xi
_{f}}=(\pi ,0,...)$.

(3) $\Theta _{fij,k}=\Theta _{iij,k}=m\Theta _{0ij,k}\neq 0$ is the
indicator of the non-Abelian statistics in the complex fermion layer, which
is generated by\textbf{\ }degeneracies in the complex fermion layer and the
relevant loops have integer quantum dimension.

(4) $\Theta _{ijk,l}\neq 0$ or $\{\Theta _{fij,k}=0,\Theta _{0ij,k}\neq 0\}$
is the indicator of the non-Abelian statistics in the BSPT layer, which is
generated by degeneracies in the BSPT layer and the relevant loops have
integer quantum dimension.

\begin{figure}[h]
\centering
\includegraphics[
width=0.5\textwidth]{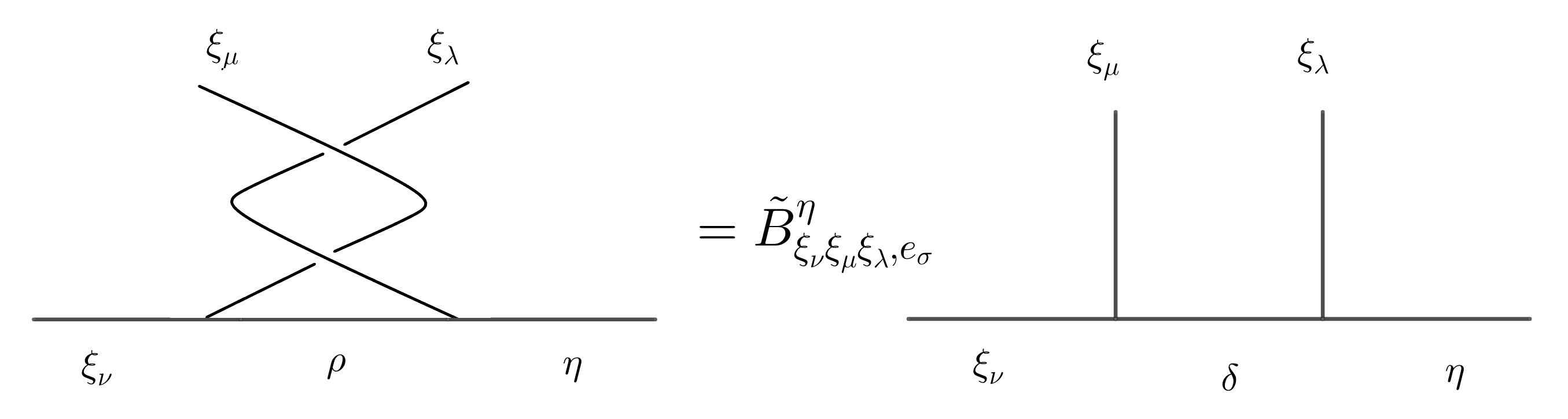}
\caption{The diagram expression of $\widetilde{B}_{\xi _{\nu }\xi _{\mu }\xi_{\lambda }, e_{\sigma }}^{\eta }$ in the standard basis.}
\label{Fig3}
\end{figure}

\begin{figure}[h]
\centering
\includegraphics[
width=0.5\textwidth]{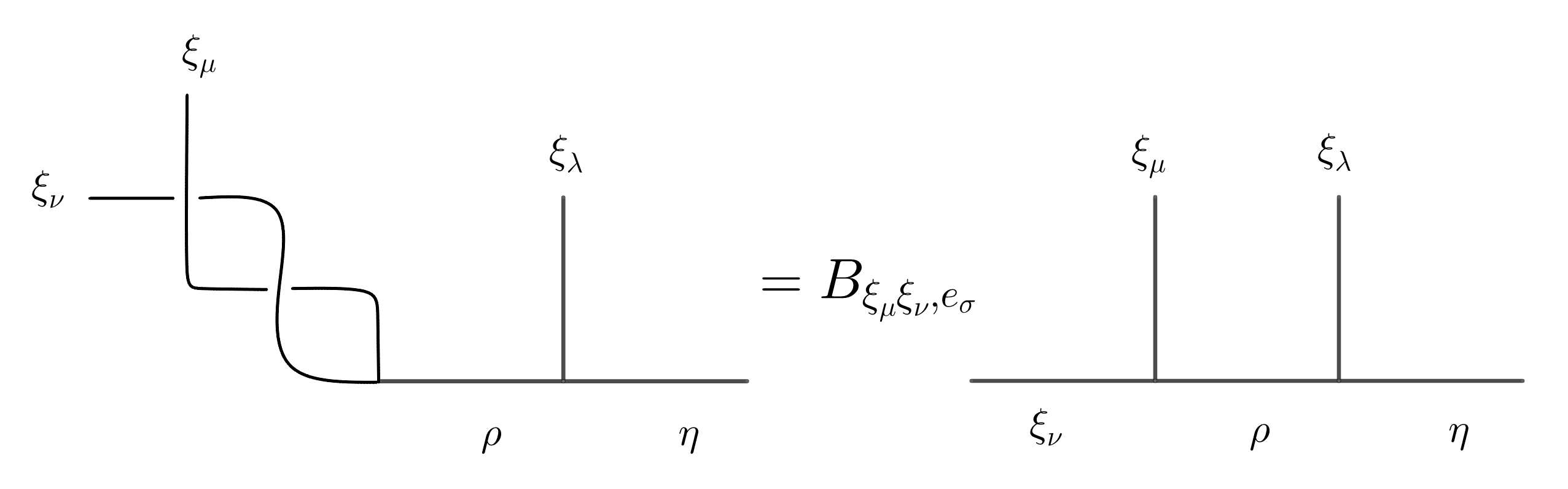}
\caption{The diagram expression of $B_{\xi _{\mu }\xi _{\nu },e_{\sigma }}$ in
the standard basis.}
\label{Fig4}
\end{figure}

We then prove that the first statistics-type indicator $\Theta _{00i,j}=\pi $
($m$ is odd) uniquely indicates the Majorana-chain layer. To proceed, we need to obtain an explicit expression of the topological invariant $\Theta _{\mu \nu \lambda ,\sigma }$ as the following (The definitions we used below are introduced in Appendix:

We assume three loops $\xi _{\mu },\xi _{\nu },\xi
_{\lambda }$ all linked to a base loop $\xi _{\sigma }$. Mathematically, let the total fusion outcome $\eta 
$ of the three loops $\xi _{\mu },\xi _{\nu },\xi _{\lambda }$ be fixed, and
the standard basis is to let $\xi _{\nu }$ firstly fuse with $\xi _{\mu }$,
then their fusion channel again fuse with $\xi _{\lambda }$. We choose the
basis of the first local fusion space $V_{\xi _{\nu }\xi _{\mu },c} $ to be diagonalized under the braiding of $\xi _{\mu }$ around $\xi
_{\nu }$, and the braiding of $\xi _{\mu }$ around $\xi _{\lambda }$ is then
generally non-diagonalized under this basis, which is expressed as:
\begin{widetext}
\begin{equation}
\widetilde{B}_{\xi _{\nu }\xi _{\mu }\xi _{\lambda },e_{\sigma }}^{\eta
}=F_{\xi _{\nu }\xi _{\mu }\xi _{\lambda },e_{\sigma }}^{\eta }B_{\xi _{\mu
}\xi _{\lambda },e_{\sigma }}(F_{\xi _{\nu }\xi _{\mu }\xi _{\lambda
},e_{\sigma }}^{\eta })^{-1}:\underset{\delta }{\oplus }(V_{\xi _{\nu }\xi
_{\mu },c}^{\delta }\otimes V_{\delta \xi _{\lambda },c}^{\eta })\rightarrow 
\underset{\rho }{\oplus }(V_{\xi _{\nu }\xi _{\mu },c}^{\rho }\otimes
V_{\rho \xi _{\lambda },c}^{\eta })\label{invar3}
\end{equation}%
\end{widetext}
where $\widetilde{B}_{\xi _{\nu }\xi _{\mu }\xi _{\lambda },e_{\sigma }}$ only braids $\xi _{\mu}$ around $\xi _{\lambda}$, while it depends on $\xi _{\nu}$, as shown in Fig. \ref{Fig3}. And $B_{\xi _{\mu }\xi _{\nu },e_{\sigma }}$ is redefined
in the same basis as $\widetilde{B}_{\xi _{\nu }\xi _{\mu }\xi _{\lambda
},e_{\sigma }}^{\eta }$:
\begin{equation}
B_{\xi _{\mu }\xi _{\nu },e_{\sigma }}:\underset{\delta }{\oplus }(V_{\xi
_{\nu }\xi _{\mu },c}^{\delta }\otimes V_{\delta \xi _{\lambda },c}^{\eta
})\rightarrow \underset{\delta }{\oplus }(V_{\xi _{\nu }\xi _{\mu
},c}^{\delta }\otimes V_{\delta \xi _{\lambda },c}^{\eta })\label{invar4}
\end{equation}%
which has the same expression as $B_{\xi _{\mu }\xi _{\nu },e_{\sigma }}:%
\underset{\delta }{\oplus }V_{\xi _{\nu }\xi _{\mu },c}^{\delta }\rightarrow 
\underset{\delta }{\oplus }V_{\xi _{\nu }\xi _{\mu },c}^{\delta }$, as
though the fusion space is extended, the basis in the extended fusion space
should keep diagonalized under the braiding of $\xi _{\mu }$ around $\xi
_{\nu }$, as shown in Fig. \ref{Fig4}.
Then $\Theta _{\mu \nu \lambda ,\sigma }$ can
be expressed through:
\begin{equation}
e^{i\Theta _{\mu \nu \lambda ,\sigma }}I=(\widetilde{B}_{\xi _{\nu }\xi
_{\mu }\xi _{\lambda },e_{\sigma }}^{\eta })^{-1}(B_{\xi _{\mu }\xi _{\nu
},e_{\sigma }})^{-1}\widetilde{B}_{\xi _{\nu }\xi _{\mu }\xi _{\lambda
},e_{\sigma }}^{\eta }B_{\xi _{\mu }\xi _{\nu },e_{\sigma }}\label{invar5}
\end{equation}%
where $I$ is the identity matrix in the vector space $\underset{\delta }{%
\oplus }(V_{\xi _{\nu }\xi _{\mu },c}^{\delta }\otimes V_{\delta \xi
_{\lambda },c}^{\eta })$.

Now we are ready to go back to the proof. {For simplicity we can consider $m=1$ only, which is due to $\mathbb{Z}_{2m}^f$ is isomorphic to $\mathbb{Z}_{2}^f \times \mathbb{Z}_{m}$, and $\mathbb{Z}_{m}$ can be absorbed into $\prod_{i}\mathbb{Z}_{N_i}$ part of $G_{f}$.  From constraints Eq. (\ref{c1}) and Eq. (\ref{c14}) in method section, $\Theta _{iii,j}=m\Theta_{0ii,j}=m^2\Theta_{00i,j}$. Therefore when $m=1$, we have the relation $\Theta_{00i,j}=\Theta _{iii,j}$. }

Firstly we show that the non-Abelian statistics
in Majorana-chain layer (i.e. the Ising type statistics) must have $\Theta
_{00i,j}=\pi $: Do a dimension reduction for the gauged Ising type FSPT
system from 3D to 2D by choosing $\xi _{j}$ as the base loop, and condense
all the bosonic quasiparticles (as the Ising type statistics is irrelavant
to the bosonic matter). The remaining 2D quasiparticles are exactly the
Ising anyons: a vortex carrying one majorana mode $\sigma $, a fermion $\psi $
and vacuum $1$, which satisfies:
\begin{equation}
e^{i\Theta _{\sigma \sigma \sigma }}I=(\widetilde{B}_{_{\sigma \sigma \sigma
}}^{\sigma })^{-1}(B_{_{\sigma \sigma }})^{-1}\widetilde{B}_{_{\sigma \sigma
\sigma }}^{\sigma }B_{_{\sigma \sigma }}=e^{i\pi }\left[ 
\begin{array}{cc}
1 & 0 \\ 
0 & 1%
\end{array}%
\right] \label{Ising1}
\end{equation}
where we have $\Theta _{\sigma \sigma \sigma }=\pi $. And the 3D topological
invariants $\Theta _{iii,j}$ is exactly equal to the 2D one $\Theta _{\sigma
\sigma \sigma }$ after dimension reduction and the condensation of all
bosons, i.e. $\Theta _{00i,j}=\Theta _{iii,j}=\Theta _{\sigma \sigma \sigma
}=\pi $. Secondly we show that $\Theta _{00i,j}=\pi $ corresponds uniquely
to the Ising type statistics: From constraint Eq. (\ref{c1}) in method section, when $m=1$, $\Theta
_{00i,j}$ can only take values $0$ or $\pi $; when $m$ is even , $\Theta
_{00i,j}$ vanishes. We assume the types of non-Abelian statistics in our
gauged FSPT\ system contain only: (1) Ising type in Majorana-chain layer (2)
fermionic type in complex fermion layer (3) bosonic type in BSPT\ layer. Solving the constraints as listed in Appendix, and examing the generating phases by
mapping to 2D model constructions after dimension reduction\cite{2DFSPTbraiding}, we find
that to construct the generating phase $\Theta _{00i,j}=\pi $,
there always exist loops with quantum dimension $\sqrt{2}$, which
is the unique property of Ising anyons in the Majorana-chain layer.

The second statistics-type indicator $\Theta _{fi,j}=\frac{N_{0i}}{\gcd
(2,N_{i})}\Theta _{0i,j}\neq 0$ ($i\neq j$) is proposed and proven in Ref.\onlinecite{threeloop2018}. Combining the results in Ref.\onlinecite{2DFSPTbraiding} and Ref.\onlinecite{threeloop2018}, we infer that $\Theta
_{fij,k}=\Theta _{iij,k}=m\Theta _{0ij,k}\neq 0$ is the indicator for the
non-Abelian statistics in the complex fermion layer. Finally $\Theta
_{ijk,l}\neq 0$ is obviously the indicator for the BSPT layer by the
definition of the topological invariant $\Theta _{\mu \nu \lambda ,\sigma }$%
. However, if we consider a special example $G_{f}=%
\mathbb{Z}
_{4}^{f}\times 
\mathbb{Z}
_{2}\times 
\mathbb{Z}
_{2}\times 
\mathbb{Z}
_{2}$, where $\Theta _{0ij,k}=\pi $\ and $\Theta _{fij,k}=2\Theta _{0ij,k}=0$
should still belong to the non-Abelian statistics in BSPT layer. Hence we
conclude that $\Theta _{ijk,l}\neq 0$ and $\{\Theta _{fij,k}=0,\Theta
_{0ij,k}\neq 0\}$ are both the indicators for the non-Abelian statistics in
BSPT layer.

By checking the linear dependence among the topological invariants, we can also determine relations between the three layers, i.e., simply stacked or absorbed.  {We summarize the group structure of our classification result by layers, i.e., classification corresponds to BSPT phase, complex fermion layer and Majorana chain layer, and whether they are non-trivial group extension (we call absorbed) or simple direct product (we call stacking), in Table II.}

Furthermore, invoking the known model construction for 2D FSPT
phases\cite{2DFSPTbraiding} and by the fact that quantum dimensions are invariant under
dimension reduction, we can find the quantum dimensions of loop-like
excitations linked to certain base loops. From the quantum dimensions, we can further show that the non-Abelian three-loop braiding statistics resulting from the Majorana chain layer is due to the unpaired Majorana modes attached to linked loops. Below we will discuss two simplest examples for such kinds of non-Abelian three-loop braiding statistics. 

\begin{widetext}

\begin{scriptsize}

\begin{tabular}{|l|l|l|l|l|l|l|}
\hline
                            & Cases                                                   & BSPT $\mathcal{B}$                                                                                                                       & Cases                             & \begin{tabular}[c]{@{}l@{}}Complex\\ fermion $\mathcal{C}$\end{tabular} & \begin{tabular}[c]{@{}l@{}}Kitaev\\ chain $\mathcal{K}$\end{tabular} & Group structure                                                       \\ \hline
\multirow{2}{*}{$B_{i}$}    & \multirow{2}{*}{If $m$ is even}                         & \multirow{2}{*}{$\mathbb{Z}_{\gcd \{N_{0}/2,N_{i}\}}\times \mathbb{Z}_{\gcd \{N_{0}/2,N_{i}\}/2}$}                                       & If $N_{i}\geq N_{0}/2$            & $\mathbb{Z}_1$                                                          & $\mathbb{Z}_1$                                                       & $\mathcal{B}$                                                         \\ \cline{4-7} 
                            &                                                         &                                                                                                                                          & If $N_{i}< N_{0}/2$               & $\mathbb{Z}_2$                                                          & $\mathbb{Z}_1$                                                       & $\mathcal{C}\ltimes \mathcal{B}$                                      \\ \hline
\multirow{7}{*}{$C_{ij}$}   & If $m$ is odd, $N_{i}=N_{j}=2$                          & $\mathbb{Z}_{2}\times \mathbb{Z}_{2}$                                                                                                    & \textbackslash{}                  & $\mathbb{Z}_1$                                                          & $\mathbb{Z}_1$                                                       & $\mathcal{B}$                                                         \\ \cline{2-7} 
                            & If $m$ is odd, $N_{i}=2$, $N_{j}=4$                     & $\mathbb{Z}_{2}\times \mathbb{Z}_{2}$                                                                                                    & \textbackslash{}                  & $\mathbb{Z}_2$                                                          & $\mathbb{Z}_1$                                                       & $\mathcal{C}\ltimes \mathcal{B}$                                      \\ \cline{2-7} 
                            & If $m$ is odd, $N_{i}=2$, $N_{j}\geq 8$                 & $\mathbb{Z}_{2}\times \mathbb{Z}_{2}$                                                                                                    & \textbackslash{}                  & $\mathbb{Z}_2$                                                          & $\mathbb{Z}_2$                                                       & ${\mathcal{K}}\ltimes \mathcal{C}\ltimes \mathcal{B}$  \\ \cline{2-7} 
                            & \multirow{2}{*}{If $m$ is odd, $4\leq N_{i}\leq N_{j}$} & \multirow{2}{*}{$\mathbb{Z}_{N_{ij}}\times \mathbb{Z}_{N_{ij}}$}                                                                         & If $N_{i}=N_{j}$                  & $\mathbb{Z}_1$                                                          & $\mathbb{Z}_2$                                                       & $\mathcal{B}\times{\mathcal{K}}$                       \\ \cline{4-7} 
                            &                                                         &                                                                                                                                          & If $N_{i}<N_{j}$                  & $\mathbb{Z}_2$                                                          & $\mathbb{Z}_2$                                                       & $(\mathcal{C}\ltimes \mathcal{B})\times {\mathcal{K}}$ \\ \cline{2-7} 
                            & \multirow{2}{*}{If $m$ is even}                         & \multirow{2}{*}{$\mathbb{Z}_{N_{ij}}\times \mathbb{Z}_{N_{ij}}\times \mathbb{Z}_{\gcd \{N_{0}/2,N_{ij}\}}\times \mathbb{Z}_{N_{0ij}/2}$} & If $N_{i}=N_{j}$                  & $\mathbb{Z}_1$                                                          & $\mathbb{Z}_1$                                                       & $\mathcal{B}$                                                         \\ \cline{4-7} 
                            &                                                         &                                                                                                                                          & If $N_{i}<N_{j}$                  & $\mathbb{Z}_2$                                                          & $\mathbb{Z}_1$                                                       & $\mathcal{C}\ltimes \mathcal{B}$                                      \\ \hline
\multirow{5}{*}{$D_{ijk}$}  & If $m$ is odd, $N_{i}=N_{j}=N_{k}=2$                    & $\mathbb{Z}_{2}\times \mathbb{Z}_{2}$                                                                                                    & \textbackslash{}                  & $\mathbb{Z}_1$                                                          & $\mathbb{Z}_1$                                                       & $\mathcal{B}$                                                         \\ \cline{2-7} 
                            & If $m$ is odd, $N_{i}=N_{j}=2$ and $N_{k}\geq 4$        & $\mathbb{Z}_{2}\times \mathbb{Z}_{2}$                                                                                                    & \textbackslash{}                  & $\mathbb{Z}_2$                                                          & $\mathbb{Z}_1$                                                       & ${\mathcal{C}}\ltimes \mathcal{B}$                     \\ \cline{2-7} 
                            & If $m$ is odd and otherwise                             & $\mathbb{Z}_{N_{ijk}}\times \mathbb{Z}_{N_{ijk}}$                                                                                        & \textbackslash{}                  & $\mathbb{Z}_2$                                                          & $\mathbb{Z}_1$                                                       & ${\mathcal{C}}\times \mathcal{B}$                      \\ \cline{2-7} 
                            & If $m$ is even, $N_{0ijk}\neq N_{0}$                    & $\mathbb{Z}_{N_{ijk}}\times \mathbb{Z}_{N_{ijk}}\times \mathbb{Z}_{N_{0ijk}}$                                                            & \textbackslash{}                  & $\mathbb{Z}_1$                                                          & $\mathbb{Z}_1$                                                       & $\mathcal{B}$                                                         \\ \cline{2-7} 
                            & If $m$ is even, $N_{0ijk}= N_{0}$                       & $\mathbb{Z}_{N_{ijk}}\times \mathbb{Z}_{N_{ijk}}\times \mathbb{Z}_{N_{0ijk}/2}$                                                          & \textbackslash{}                  & $\mathbb{Z}_2$                                                          & $\mathbb{Z}_1$                                                       & ${\mathcal{C}}\ltimes \mathcal{B}$                     \\ \hline
\multirow{2}{*}{$E_{ijkl}$} & If $m$ is odd                                           & \multirow{2}{*}{$\mathbb{Z}_{N_{ijkl}}$}                                                                                                 & \multirow{2}{*}{\textbackslash{}} & \multirow{2}{*}{$\mathbb{Z}_1$}                                         & \multirow{2}{*}{$\mathbb{Z}_1$}                                      & \multirow{2}{*}{$\mathcal{B}$}                                        \\ \cline{2-2}
                            & If $m$ is even                                          &                                                                                                                                          &                                   &                                                                         &                                                                      &                                                                       \\ \hline
\end{tabular}

\end{scriptsize}

\begin{center}
{
Table II. Layer group structure of the classification group of 3D FSPT phases with finite unitary Abelian
symmetry groups (We assume $N_{i}\leq
N_{j}\leq N_{k}\leq N_{l}$ without loss of generality). The classification groups of BSPT layer, complex layer and Majorana chain layer are denoted as $\mathcal{B}$, $\mathcal{C}$ and $\mathcal{K}$ respectively. As the group structure depends on further cases beyond Table I, we list the further cases in column 4. We denote the simple direct product as $\times$, and non-trivial group extension as $\ltimes$. And the classification groups with non-Abelian braiding statistics are denoted in red color.}
\end{center}

\end{widetext}

\subsection{Simplest examples for non-Abelian three-loop braiding statistics}
\subsubsection{$G^f=
\mathbb{Z}
_{2}^{f}\times 
\mathbb{Z}
_{2}\times 
\mathbb{Z}
_{8}$}

Firstly, we recall the stacking group classification of FSPT phases:
\begin{equation}
H_{stack}=A\times \underset{i}{\prod }B_{i}\times \underset{i<j}{\prod }%
C_{ij} \label{H3}
\end{equation}
where from Table I we know that: $A$ protected by $%
\mathbb{Z}
_{2}^{f}$ is trivial, $B_{1}$ and $B_{2}$\ protected by $%
\mathbb{Z}
_{2}^{f}\times 
\mathbb{Z}
_{2}$ and $%
\mathbb{Z}
_{2}^{f}\times 
\mathbb{Z}
_{8}$ respectively are trivial, while $C_{12}$ protected by $%
\mathbb{Z}
_{2}^{f}\times 
\mathbb{Z}
_{2}\times 
\mathbb{Z}
_{8}$ is nontrivial. Therefore the classification of FSPT phases for the
symmetry group is $H_{stack}=C_{12}$. Then we explicitly show the
calculation of $C_{12}$: Invoking the known 2D results and combining with
the 3D constraints $N_{\sigma }\Theta _{\mu ,\sigma }=0$, $N_{\sigma }\Theta
_{\mu \nu ,\sigma }=0$, $N_{\sigma }\Theta _{\mu \nu \lambda ,\sigma }=0$,
the generating phases for the subsets (C1), (C2), (C3), (C4) and (C5) are:
\begin{equation}
(\Theta _{ij,0},\Theta _{0ij,0})=(\frac{2\pi }{N_{ij}},0)\times a+(0,\frac{%
2\pi }{N_{0ij}})\times b=(\pi a,\pi b)
\end{equation}
\begin{equation}
(\Theta _{ij,i},\Theta _{0ij,i})=(\frac{2\pi }{N_{ij}},0)\times c+(0,\frac{%
2\pi }{N_{0ij}})\times d=(\pi c,\pi d)
\end{equation}
\begin{equation}
(\Theta _{ij,j},\Theta _{0ij,j})=(\frac{2\pi }{N_{ij}},0)\times e+(0,\frac{%
2\pi }{N_{0ij}})\times f=(\pi e,\pi f)
\end{equation}
\begin{eqnarray}
(\Theta _{i,j},\Theta _{0i,j},\Theta _{00i,j})&=&(\frac{\pi }{2N_{i}},-\frac{%
\pi }{N_{0i}},\pi )\times g+(0,\frac{4\pi }{N_{0i}},0)\nonumber\\ &=&(\frac{\pi }{4},-%
\frac{\pi }{2},\pi )g
\end{eqnarray}
\begin{eqnarray}
(\Theta _{j,i},\Theta _{0j,i},\Theta _{00j,i})&=&(\frac{\pi }{N_{j}},\frac{%
2\pi }{N_{0j}},0)\times N_{j}h+(0,\frac{4\pi }{N_{0j}},\pi )\times i\nonumber\\ &=&(\pi
h,0,\pi i)
\end{eqnarray}%
where $a,b,c,d,e,f,g,h,i$ are all integers. By the constraint $\Theta
_{0ij,0}=\Theta _{00i,j}=-\Theta _{0ij,i}=-\Theta _{00j,i}=-\Theta _{0ij,j}$%
, we have $b=d=f=g=i$ (mod 2). By the constraint $\Theta _{ij,0}+\Theta
_{oj,i}+4\Theta _{0i,j}=0$, we have $a=0$ (mod 2). By the constraint $\Theta
_{ij,i}=-4\Theta _{i,j}$, we have $c=-g$ (mod 8). By the constraint $\Theta
_{ij,j}=-\Theta _{j,i}$, we have $e=-h$ (mod 2).

Combining all the constraints: $a=0$ (mod 2)$,b=d=f=g=i=-c$ (mod 8), $e=-h$
(mod 2), i.e. the generating phases are:
\begin{eqnarray}
&&(\Theta _{0ij,0},\Theta _{ij,i},\Theta _{0ij,i},\Theta _{0ij,j},\Theta
_{i,j},\Theta _{0i,j},\Theta _{00i,j},\Theta _{00j,i})\nonumber\\&=&(\pi ,\pi ,\pi ,\pi ,%
\frac{\pi }{4},-\frac{\pi }{2},\pi ,\pi )
\end{eqnarray}
\begin{equation}
(\Theta _{ij,j},\Theta _{j,i})=(\pi ,\pi )
\end{equation}%
while all other topological invariants vanish:
\begin{equation}
\Theta _{0,0}=0
\end{equation}
\begin{eqnarray}
&&(\Theta _{0,i},\Theta _{i,0},\Theta _{0i,0},\Theta _{00i,0},\Theta
_{i,i},\Theta _{0i,i},\Theta _{00i,i})
\nonumber\\&=&(0,0,0,0,0,0,0)
\end{eqnarray}
\begin{eqnarray}
&&(\Theta _{0,j},\Theta _{j,0},\Theta _{0j,0},\Theta _{00j,0},\Theta
_{j,j},\Theta _{0j,j},\Theta _{00j,j})
\nonumber\\&=&(0,0,0,0,0,0,0)
\end{eqnarray}
\begin{equation}
(\Theta _{ij,0},\Theta _{0j,i})=(0,0)
\end{equation}

Hence in this case the classification is $%
\mathbb{Z}
_{8}\times 
\mathbb{Z}
_{2}$, which is a $%
\mathbb{Z}
_{2}$ complex fermion layer absorbed into a $%
\mathbb{Z}
_{2}\times 
\mathbb{Z}
_{2}$\ BSPT layer, together\ forming a $%
\mathbb{Z}
_{4}\times 
\mathbb{Z}
_{2}$ classification, and then a $%
\mathbb{Z}
_{2}$ Majorana-chain layer again absorbed into the $%
\mathbb{Z}
_{4}\times 
\mathbb{Z}
_{2}$ above, as the complex fermion layer indicator is $\Theta
_{fi,j}=\Theta _{0i,j}=-\frac{\pi }{2}$.

Conveniently we can view the "$%
\mathbb{Z}
_{8}$" part of the classification being generated by:
\begin{equation}
\Theta _{i,j}=\{\frac{\pi }{4},\frac{\pi }{2},\frac{3}{4}\pi ,\pi ,\frac{5}{4%
}\pi ,\frac{3}{2}\pi ,\frac{7}{4}\pi ,0\}
\end{equation}%
where $\Theta _{i,j}=\{0,\pi \}$ correspond to Abelian BSPT phases, $\Theta _{i,j}=\{%
\frac{\pi }{2},\frac{3}{2}\pi \}$ are Abelian FSPT phases (contain both BSPT
layer and complex fermion layer), and $\Theta _{i,j}=\{\frac{\pi }{4},\frac{3%
}{4}\pi ,\frac{5}{4}\pi ,\frac{7}{4}\pi \}$ are non-Abelian FSPT\ phases
(contain all BSPT layer, complex fermion layer and Majorana chain layer).
Recall that $\Theta _{00i,j}=\pi $ ($m$ is odd) is the indicator of the
Majorana chain layer.\ And the four non-Abelian FSPT phases all have $(\Theta
_{00i,j},\Theta _{00j,i})=(\pi ,\pi )$, which means that loops $\xi _{i}$
and $\xi _{j}$ each carry one unpaired Majorana mode simultaneously and both
have quantum dimension $\sqrt{2}$, which is the origin of the non-Abelian
statistics in Majorana chain layer. On the other hand, the "$%
\mathbb{Z}
_{2}$" part of the classification is generated by:
\begin{equation}
\Theta _{j,i}=\{0,\pi \}
\end{equation}%
where $\Theta _{j,i}=\pi $ is a non-trivial BSPT phase, and $\Theta _{j,i}=0$
is a trivial BSPT phase.

We can also understand the 3D braiding statistics by doing a dimension
reduction from 3D to 2D and applying the known model construction for 2D
generating phases\cite{2DFSPTbraiding}. Firstly we choose $\xi _{j}$ always to be the base
loop, and the 2D system after dimension reduction has symmetry $%
\mathbb{Z}
_{2}^{f}\times 
\mathbb{Z}
_{2}$, which has only one generating phase $(\Theta _{i},\Theta _{0i},\Theta
_{00i})=(\frac{\pi }{4},-\frac{\pi }{2},\pi )$, i.e. the subset (C4) in
category C. And it can be realized by a two-layer model construction: the
first layer $a$ is a charge-2 superconductor with chiral central charge $-%
\frac{1}{2}$ (Ising type), while the second layer $b$ is a charge-2
superconductor with chiral central charge $\frac{1}{2}$ (Ising type). The 2D
vortex $\xi _{0}$ is composited by a unit-flux vortex in layer $a$\ and a
unit-flux vortex in layer $b$, which therefore has quantum dimension $2$.
The 2D vortex $\xi _{i}$ is composited only by a unit-flux vortex in layer $%
b $, which therefore has quantum dimension $\sqrt{2}$. As the quantum
dimensions of loops are invariant under dimension reudction, we conclude
that for non-Abelian FSPT\ phases, with $\xi _{j}$ all being base loops,
loop $\xi _{0}$ has quantum dimension $2$ and loop $\xi _{i}$ has quantum
dimension $\sqrt{2}$.

Secondly we choose $\xi _{i}$ always to be the base loop, and the 2D system
after dimension reduction has symmetry $%
\mathbb{Z}
_{2}^{f}\times 
\mathbb{Z}
_{8}$, which has two generating phases $(\Theta _{j},\Theta _{0j},\Theta
_{00j})=(\frac{\pi }{8},\pi ,0)$ and $(\Theta _{j},\Theta _{0j},\Theta
_{00j})=(0,0,\pi )$, where the first one is\ trivialized to a $%
\mathbb{Z}
_{2}$ BSPT in 3D, and both constitute the subset (C5) in category C. Only
the second generating phase corresponds to non-Abelian statistics and can be
realized by a three-layer model construction: the first layer $a$ is a
charge-2 superconductor with chiral central charge $-\frac{1}{2}$ (Ising
type), the second layer $b$ is a charge-8 superconductor with chiral central
charge $0$ (Abelian layer), and the third layer $c$ is a charge-2
superconductor with chiral central charge $\frac{1}{2}$ (Ising type). The 2D
vortex $\xi _{0}$ is composited by a unit flux in layer $a$, four times of
unit flux in layer $b$, and a unit flux in layer $c$, which therefore has
quantum dimension $2$. The 2D vortex $\xi _{j}$ is composited only by a unit
flux in layer $b$ and a unit flux in layer $c$, which therefore has quantum
dimension $\sqrt{2}$.

Thirdly we do not specify the base loop, and let the 2D system after
dimension reduction have the full symmetry $%
\mathbb{Z}
_{2}^{f}\times 
\mathbb{Z}
_{2}\times 
\mathbb{Z}
_{8}$, which has two generating phases $(\Theta _{ij,0},\Theta
_{0ij,0})=(\pi ,0)$ and $(\Theta _{ij,0},\Theta _{0ij,0})=(0,\pi )$ (or $%
(\Theta _{ij,i},\Theta _{0ij,i})=(\pi ,0)$ and $(\Theta _{ij,i},\Theta
_{0ij,i})=(0,\pi )$, $(\Theta _{ij,0},\Theta _{0ij,0})=(\pi ,0)$ and $%
(\Theta _{ij,0},\Theta _{0ij,0})=(0,\pi )$), i.e. the subset (C1) (or (C2),
(C3)) in category C. Only the second generating phase corresponds to
non-Abelian statistics and can be realized by a four-layer model
construction: the first layer $a$ is a charge-2 superconductor with chiral
central charge $-\frac{1}{2}$ (Ising type), the second layer $b$ is a
charge-2 superconductor with chiral central charge $0$ (Abelian layer), the
third layer $c$ is a charge-8 superconductor with chiral central charge $0$
(Abelian layer), and the fourth layer $d$ is a charge-2 superconductor with
chiral central charge $\frac{1}{2}$ (Ising type). The 2D vortex $\xi _{0}$
is composited by a unit flux in layer $a$, a unit flux in layer $b$, four
times of unit flux in layer $c$, and a unit flux in layer $d$, which
therefore has quantum dimension $2$. The vortex $\xi _{i}$ is composited by
a unit flux in layer $b$ and a unit flux in layer $d$, which therefore has
quantum dimension $\sqrt{2}$. Similarly, the vortex $\xi _{j}$ is composited
by a unit flux in layer $c$ and a unit flux in layer $d$, which also has
quantum dimension $\sqrt{2}$. In conclusion, we find that no matter how we
do the dimension reduction, the quantum dimensions of the loops coincide,
i.e. in our three-loop braiding system with full symmetry $%
\mathbb{Z}
_{2}^{f}\times 
\mathbb{Z}
_{2}\times 
\mathbb{Z}
_{8}$, for those non-Abelian FSPT phases, the loop $\xi _{0}$ has quantum
dimension $2$, and loops $\xi _{i}$ and $\xi _{j}$ both have quantum
dimension $\sqrt{2}$, which means that loops $\xi _{i}$ and $\xi _{j}$ each
carry an unpaired Majorana mode.

\subsubsection{$G^f=%
\mathbb{Z}
_{2}^{f}\times 
\mathbb{Z}
_{4}\times 
\mathbb{Z}
_{4}$}

Similarly in the stacking group classification, $A,B_{1},B_{2}$ are all
trivial, and we only need to consider $H_{stack}=C_{12}$. Invoking the known
2D results and combining with the 3D constraints $N_{\sigma }\Theta _{\mu
,\sigma }=0$, $N_{\sigma }\Theta _{\mu \nu ,\sigma }=0$, $N_{\sigma }\Theta
_{\mu \nu \lambda ,\sigma }=0$, the generating phases for the subsets (C1),
(C2), (C3), (C4) and (C5) are:
\begin{eqnarray}
(\Theta _{ij,0},\Theta _{0ij,0})&=&(\frac{2\pi }{N_{ij}},0)\times \frac{N_{ij}%
}{2}a+(0,\frac{2\pi }{N_{0ij}})\times \frac{N_{0ij}}{2}b
\nonumber\\&=&(\pi a,\pi b)
\end{eqnarray}
\begin{equation}
(\Theta _{ij,i},\Theta _{0ij,i})=(\frac{2\pi }{N_{ij}},0)\times c+(0,\frac{%
2\pi }{N_{0ij}})\times d=(\frac{\pi }{2}c,\pi d)
\end{equation}
\begin{equation}
(\Theta _{ij,j},\Theta _{0ij,j})=(\frac{2\pi }{N_{ij}},0)\times e+(0,\frac{%
2\pi }{N_{0ij}})\times f=(\frac{\pi }{2}e,\pi f)
\end{equation}
\begin{eqnarray}
(\Theta _{i,j},\Theta _{0i,j},\Theta _{00i,j})&=&(\frac{\pi }{N_{i}},\frac{%
2\pi }{N_{0i}},0)\times 2g+(0,\frac{2\pi }{N_{0i}},\pi )\times h \nonumber \\ &=&(\frac{\pi 
}{2}g,\pi h,\pi h)
\end{eqnarray}
\begin{eqnarray}
(\Theta _{j,i},\Theta _{0j,i},\Theta _{00j,i})&=&(\frac{\pi }{N_{j}},\frac{%
2\pi }{N_{0j}},0)\times 2l+(0,\frac{2\pi }{N_{0j}},\pi )\times m\nonumber\\ &=&(\frac{\pi 
}{2}l,\pi m,\pi m)
\end{eqnarray}

By the constraint $\Theta _{0ij,0}=\Theta _{00i,j}=-\Theta _{0ij,i}=-\Theta
_{00j,i}=-\Theta _{0ij,j}$, we have $b=d=f=h=m$ (mod 2). By the constraint $%
\Theta _{ij,0}+\Theta _{0j,i}+\Theta _{0i,j}=0$, we have $a=0$ (mod 2). By
the constraint $\Theta _{ij,i}=-\Theta _{i,j}$, we have $c=-g$ (mod 4). By
the constraint $\Theta _{ij,j}=-\Theta _{j,i}$, we have $e=-l$ (mod 4).

Combining all the constraints: $a=0$ (mod 2), $b=d=f=h=m$ (mod 2), $c=-g$
(mod 4), $e=-l$ (mod 4), i.e. the generating phases are:
\begin{eqnarray}
&&(\Theta _{0ij,0},\Theta _{0ij,i},\Theta _{0ij,j},\Theta _{0i,j},\Theta
_{00i,j},\Theta _{0j,i},\Theta _{00j,i})\nonumber\\&=&(\pi ,\pi ,\pi ,\pi ,\pi ,\pi ,\pi )
\end{eqnarray}
\begin{equation}
(\Theta _{ij,i},\Theta _{i,j})=(\frac{\pi }{2},\frac{\pi }{2})
\end{equation}
\begin{equation}
(\Theta _{ij,j},\Theta _{j,i})=(\frac{\pi }{2},\frac{\pi }{2})
\end{equation}%
while all other topological invariants vanish.

Hence in this case the classification is $%
\mathbb{Z}
_{4}\times 
\mathbb{Z}
_{4}\times 
\mathbb{Z}
_{2}$ , which is a $%
\mathbb{Z}
_{4}\times 
\mathbb{Z}
_{4}$\ BSPT simply stacking with a $%
\mathbb{Z}
_{2}$\ "Majorana chain layer absorbed in complex fermion layer", as the
complex fermion layer indicator is $\Theta _{fi,j}=\Theta _{0i,j}=\pi $.

The "$%
\mathbb{Z}
_{2}$" part of the classification can be viewed to be generated by:
\begin{equation}
(\Theta _{0i,j},\Theta _{00i,j})=(\pi ,\pi )\text{ or }(\Theta
_{0j,i},\Theta _{00j,i})=(\pi ,\pi )
\end{equation}%
while all other $\pi $ valued topological invariants are related by the
anti-symmetric constraint of $\Theta _{\mu \nu \lambda ,\sigma }$. We do a
dimension reduction by always choosing $\xi _{j}$ as the base loop, and the
2D system has symmetry $%
\mathbb{Z}
_{2}^{f}\times 
\mathbb{Z}
_{4}$. We find that $(\Theta _{0i},\Theta _{00i})=(\pi ,\pi )$ is exactly
the second generating phase for this 2D FSPT system, which can be realized
by a three-layer model construction\cite{2DFSPTbraiding}: the first layer $a$ is a charge-2
superconductor with chiral central charge $\frac{3}{2}$ (Ising type), the
second layer $b$ is a charge-4 superconductor with chiral central charge $-2$
(Abelian layer), and the third layer $c$ is a charge-2 superconductor with
chiral central charge $\frac{1}{2}$ (Ising type). The 2D vortex $\xi _{0}$
is composited by a unit flux in layer $a$, two times of unit flux in layer $%
b $, and a unit flux in layer $c$, which therefore has quantum dimension $2$%
. The vortex $\xi _{i}$ is composited by a unit flux in layer $b$ and a unit
flux in layer $c$, which therefore has quantum dimension $\sqrt{2}$. As the
quantum dimensions of the loops are invariant under dimension reduction, and
the symmetry groups of $\xi _{i}$ and $\xi _{j}$ are both $%
\mathbb{Z}
_{4}$ so that it is free to choose which is $%
\mathbb{Z}
_{N_{i}}$ and which is $%
\mathbb{Z}
_{N_{j}}$, we conclude that in our gauged 3D FSPT systems, loop $\xi _{0}$
has quantum dimension $2$ and both loop $\xi _{i}$ and $\xi _{j}$ have
quantum dimension $\sqrt{2}$.

Then we can again check the quantum dimension
of loops by doing the dimension reduction without specifying the base loop,
and the 2D system has the full symmetry $%
\mathbb{Z}
_{2}^{f}\times 
\mathbb{Z}
_{4}\times 
\mathbb{Z}
_{4}$. The second non-Abelian generating phase $(\Theta _{ij,0},\Theta
_{0ij,0})=(0,\pi )$ (or $(\Theta _{ij,i},\Theta _{0ij,i})$, $(\Theta
_{ij,j},\Theta _{0ij,j})$) can also be realized by a four-layer construction
similarly as in the first example. Then the quantum dimension of $\xi _{0}$
will still be found as $2$, and the quantum dimensions of $\xi _{i}$ and $%
\xi _{j}$ as both $\sqrt{2}$. Therefore in our construction the nontrivial
non-Abelian FSPT phase in the $%
\mathbb{Z}
_{2}$ classification is due to the unpaired Majorana modes attached on $\xi
_{i}$ and $\xi _{j}$.

\section{Conclusions and discussions}

In summary, we obtain the classification of 3D FSPT phases with arbitrary finite
unitary Abelian total symmetry $G^f$, by gauging the symmetry and studying the
topological invariants $\{\Theta _{\mu ,\sigma },\Theta _{\mu \nu ,\sigma
},\Theta _{\mu \nu \lambda ,\sigma }\}$ defined through the braiding
statistics of loop-like excitations in certain three-loop braiding
processes and solving the corresponding constraints for these topological invariants. 
We further compare this result with the classification obtained by
the general group supercohomology theory in Ref\cite{WangGu2018} and find a systematical agreement. In particular, we can realize any set of allowed values of topological invariants corresponding to a distinguished FSPT phase. Moreover, from several special topological invariants, we can further identify different origins of Non-Abelian three-loop braiding statistics from the corresponding FSPT constructions, i.e.,  the Majorana chain layer, and complex fermion layer and BSPT layer.
Specifically, we argue that the
non-Abelian statistics in the Majorana chain layer is due to the unpaired
Majorana modes attached on loops. 

For future study, it remains
unknown how to apply the braiding statistics method to SPT phases with
antiunitary symmetry such as the time reversal symmetry, as we do not know
how to gauge an antiunitary symmetry. It is expected to generalize the
Abelian total symmetry groups $G^f$ to general non-Abelian symmetry groups and have a complete understanding of topological invariants for FSPT phases in 3D. 
Of course, how to use Non-Abelian three-loop braiding statistics to realize topological quantum computation would be anther fascinating future direction. Potential application in fundamental physics was also discussed in Ref.~\onlinecite{neutrino}, it was conjectured that elementary particles could be further divided into topological Majorana modes attached on linked loops and such a scenario naturally explains the origin of three generations of elementary particles. 

\section{Method}
\label{sec:invariants}

In this section, we define the topological invariants $\{\Theta_{\mu,\sigma}, \Theta_{\mu\nu, \sigma}, \Theta_{\mu\nu\lambda, \sigma}\}$ through the three-loop braiding statistics. Then, we discuss the 14 constraints on the topological invariants. 

\subsection{Definitions of topological invariants}

Generally speaking, the full set of braiding statistics among particles and loops is very complicated, in particular when the braiding statistics are non-Abelian. Here, we define a subset of the braiding
statistics data, which we call \textit{topological invariants}.  They are Abelian phase factors associated with certain composite three-loop braiding processes, and thereby are easier to deal with. Yet, this subset still contains enough information to distinguish all different FSPT
phases, as we will show later.

We will define three types of topological invariants, denoted by $\Theta _{\mu ,\sigma }$, $\Theta _{\mu \nu ,\sigma
}$ and $\Theta _{\mu \nu \lambda ,\sigma }$ respectively.  The definitions are straightforward generalizations of the 2D counterparts given in Ref.~\onlinecite{2DFSPTbraiding}. To do that, we introduce a notation. Let  $\xi _{\mu }$ be a loop that carries the type-$\mu $ unit flux,
i.e.,  $\phi _{\xi _{\mu }}=\frac{2\pi }{N_{\mu }}e_{\mu }$, where $e_{\mu
}=(0,...,1,...,0)$ with the $\mu $-th entry being 1 and all other entries being 0. Then, we define $\Theta _{\mu ,\sigma }$, $\Theta _{\mu \nu ,\sigma
}$, and $\Theta _{\mu \nu \lambda ,\sigma }$  as follows.  These definitions work for all $N_\mu$, not limited to the special values in Eq.~\eqref{eq:specialN}.

(i) We define 
\begin{equation}
\Theta _{\mu ,\sigma }=\tilde{N}_{\mu }\theta _{\xi _{\mu },e_{\sigma
}}
\label{invar0}
\end{equation}
where
\begin{align}
\tilde{N}_{0}&=\left\{\begin{array}{ll}
2m, & \text{if $m$ is even}\\
m, & \text{if $m$ is odd}
\end{array}\right.\nonumber\\
\tilde{N}_{i}& =\left\{\begin{array}{ll}
N_i, & \text{if $N_i$ is even}\\
2N_i, & \text{if $N_i$ is odd}
\end{array}\right.\nonumber
\end{align}

The quantity $\theta _{\xi _{\mu },e_{\sigma }}$ is the topological spin of the loop $%
\xi _{\mu }$, when it is linked to another loop $\xi _{\sigma }$. It is defined as\cite{Kitaev2006}:
\begin{eqnarray}
e^{i\theta _{\xi _{\mu },e_{\sigma }}}&=&\frac{1}{d_{\xi _{\mu },e_{\sigma }}}%
\underset{\delta }{\sum }d_{\delta ,e_{\sigma }}\text{tr}(R_{\xi _{\mu }\xi
_{\mu },e_{\sigma }}^{\delta })
\label{topospin}
\end{eqnarray}
where $R_{\xi _{\mu }\xi _{\mu },e_{\sigma }}^{\delta }$ is the $R$-matrix between two $\xi_\mu$ loops in the $\delta$ fusion channel, and all loops are linked to $\xi_\sigma$ (see Appendix for details).

(ii) We define $\Theta _{\mu \nu ,\sigma }$ as the phase associated with braiding $\xi
_{\mu }$ around $\xi _{\nu }$ for $N^{\mu \nu }$ times, when both are linked to the
base loop $\xi _{\sigma }$. Here, $N^{\mu\nu}$ is the least common multiple of $N_\mu$ and $N_\nu$.  In terms of formulas, we have the following expression 
\begin{equation}
e^{i\Theta _{\mu \nu ,\sigma }}I=(B_{\xi _{\mu }\xi _{\nu },e_{\sigma
}})^{N^{\mu \nu }}
\label{invar2}
\end{equation}%
where $B_{\xi_\mu\xi_\nu,e_\sigma}$ denotes the unitary operator associated with braiding $\xi_\mu$ around $\xi_\mu$ only once, while both are linked to $\xi_\sigma$, and $I$ is the identity operator.  The operator $B_{\xi_\mu\xi_\nu,e_\sigma}$ can be expressed in term of $R$ matrices, and $F$ matrices if needed, once we choose a basis for the fusion spaces. 


(iii) We define $\Theta _{\mu \nu \lambda ,\sigma }$ as follows. Consider three loops $\xi _{\mu },\xi _{\nu },\xi
_{\lambda }$ all linked to a base loop $\xi _{\sigma }$. Then, $\Theta_{\mu\nu\lambda,\sigma}$ is the phase associated with braiding $\xi _{\mu }$ around $\xi _{\nu }$ first, then around $\xi _{\lambda }$, then
around $\xi _{\nu }$ in opposite direction and finally around $\xi _{\lambda
}$ in opposite direction.

For the topological invariants $\{\Theta _{\mu ,\sigma
},\Theta _{\mu \nu ,\sigma },\Theta _{\mu \nu \lambda ,\sigma }\}$ to be
well-defined, we need to show that (1) The corresponding
braiding processes indeed lead to Abelian phases  and (2) the Abelian phases only depend on the gauge flux of the loops, i.e. independent of charge attachments.  The proofs are the same as those for the 2D
topological invariants $\{\Theta _{\mu },\Theta _{\mu \nu },\Theta _{\mu \nu
\lambda }\}$, so we do not repeat them here and instead refer the readers to Ref.~\onlinecite{2DFSPTbraiding}. (The only addition for 3D is that one needs to carry the base loop index $\sigma$ in every step of the proofs). The reason that the proofs are identical is that the 3D invariants $\{\Theta _{\mu ,\sigma },\Theta _{\mu \nu ,\sigma
},\Theta _{\mu \nu \lambda ,\sigma }\}$ can be related to the 2D invariants $\{\Theta _{\mu
},\Theta _{\mu \nu },\Theta _{\mu \nu \lambda }\}$ by dimension reduction.\cite{topoinvar15}

\subsection{Constraints of topological invariants}
\label{sec:constraints}

The topological invariants $\{\Theta _{\mu ,\sigma },\Theta
_{\mu \nu ,\sigma },\Theta _{\mu \nu \lambda ,\sigma }\}$ should satisfy certain constraints. We claim that they  satisfy the
following 14 constraints, Eqs.~\eqref{c1}-\eqref{c7} and Eqs.~\eqref{c14}-\eqref{c13}. While we are not able to prove all the constraints, we believe they are rather complete. At least, the solutions to these constraints are all realized in the layer construction of FSPT phases (see Appendix). We divide 14 constraints into two groups. 

\textbf{Group I}: Seven constraints that follow from the 2D counterparts: 
\begin{equation}
\Theta _{\mu \mu \nu ,\sigma }=\Theta _{\nu \nu \mu ,\sigma }=m\Theta _{0\mu
\nu ,\sigma }  \label{c1}
\end{equation}
\begin{equation}
\Theta _{\mu \nu ,\sigma }=\Theta _{\nu \mu ,\sigma }\label{c2}
\end{equation}
\begin{equation}
N_{\mu \nu }\Theta _{\mu \nu ,\sigma }=\mathcal{F}(N^{\mu \nu })\Theta _{\mu
\mu \nu ,\sigma }  \label{c3}
\end{equation}
\begin{eqnarray}
\frac{N_{i}}{2}\Theta _{ii,\sigma }&=&\frac{N_{0i}}{2}\Theta _{0i,\sigma }+[%
\frac{N_{i}}{2}\mathcal{F}(m)+m\mathcal{F}(\frac{N_{i}}{2})]\Theta
_{00i,\sigma }\nonumber\\&&
(N_{i}
\text{ is even})  
\label{c4}
\end{eqnarray}
\begin{equation}
\Theta _{ii,\sigma }=\left\{\begin{array}{l}
2\Theta _{i,\sigma }+\mathcal{F}%
(N_{i})\Theta _{iii,\sigma }\text{, \ \ if }N_{i}\text{ is even}\\ \Theta
_{i,\sigma }\text{, \ \ if }N_{i}\text{ is odd}
\end{array}\right. \label{c5}
\end{equation}
\begin{equation}
\Theta _{00,\sigma }=\left\{\begin{array}{l}
2\Theta _{0,\sigma }\text{, \ \ if }m%
\text{ is even} \\ 4\Theta _{0,\sigma }+\Theta _{000,\sigma }\text{, \ \ if }m%
\text{ is odd} 
\end{array}\right. \label{c6}
\end{equation}
\begin{equation}
\left\{\begin{array}{l}
\frac{m}{2}\Theta _{0,\sigma }=0\text{, \ \ if }m\text{ is
even} \\ {m\Theta _{0,\sigma }+\frac{m^{2}-1}{8}\Theta _{000,\sigma }=0\text{, \
\ if }m\text{ is odd}} 
\end{array}\right. \label{c7}
\end{equation}
where $\mathcal{F}(N)=\frac{1}{2}N(N-1)$.

The constraints Eq. (\ref{c1}) to Eq. (\ref{c7}) are exactly the 2D fermionic
constraints in Ref.~\onlinecite{2DFSPTbraiding} with a base loop inserted. Since the 3D topological
invariants $\{\Theta _{\mu ,\sigma },\Theta _{\mu \nu ,\sigma },\Theta _{\mu
\nu \lambda ,\sigma }\}$\ are related to the 2D ones $\{\Theta _{\mu
},\Theta _{\mu \nu },\Theta _{\mu \nu \lambda }\}$ by dimension reduction, the 3D topological invariants satisfy all the 2D constraints.

We briefly explain the meaning of the above constraints. For constraint Eq. (\ref{c1}), firstly we notice a fact that $N$ copies of the topological invariant  $\Theta _{\mu \nu \lambda ,\sigma }$ are equivalent to do the braiding process for $N$ copies the type-$\mu$ loop, or type-$\nu$, type-$\lambda$ loop, expressed as:
\begin{equation}
N\Theta _{\mu \nu \lambda ,\sigma }=\Theta _{[N\xi _{\mu }] \nu \lambda ,\sigma }=\Theta _{\mu [N\xi _{\nu }] \lambda ,\sigma }=\Theta _{\mu \nu [N\xi _{\lambda }],\sigma }\label{fact}
\end{equation}
where $[N\xi _{\mu }]$ means $N$ copies the type-$\mu$ loop, which can be obtained directly by the definition of $\Theta _{\mu \nu \lambda ,\sigma }$. Then by this fact, the expression $m\Theta _{0\mu
\nu ,\sigma }$ can be rewritten as\cite{2DFSPTbraiding}:
\begin{equation}
e^{im\Theta _{0\mu \nu ,\sigma }}=e^{i\Theta _{[m\xi _{0}]\mu \nu ,\sigma}}
=e^{i\Theta _{f\mu \nu ,\sigma } }
\label{e1}
\end{equation}
where $f$ is the fermion-parity loop. And constraint Eq. (\ref{c1}) illustrates an equivalence $\Theta _{f\mu \nu ,\sigma }=\Theta _{\mu \mu \nu ,\sigma }$, explicitly proved in the appendix of Ref.~\onlinecite{2DFSPTbraiding}. Moreover, as the positions of type-$\mu$ and type-$\nu$ loops are symmetric in   $\Theta _{f\mu \nu ,\sigma }$, the equality can be extended to $\Theta _{\nu \nu \mu ,\sigma }$.
The constraint Eq. (\ref{c2}) simply points out that the type-$\mu$ and type-$\nu$ loops are symmetric in a three-loop braiding process. The constraints Eq. (\ref{c3}) and Eq. (\ref{c4}) are obtained by rearranging the order of certain braiding processes, where the rearrangements give rise to the non-Abelian phase factors $\Theta _{\mu
\mu \nu ,\sigma }$ and $\Theta _{00i,\sigma }$.
For constraints Eq. (\ref{c5}) and Eq. (\ref{c6}), there are two corollaries relating the type-$\mu$ loop and its anti-loop\cite{2DFSPTbraiding}:
\begin{equation}
\Theta _{\mu \mu ,\sigma }+\Theta _{\mu \overline{\mu} ,\sigma }=\mathcal{F}(N_{\mu})\Theta _{\mu \mu \mu,\sigma }
 \label{coro1}
\end{equation}
\begin{equation}
\Theta _{\mu \overline{\mu} ,\sigma }=-2N_{\mu}\theta _{\xi _{\mu },e_{\sigma}}
 \label{coro2}
\end{equation}
where $\overline{\mu}$ denotes for the anti-loop $\overline{\xi}_{\mu}$ with gauge flux $\phi_{\overline{\xi}_{\mu}}=-\phi_{\xi_{\mu}}$. Combining the two corollaries and inducing the definition of $\Theta _{\mu ,\sigma }$ exactly give constraints Eq. (\ref{c5}) and Eq. (\ref{c6}). And the constraint Eq. (\ref{c7}) obtained by demanding the chiral central charge vanishes for FSPT phases.

\textbf{Group II}: Seven constraints that are intrinsically 3D:
\begin{equation}
\Theta _{\mu \nu \lambda ,\sigma }=sgn(\widehat{p})\Theta _{\widehat{p}%
\left( \mu \right) \widehat{p}(\nu )\widehat{p}(\lambda ),\widehat{p}(\sigma
)}  \label{c14}
\end{equation}
\begin{equation}
N_{\mu \nu \lambda \sigma }\Theta _{\mu \nu \lambda ,\sigma }=0  \label{c8}
\end{equation}
\begin{equation}
N_{\sigma }\Theta _{\mu \nu ,\sigma }=0  \label{c9}
\end{equation}
\begin{equation}
N_{\sigma }\Theta _{\mu ,\sigma }=0 \label{c10}
\end{equation}
\begin{equation}
\Theta _{\mu ,\mu }=0 \label{c11}
\end{equation}
\begin{equation}
\frac{N^{\mu \nu \sigma }}{N^{\mu \nu }}\Theta _{\mu \nu ,\sigma }+\frac{%
N^{\mu \nu \sigma }}{N^{\nu \sigma }}\Theta _{\nu \sigma ,\mu }+\frac{N^{\mu
\nu \sigma }}{N^{\sigma \mu }}\Theta _{\sigma \mu ,\nu }=0  \label{c12}
\end{equation}
\begin{equation}
\frac{N^{\mu \sigma }}{\widetilde{N}_{\mu }}\Theta _{\mu ,\sigma }+\Theta
_{\mu \sigma ,\mu }=0\text{ \ \ (}N^{\mu \sigma }\text{ is even)} 
\label{c13}
\end{equation}
where $sgn(\widehat{p})=(-1)^{N(\widehat{%
p})}$ and $N(\widehat{p})$ is the number of permutations for the four
indices $\mu ,\nu ,\lambda ,\sigma $.

The constrants Eq. (\ref{c14}) to Eq. (\ref{c13}) are newly involved 3D constraints
(Specially Eq. (\ref{c8}) is a 2D constraint $N_{\mu \nu \lambda }\Theta _{\mu \nu
\lambda ,\sigma }=0$ combined with a 3D constraint $N_{\sigma }\Theta _{\mu
\nu \lambda ,\sigma }=0$), which can be traced from 3D bosonic non-Abelian
case\cite{topoinvar15} and 3D fermionic Abelian case\cite{threeloop2018}. However, we need to prove that these 3D constraints still hold in 3D fermionic non-Abelian case. 

Firstly, we argue that the constraints Eqs. (\ref{c12})(\ref{c13}) proved in Abelian case still hold in non-Abelian case. The constraint Eq. (\ref{c12}) is called the cyclic relation. Imagining that we create $%
N^{\mu \nu \sigma }$ identical three-loop systems with identical fusion
channel and identical total charge. By anyon charge conservation, after
braiding and fusion, the total charge should still be $N^{\mu \nu \sigma
}Q_{link}$, where $Q_{link}$ is the total charge for a single three-loop
system. Then the next step of the proof is similar to the Abelian case\cite{threeloop2018},
where the difference is that the "vertical" fusions may have multiple fusion
channels (differ only by charges). But we do not need to care about the
charges attached on the resultant loop after fusion, as finally the total
charge should still be $N^{\mu \nu \sigma }Q_{link}$, by which we fall into the same result as the proof in Ref\cite{threeloop2018}. And constraint Eq. (\ref{c13}) is actually the cyclic relation Eq. (\ref{c12}) divided by half on both sides (mod $2\pi$), which then involves fermionic statistics and hence an intrinsic fermionic constraint. It can be argued that it holds in non-Abelian case in a similar manner.

Then we can rigorously prove the constraints Eq. (\ref{c8}) to Eq. (\ref{c10}). The prerequisite to prove them is to assume a 3D "vertical" fusion rule, which naturally gives the linear properties of the topological invariants, explicitly shown in Appendix.

However, the constraints Eq. (\ref{c14}) and Eq. (\ref{c11}) are left unproven. For constraint Eq. (\ref{c14}), it is a generalization of the 2D constraint $\Theta _{\mu \nu \lambda}=sgn(\widehat{p})\Theta _{\widehat{p}%
\left( \mu \right) \widehat{p}(\nu )\widehat{p}(\lambda )}$, where the 2D version can be easily proved by a Borromean ring configuration\cite{topoinvar15}. While here we generalize the totally anti-symmetric property for the indices of $\Theta _{\mu \nu \lambda ,\sigma }$ to the base loop. And the constraint Eq. (\ref{c11}) is simply a conjecture, which means that the topological invariant $\Theta _{\mu ,\sigma }$ vanishes if the two linked loops fall into the same type.

\section*{Acknowledgements}
This work is supported by Hong Kong's Research Grants Council
(ECS 21301018, GRF No.14306918, ANR/RGC Joint Research Scheme
No. A-CUHK402/18).

\section*{Author contributions}
Jingren Zhou and Qingrui Wang carried out the calculations; Chenjie Wang and Zhengcheng Gu supervised the project; Jingren Zhou, Chenjie Wang and Zhengcheng Gu wrote the manuscript. Jingren Zhou and Qingrui Wang prepared Appendix. 

\section*{Competing interests}
The authors declare no competing interests.

\appendix
\section{Some Basic Properties of the Fusion Rule}

There are several properties for the generally nontrivial fusion rule $%
\alpha \times \beta =\underset{\delta }{\sum }N_{\alpha \beta }^{\delta
}\delta $ in our gauged FSPT system, where $\alpha ,\beta ,\delta $ are all
loop-like excitations, and the proofs are the same as the bosonic case given
in Ref\cite{topoinvar15}. The properties are:

(1) For any fusion channel $\delta $:
\begin{equation}
\phi _{\delta }=\phi _{\alpha }+\phi _{\beta }
\end{equation}%
which means that different fusion channels only differ by their attached
charges.

(2) When a loop $\alpha $ is fused with a charge $q$, there is exactly one
fusion outcome:
\begin{equation}
q\times \alpha =\alpha ^{\prime }
\end{equation}

(3) The fusion multiplicity $N_{\alpha \overline{\alpha} }^{q}=0,1$, where $q$ is any charge in the fusion channels of $\alpha$ and $\overline{\alpha} $.

(4) If $\phi _{\alpha ^{\prime }}=\phi _{\alpha }$ and $\phi _{\beta
^{\prime }}=\phi _{\beta }$, then there exist charges $q_{1}$ and $q_{2}$
such that $\alpha ^{\prime }=\alpha \times q_{1}$, $\beta ^{\prime }=\beta
\times q_{2}$ and $\delta ^{\prime }=\delta \times q_{1}\times q_{2}$.

\section{Some Basic Definitions}
\label{app:definitions}

Define: The \textit{fusion space} $V_{\alpha \beta ,c}^{\delta }$ that fuses
two loops $\alpha ,\beta $ into a single fusion channel $\delta $ with base
loop $c$, is a Hilbert space spanned by the set of orthogonal basis\cite{Kitaev2006,preskill1999}:
\begin{equation}
\{\left\vert \alpha \beta ,c;\delta ,\mu \right\rangle |\mu =1,...,N_{\alpha
\beta ,c}^{\delta }\}
\end{equation}%
which can be simplified as $\{\left\vert \alpha \beta ,c;\delta
\right\rangle \}$ as $N_{\alpha \beta ,c}^{\delta }$ is always $1$ in our
theory. And the full Hilbert space for the fusion of $\alpha ,\beta $ with
base loop $c$ is:
\begin{equation}
V_{\alpha \beta ,c}\cong \underset{\delta }{\oplus }V_{\alpha \beta
,c}^{\delta }
\end{equation}

Accordingly the \textit{spliting space} for a single fusion channel $\delta $
is spanned by the dual basis:
\begin{equation}
\{\left\langle \alpha \beta ,c;\delta \right\vert \}
\end{equation}

Define: Consider a local system involving only two loops $\alpha ,\beta $
both linked to a base loop $\gamma $, and their fusion outcome $\delta $ is
known. The \textit{Abelian} $R$\textit{-symbol} $R_{\alpha \beta ,c}^{\delta }$ that
exchanges two loops $\alpha ,\beta $, during which their fusion channel $%
\delta $ is fixed, is defined as a map\cite{Kitaev2006,preskill1999}:
\begin{equation}
R_{\alpha \beta ,c}^{\delta }:V_{\alpha \beta ,c}^{\delta }\rightarrow
V_{\beta \alpha ,c}^{\delta }
\end{equation}
\begin{equation}
\left\vert \beta \alpha ,c;\delta \right\rangle =R_{\alpha \beta ,c}^{\delta
}\left\vert \alpha \beta ,c;\delta \right\rangle
\end{equation}%
which is a basis-dependent pure phase, as $\left\vert \alpha \beta ,c;\delta
\right\rangle $ may differ $\left\vert \beta \alpha ,c;\delta \right\rangle $
by a gauge transformation. Specially, the $R$-symbol $R_{\alpha \alpha
,c}^{\delta }:V_{\alpha \alpha ,c}^{\delta }\rightarrow V_{\alpha \alpha
,c}^{\delta }$ exchanging two identical loops is basis-independent.

Define: The \textit{non-Abelian} $R$\textit{-symbol} $R_{\alpha \beta ,c}$ is defined as a matrix:
\begin{equation}
R_{\alpha \beta ,c}:\underset{\delta }{\oplus }V_{\alpha \beta ,c}^{\delta
}\rightarrow \underset{\delta }{\oplus }V_{\beta \alpha ,c}^{\delta }
\end{equation}
which can be diagonalized by choosing a proper basis if there is no other fusion process involved:
\begin{equation}
R_{\alpha \beta ,c}=\left[ 
\begin{array}{ccc}
R_{\alpha \beta ,c}^{\delta _{1}} & 0 &  \\ 
0 & R_{\alpha \beta ,c}^{\delta _{2}} &  \\ 
&  & \ddots%
\end{array}%
\right]
\end{equation}
where $\delta_{1},\delta_{2},...$ are all the possible fusion channels of $\alpha$ and $\beta$.

Example: For 2D Ising anyons\cite{Kitaev2006}, which contain anyon types $\{1,\sigma,\psi\}$,
\begin{equation}
R_{\sigma \sigma}=\mathcal{X}e^{-i\nu\pi/8}\left[ 
\begin{array}{cc}
1 &  0\\ 
0 & i%
\end{array}%
\right]
\end{equation}
where $\mathcal{X}$ is the Frobenius-Schur indicator, and the Chern number $\nu$ is odd (mod 16) for non-Abelian Ising anyons.

Define: For the same system above, similarly the \textit{Abelian} $B$\textit{-symbol} $%
B_{\alpha \beta ,c}^{\delta }$ that braids loop $\alpha $ around $\beta $
linked to a base loop $\gamma $\ is defined as:
\begin{equation}
B_{\alpha \beta ,c}^{\delta }=R_{\beta \alpha ,c}^{\delta }R_{\alpha \beta
,c}^{\delta }:V_{\alpha \beta ,c}^{\delta }\rightarrow V_{\alpha \beta
,c}^{\delta }
\end{equation}%
which is basis-independent as it maps between the same fusion space.

Define: The \textit{non-Abelian} $B$\textit{-symbol} $B_{\alpha \beta ,c}$ is defined as:
\begin{equation}
B_{\alpha \beta ,c}:\underset{\delta }{\oplus }V_{\alpha \beta ,c}^{\delta
}\rightarrow \underset{\delta }{\oplus }V_{\alpha \beta ,c}^{\delta }
\end{equation}
which can be diagonalized by choosing a proper basis if there is no other fusion process involved:
\begin{equation}
B_{\alpha \beta ,c}=\left[ 
\begin{array}{ccc}
B_{\alpha \beta ,c}^{\delta _{1}} & 0 &  \\ 
0 & B_{\alpha \beta ,c}^{\delta _{2}} &  \\ 
&  & \ddots%
\end{array}%
\right]
\end{equation}

Example: For 2D Ising anyons,
\begin{equation}
B_{\sigma \sigma}=e^{-i\nu\pi/4}\left[ 
\begin{array}{cc}
1 &  0\\ 
0 & -1%
\end{array}%
\right]
\end{equation}

Define: Consider a local system involving three loops $\alpha ,\beta
,\epsilon $ all linked to a base loop $\gamma $, whose total fusion outcome $%
\eta $ is known. The $F$\textit{-symbol} $F_{\epsilon \alpha \beta ,c}^{\eta
}$ that maps between two different fusion ways, is defined as a generally
non-diagonalized matrix\cite{Kitaev2006,preskill1999}:
\begin{equation}
F_{\epsilon \alpha \beta ,c}^{\eta }:\underset{\delta }{\oplus }(V_{\alpha
\beta ,c}^{\delta }\otimes V_{\delta \epsilon ,c}^{\eta })\rightarrow 
\underset{\rho }{\oplus }(V_{\epsilon \alpha ,c}^{\rho }\otimes V_{\rho
\beta ,c}^{\eta })
\end{equation}
\begin{equation}
\left\vert \rho \beta ,c;\eta \right\rangle \left\vert \alpha \epsilon
,c;\rho \right\rangle =\underset{\delta }{\sum }(F_{\epsilon \alpha \beta
,c}^{\eta })_{\delta }^{\rho }\left\vert \delta \epsilon ,c;\eta
\right\rangle \left\vert \alpha \beta ,c;\delta \right\rangle
\end{equation}

Example: For 2D Ising anyons,
\begin{equation}
F_{\sigma \sigma \sigma }^{\sigma }=\frac{\mathcal{X}}{\sqrt{2}}
\left[ 
\begin{array}{cc}
1 & 1 \\ 
1 & -1%
\end{array}%
\right]
\end{equation}

\begin{figure}
\centering
\includegraphics[
width=0.5\textwidth]{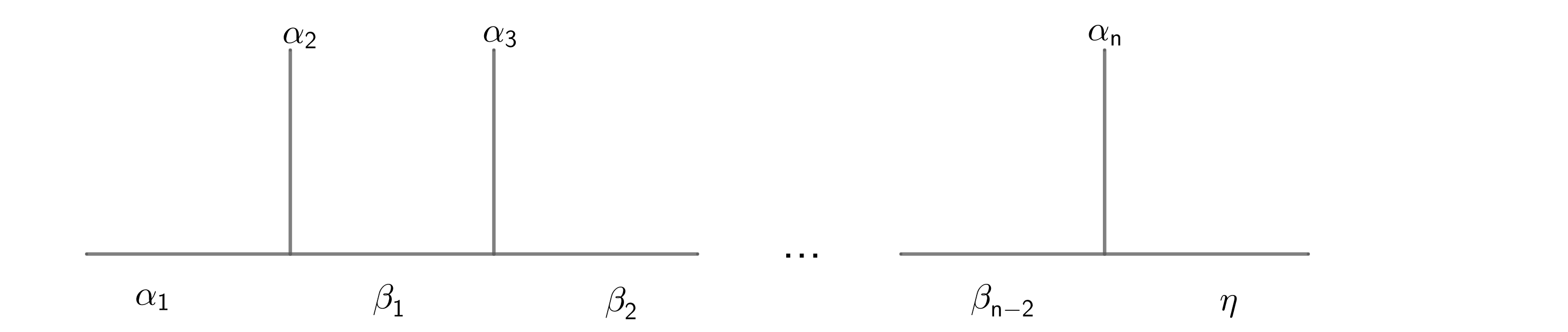}
\caption{The diagram expression of the standard basis.}
\label{Fig5}
\end{figure}

Define: Consider $n$ loops $\alpha _{1},\alpha _{2},...,\alpha _{n}$ all
linked to a base loop $\gamma $, where the total fusion outcome $\eta $ of
the $n$ loops is known. Then we define a \textit{standard basis} in the
total fusion space by specifying a particular fusion order\cite{preskill1999}. For
example, firstly fusing $\alpha _{1}$ and $\alpha _{2}$, then fusing the
result with $\alpha _{3}$, then fusing the result with $\alpha _{4}$, and so
on. The total fusion space can therefore be decomposited as:
\begin{equation}
V_{\alpha _{1},...,\alpha _{n},c}^{\eta }\cong \underset{\beta
_{1},...,\beta _{n-2}}{\oplus }V_{\alpha _{1}\alpha _{2},c}^{\beta
_{1}}\otimes V_{\beta _{1}\alpha _{3},c}^{\beta _{2}}\otimes ...\otimes
V_{\beta _{n-2}\alpha _{n},c}^{\eta }
\end{equation}%
which is equivalently expressed by the diagram in Fig.\ref{Fig5}.

Define: Consider a local system involving three loops $\alpha ,\beta
,\epsilon $ all linked to a base loop $\gamma $, where the total fusion
outcome of the three loops $\eta $ is known. The $\widetilde{R}$\textit{%
-matrix} $\widetilde{R}_{\epsilon \alpha \beta ,c}^{\eta }$ that exchanges
two loops $\alpha ,\beta $, while it is diagonalized in the fusion space of $%
\epsilon $ and $\alpha $, is defined as a generally non-diagonalized matrix
\cite{pachos2012}:
\begin{equation}
\widetilde{R}_{\epsilon \alpha \beta ,c}^{\eta }:\underset{\delta }{\oplus }%
(V_{\epsilon \beta ,c}^{\delta }\otimes V_{\delta  \alpha ,c}^{\eta})
\rightarrow \underset{\rho }{\oplus }(V_{\epsilon \beta ,c}^{\rho }\otimes
V_{\rho \alpha ,c}^{\eta })
\end{equation}
\begin{equation}
\widetilde{R}_{\epsilon \alpha \beta ,c}^{\eta }=F_{\epsilon \beta \alpha
,c}^{\eta }R_{\alpha \beta ,c}(F_{\epsilon \alpha \beta ,c}^{\eta })^{-1}
\end{equation}

Define: For the same system above, similarly the $\widetilde{B}$\textit{%
-matrix} $\widetilde{B}_{\epsilon \alpha \beta ,c}^{\eta }$ that braids loop 
$\alpha $ around $\beta $, while it is diagonalized in the fusion space of $%
\epsilon $ and $\alpha $, is defined as a generally non-diagonalized matrix:
\begin{widetext}
\begin{equation}
\widetilde{B}_{\epsilon \alpha \beta ,c}^{\eta }:\underset{\delta }{\oplus }%
(V_{\epsilon \alpha ,c}^{\delta }\otimes V_{\delta \beta ,c}^{\eta
})\rightarrow \underset{\rho }{\oplus }(V_{\epsilon \alpha ,c}^{\rho
}\otimes V_{\rho \beta ,c}^{\eta })
\end{equation}
\begin{equation}
\widetilde{B}_{\epsilon \alpha \beta ,c}^{\eta }=\widetilde{R}_{\beta \alpha
,c}\widetilde{R}_{\alpha \beta ,c}=F_{\epsilon \alpha \beta ,c}^{\eta
}R_{\beta \alpha ,c}(F_{\epsilon \beta \alpha ,c}^{\eta })^{-1}F_{\epsilon
\beta \alpha ,c}^{\eta }R_{\alpha \beta ,c}(F_{\epsilon \alpha \beta
,c}^{\eta })^{-1}=F_{\epsilon \alpha \beta ,c}^{\eta }B_{\alpha \beta
,c}(F_{\epsilon \alpha \beta ,c}^{\eta })^{-1}
\end{equation}
\end{widetext}

Example: For 2D Ising anyons,
\begin{equation}
\widetilde{B}_{\sigma \sigma \sigma }^{\sigma }:(V_{\sigma \sigma
}^{1}\otimes V_{1\sigma }^{\sigma })\oplus (V_{\sigma \sigma }^{\psi
}\otimes V_{\psi \sigma }^{\sigma })\rightarrow (V_{\sigma \sigma
}^{1}\otimes V_{1\sigma }^{\sigma })\oplus (V_{\sigma \sigma }^{\psi
}\otimes V_{\psi \sigma }^{\sigma })
\end{equation}
\begin{equation}
\widetilde{B}_{\sigma \sigma \sigma }^{\sigma }=e^{-i\nu\pi /4}\left[ 
\begin{array}{cc}
0 & 1 \\ 
1 & 0%
\end{array}%
\right]
\end{equation}

Define: The \textit{fusion matrix} for a loop $\alpha $ linked to a base
loop $\gamma $\ is defined as\cite{anyoncondensation2007,anyoncondensation2010}:
\begin{equation}
\widehat{N}_{\alpha ,c}=(N_{\alpha \beta ,c}^{\delta }:\beta ,\delta \in M)
\end{equation}%
where $M$ is a finite set called \textit{superselection sectors}, which is
the set of all distinguishable particle types in a theory.

Define: The \textit{quantum dimension} of a loop $\alpha $ linked to a base
loop $\gamma $ is defined as the largest eigenvalue of the fusion matrix $%
\widehat{N}_{\alpha ,c}$, which can be understood thorugh a key property
\cite{Kitaev2006}:
\begin{equation*}
d_{\alpha ,c}d_{\beta ,c}=\underset{\delta }{\sum }N_{\alpha \beta
,c}^{\delta }d_{\delta ,c}
\end{equation*}%
which implies that the fusion matrix $\widehat{N}_{\alpha ,c}$ has an
eigenvector $v=(d_{\delta ,c}:\delta \in M)$ and the corresponding
eigenvalue is $d_{\alpha ,c}$. According to Perron-Frobenius theorem, $%
d_{\alpha ,c}$ is the largest eigenvalue of $\widehat{N}_{\alpha ,c}$.
Intuitively, quantum dimension is the intrinsic degree of freedom carried by
an anyon.

\section{Proof of the Constraints}
\label{app:proof}

\subsection{3D "Vertical"\ Fusion Rule}

\begin{figure}
\centering
\includegraphics[
width=0.35\textwidth]{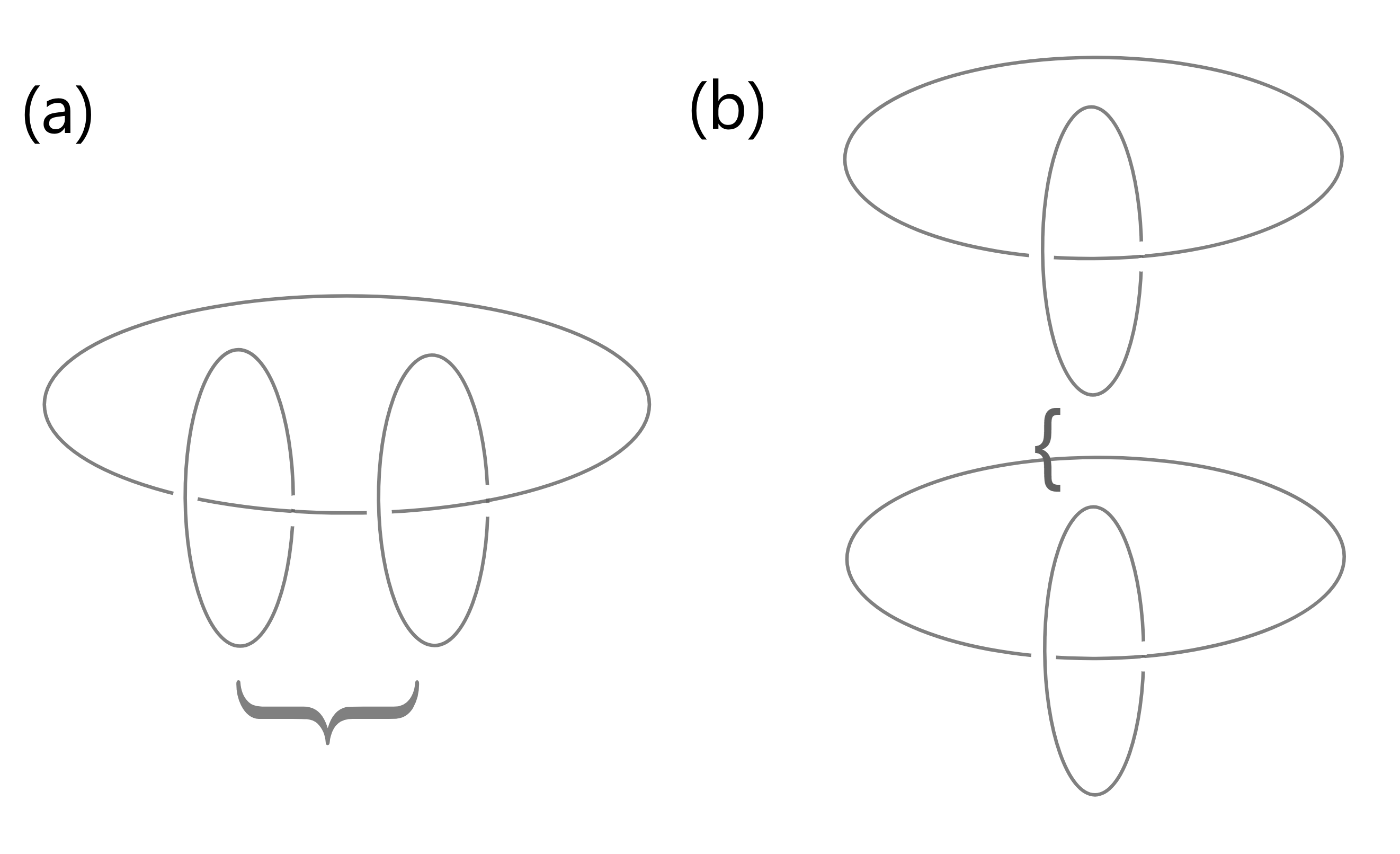}
\caption{T(a) The "horizontal" fusion. (b) The "vertical fusion".}
\label{Fig6}
\end{figure}

In order to prove some of the newly involved 3D constraints, we need to
consider a new kind of "vertical" fusion in analogy to the original
"horizontal fusion", as shown in Fig.\ref{Fig6} (a) and (b). Consider two Hopf-link
systems, where the two loops in dfferent systems are in the same type. Then
the \textit{3D "vertical" fusion rule} has the form:
\begin{equation}
\xi _{\mu ,\sigma _{1}}^{1}\circ \xi _{\mu ,\sigma _{2}}^{2}=\xi _{\mu
,(\sigma _{1}+\sigma _{2})}^{\prime }+\xi _{\mu ,(\sigma _{1}+\sigma
_{2})}^{\prime \prime }+...
\end{equation}%
where the "vertical" fusion is denoted as "$\circ $". The fusion outcomes
have the same flux but different attached charges $Q$, and the $+$ in $%
(\sigma _{1}+\sigma _{2})$ means only putting two loops together, which
applies when fusing the loops or not does not matter as the charges attached
on a base loop do not affect the three-loop braiding process. And this 
"vertical" fusion rule can be understood in a way that the two loops that
are about to fuse annihilate at a point (as particle and antiparticle) to
vacuum or some charge $Q$. And if the fusion outcome is a charge $Q$, it
will be attached to the loop after fusion.

\begin{figure}
\centering
\includegraphics[
width=0.2\textwidth]{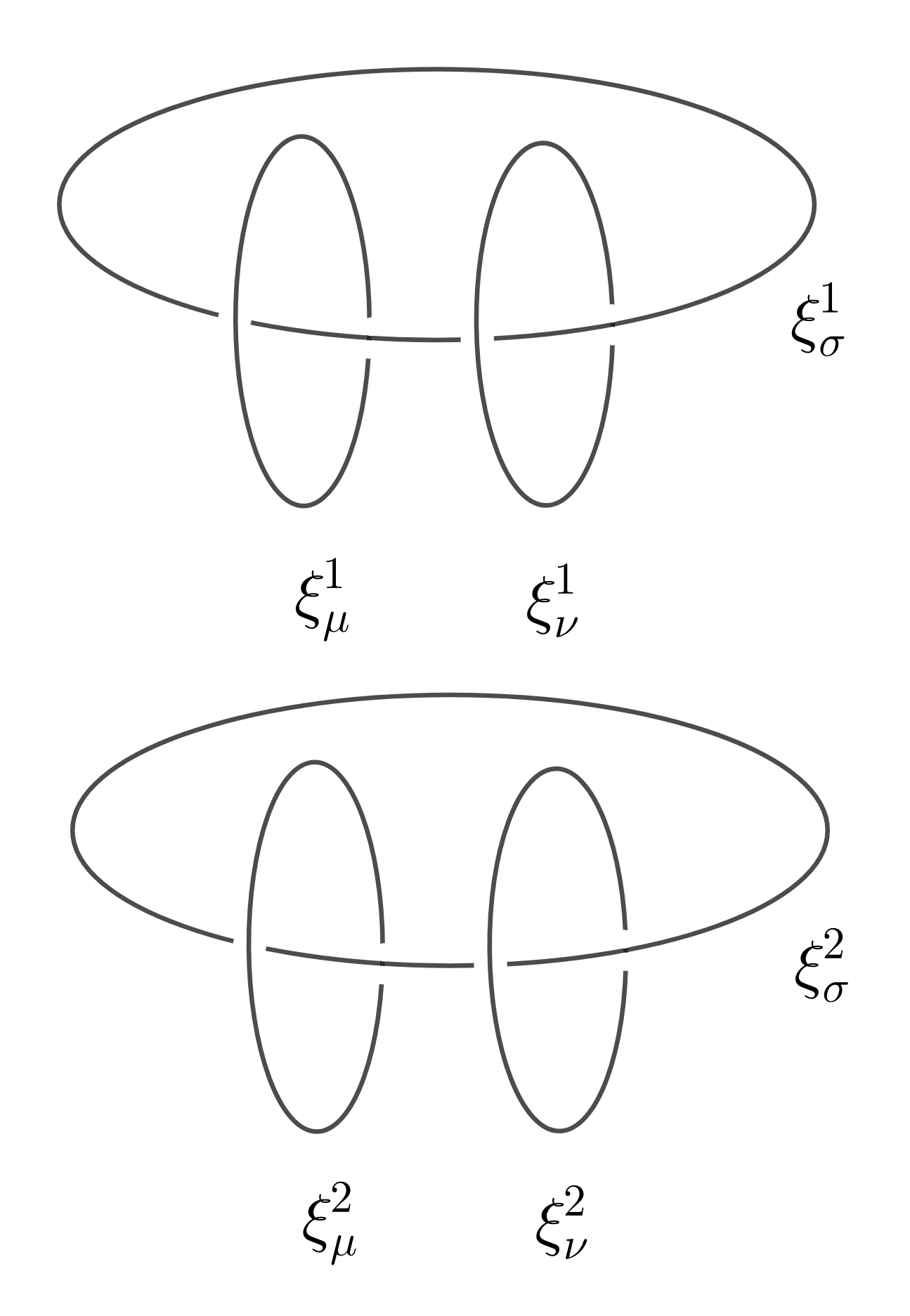}
\caption{The "vertical" fusion of two three-loop systems.}
\label{Fig7}
\end{figure}

First we would like to mention that the expression of the topological invariant $\Theta _{\mu \nu ,\sigma }$ can be further written as:
\begin{equation}
e^{i\Theta _{\mu \nu ,\sigma }}I=(B_{\xi _{\mu }\xi _{\nu },e_{\sigma
}})^{N^{\mu \nu }}=(B_{\xi _{\mu }\xi _{\nu },e_{\sigma }}^{\delta
})^{N^{\mu \nu }}I\label{invar2}
\end{equation}%
where the fusion channel $\delta $ is arbitrary, as the result is the same
for all fusion channels\cite{2DFSPTbraiding}, and $I$ is the identity matrix in the fusion
space $\underset{\delta }{\oplus }V_{\xi _{\mu }\xi _{\nu },c}^{\delta }$.

Then we consider two three-loop systems, where the three loops in different
systems are all in the same type as shown in Fig.\ref{Fig7}. Specifically, before the "vertical" fusions, we
choose to fix the fusion channel for each three-loop system. Thereby the braiding operator for
the whole system before "vertical" fusions are:
\begin{equation}
B_{\xi _{\mu }^{1}\xi _{\nu }^{1},e_{\sigma }^{1}}^{\delta _{1}}B_{\xi _{\mu}^{2}\xi _{\nu }^{2},e_{\sigma }^{2}}^{\delta _{2}}   \label{B1}
\end{equation}

While after the "vertical" fusions, the braiding operator is:
\begin{equation}
\left[ 
\begin{array}{ccc}
B_{\xi _{\mu }^{\prime }\xi _{\nu }^{\prime },(e_{\sigma }^{1}+e_{\sigma
}^{2})}^{\delta ^{\prime }} & 0 &  \\ 
0 & B_{\xi _{\mu }^{\prime \prime }\xi _{\nu }^{\prime \prime },(e_{\sigma
}^{1}+e_{\sigma }^{2})}^{\delta ^{\prime \prime }} &  \\ 
&  & \ddots%
\end{array}%
\right]   \label{B2}
\end{equation}%
where the vertical fusions $\xi _{\mu ,\sigma _{1}}^{1}\circ \xi _{\mu
,\sigma _{2}}^{2}=\xi _{\mu ,(\sigma _{1}+\sigma _{2})}^{\prime }+\xi _{\mu
,(\sigma _{1}+\sigma _{2})}^{\prime \prime }+...$ and $\xi _{\nu ,\sigma
_{1}}^{1}\circ \xi _{\nu ,\sigma _{2}}^{2}=\xi _{\nu ,(\sigma _{1}+\sigma
_{2})}^{\prime }+\xi _{\nu ,(\sigma _{1}+\sigma _{2})}^{\prime \prime }+...$
both generally have multiple fusion outcomes. According to the 4th property
in section I, the fusion outcomes of the two loops after "vertical" fusions 
$\xi _{\mu ,(\sigma _{1}+\sigma _{2})}$ and $\xi _{\nu ,(\sigma _{1}+\sigma
_{2})}$ are also multiple. And we can choose a particular basis in the fusion space such that the braiding operator is diagonalized.

Then we do the braiding processes for both cases (before and after "vertical" fusions) for $N^{\mu \nu }$
times, we obtain an equation:
\begin{equation}
(B_{\xi _{\mu }^{1}\xi _{\nu }^{1},e_{\sigma }^{1}}^{\delta _{1}})^{N^{\mu
\nu }}(B_{\xi _{\mu }^{2}\xi _{\nu }^{2},e_{\sigma }^{2}}^{\delta
_{2}})^{N^{\mu \nu }}=(B_{\xi _{\mu }^{\prime }\xi _{\nu }^{\prime
},(e_{\sigma }^{1}+e_{\sigma }^{2})}^{\delta ^{\prime }})^{N^{\mu \nu }} 
\label{B3}
\end{equation}%
where $B_{\xi _{\mu }^{\prime }\xi _{\nu }^{\prime },(e_{\sigma
}^{1}+e_{\sigma }^{2})}^{\delta ^{\prime }}$ is any of the diagonalized
entry in the matrix (\ref{B2}). The eqn. (\ref{B3}) is equivalent to the claim that:

\textit{The }$N^{\mu \nu }$\textit{\ times of braiding as a whole commutes
with the "vertical" fusions.}

\begin{figure}
\centering
\includegraphics[
width=0.25\textwidth]{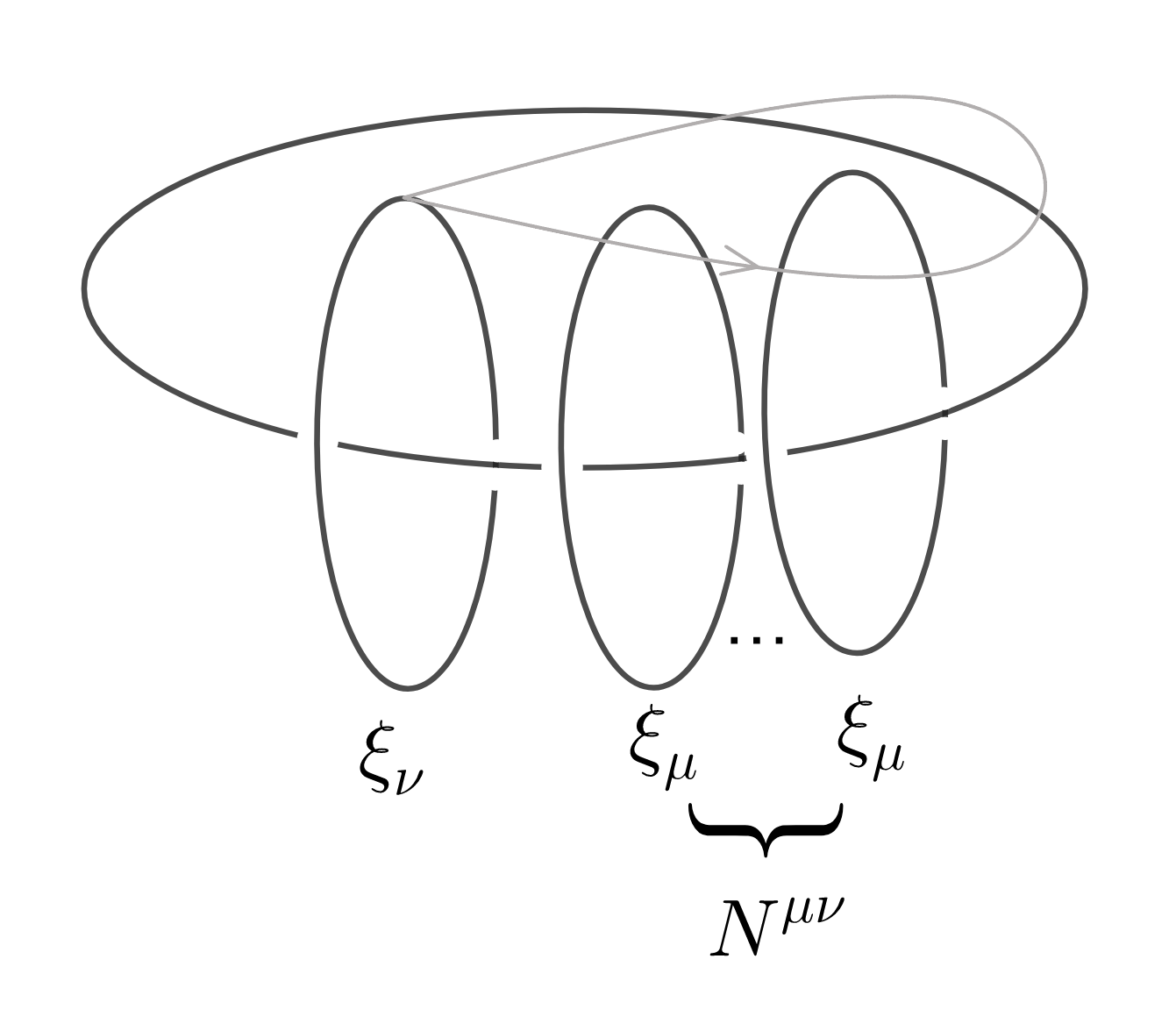}
\caption{The $N^{\mu \nu }$ loops as a whole can be viewed as an Abelian
object.}
\label{Fig8}
\end{figure}

The proof of eqn.(\ref{B3}) is given as the following: Firstly the $N^{\mu \nu }$
times of braiding can be equivalently viewed as a successive braiding of $%
N^{\mu \nu }$ identical loops. As the $N^{\mu \nu }$ times of braiding
eliminates the difference between different fusion channels, the $N^{\mu \nu
}$ loops as a whole is actually an Abelian object, as shown in Fig.\ref{Fig8}. And
the remaining proof is similar as the Fig.6 in Ref\cite{threeloop}.

Notice that the whole argument does not violate the conservation of anyon
charge, as we have only specified the fusion channels but not the total
charge of the initial state. And the exchanging operator for a loop with its
anti-loop, i.e. the $R$-operator in the vacuum fusion channel, has a similar
property if we do the exchanging processes for $\tilde{N}_{\mu }$ times:
\begin{equation}
(R_{\xi _{\mu }^{1}\overline{\xi }_{\mu }^{1},e_{\sigma }^{1}}^{0})^{\tilde{N%
}_{\mu }}(R_{\xi _{\mu }^{2}\overline{\xi }_{\mu }^{2},e_{\sigma
}^{2}}^{0})^{\tilde{N}_{\mu }}=(R_{\xi _{\mu }^{\prime }\overline{\xi }_{\mu
}^{\prime },(e_{\sigma }^{1}+e_{\sigma }^{2})}^{0})^{\tilde{N}_{\mu }} 
\label{R1}
\end{equation}%
where although the fusion channels of the two $\xi _{\mu }^{1}$ loops or two 
$\xi _{\mu }^{2}$ loops are both $0$, the total fusion outcome of the four
loops may not be $0$, i.e. generally the right-hand side of (\ref{R1}) should be $%
(R_{\xi _{\mu }^{\prime }\overline{\xi }_{\mu }^{\prime },(e_{\sigma
}^{1}+e_{\sigma }^{2})}^{\delta ^{\prime }})^{\tilde{N}_{\mu }}$. But as $%
(R_{\xi _{\mu }^{\prime }\overline{\xi }_{\mu }^{\prime },(e_{\sigma
}^{1}+e_{\sigma }^{2})}^{\delta ^{\prime }})^{\tilde{N}_{\mu }}=(R_{\xi
_{\mu }^{\prime }\overline{\xi }_{\mu }^{\prime },(e_{\sigma }^{1}+e_{\sigma
}^{2})}^{0})^{\tilde{N}_{\mu }}$ due to the $\tilde{N}_{\mu }$ times of
exchanging, we can write $(R_{\xi _{\mu }^{\prime }\overline{\xi }_{\mu
}^{\prime },(e_{\sigma }^{1}+e_{\sigma }^{2})}^{0})^{\tilde{N}_{\mu }}$ at
the right-hand side of (\ref{R1}) safely.

\subsection{Linear Properties of the Topological Invariants}

The linear properties of the braiding processes that are useful in proving
the newly involved 3D constraints are:
\begin{equation}
\Theta _{(\mu _{1}\circ \mu _{2}),(\sigma _{1}+\sigma _{2})}=\Theta _{\mu
_{1},\sigma _{1}}+\Theta _{\mu _{2},\sigma _{2}}  \label{P1}
\end{equation}
\begin{equation}
\Theta _{(\mu _{1}\circ \mu _{2})(\nu _{1}\circ \nu _{2}),(\sigma
_{1}+\sigma _{2})}=\Theta _{\mu _{1}\nu _{1},\sigma _{1}}+\Theta _{\mu
_{2}\nu _{2},\sigma _{2}} \label{P2}
\end{equation}
\begin{equation}
\Theta _{(\mu _{1}\circ \mu _{2})(\nu _{1}\circ \nu _{2})(\lambda _{1}\circ
\lambda _{2}),(\sigma _{1}+\sigma _{2})}=\Theta _{\mu _{1}\nu _{1}\lambda
_{1},\sigma _{1}}+\Theta _{\mu _{2}\nu _{2}\lambda _{2},\sigma _{2}} 
\label{P3}
\end{equation}%
which means that all the topological invariants are linear under "vertical"
fusions. We firstly prove (\ref{P2}) as the following: The the right-hand side of
(\ref{P2}) is:
\begin{eqnarray}
e^{i(\Theta _{\mu _{1}\nu _{1},\sigma _{1}}+\Theta _{\mu _{2}\nu _{2},\sigma
_{2}})}I&=&(B_{\xi _{\mu }^{1}\xi _{\nu }^{1},e_{\sigma }^{1}}^{_{\delta
_{1}}})^{N^{\mu \nu }}(B_{\xi _{\mu }^{2}\xi _{\nu }^{2},e_{\sigma
}^{2}}^{\delta _{2}})^{N^{\mu \nu }}\nonumber\\ &=&(B_{\xi _{\mu }^{\prime }\xi _{\nu
}^{\prime },(e_{\sigma }^{1}+e_{\sigma }^{2})}^{\delta ^{\prime }})^{N^{\mu
\nu }}
\end{eqnarray}%
where we have applied the eqn.(\ref{B3}), and $B_{\xi _{\mu }^{\prime }\xi _{\nu
}^{\prime },(e_{\sigma }^{1}+e_{\sigma }^{2})}^{\delta ^{\prime }}$ can be
any of the diagonalized entry in the matrix after fusion (\ref{B2})\ introduced
above. The left-hand side of (\ref{P2}) is:
\begin{eqnarray}
e^{i\Theta _{(\mu _{1}\circ \mu _{2})(\nu _{1}\circ \nu _{2}),(\sigma
_{1}+\sigma _{2})}}I&=&(B_{(\xi _{\mu }^{1}\circ \xi _{\mu }^{2})(\xi _{\nu
}^{1}\circ \xi _{\nu }^{2}),(e_{\sigma }^{1}+e_{\sigma }^{2})}^{\delta
^{\prime \prime }})^{N^{\mu \nu }}\nonumber\\ &=&(B_{\xi _{\mu }^{\prime \prime }\xi _{\nu
}^{\prime \prime },(e_{\sigma }^{1}+e_{\sigma }^{2})}^{\delta ^{\prime
\prime }})^{N^{\mu \nu }}
\end{eqnarray}%
where $B_{\xi _{\mu }\xi _{\nu },(e_{\sigma }^{1}+e_{\sigma }^{2})}^{\delta
^{\prime \prime }}$ can also be any of the entry in the same diagonalized
matrix. Then the difference between $B_{\xi _{\mu }^{\prime }\xi _{\nu
}^{\prime },(e_{\sigma }^{1}+e_{\sigma }^{2})}^{\delta ^{\prime }}$ and $%
B_{\xi _{\mu }^{\prime \prime }\xi _{\nu }^{\prime \prime },(e_{\sigma
}^{1}+e_{\sigma }^{2})}^{\delta ^{\prime \prime }}$ can be eliminiated by
the $N^{\mu \nu }$ times of braiding, which is shown in eqn.(15) of Ref\cite{topoinvar15}.

Then (\ref{P1}) and (\ref{P3}) can be proved similarly, as $\Theta _{\mu ,\sigma }$ and $%
\Theta _{\mu \nu \lambda ,\sigma }$ are also defined so as to eliminate the
effect caused by difference charge attachments.

\subsection{Partial Proof of the Constraints}

We can rigorously prove the following constraints:
\begin{equation}
N_{\mu \nu \lambda \sigma }\Theta _{\mu \nu \lambda ,\sigma }=0  \label{c8}
\end{equation}
\begin{equation}
N_{\sigma }\Theta _{\mu \nu ,\sigma }=0  \label{c9}
\end{equation}
\begin{equation}
N_{\sigma }\Theta _{\mu ,\sigma }=0 \label{c10}
\end{equation}

We firstly prove (\ref{c9}) as the following: By the property (\ref{P2}), we have:
\begin{eqnarray}
e^{iN_{\sigma }\Theta _{\mu \nu ,\sigma }}I&=&e^{i(\Theta _{\mu _{1}\nu
_{1},\sigma _{1}}+...+\Theta _{\mu _{N_{\sigma }}\nu _{N_{\sigma }},\sigma
_{N_{\sigma }}})}I\nonumber\\ &=&e^{i\Theta _{(\mu _{1}\circ ...\circ \mu _{N_{\sigma
}})(\nu _{1}\circ ...\circ \nu _{N_{\sigma }}),(\sigma _{1}+...+\sigma
_{N_{\sigma }})}}I\nonumber\\ &=&(B_{\xi _{\mu }\xi _{\nu }}^{\delta })^{N^{\mu \nu }}=I
\end{eqnarray}%
where we realize the phase $N_{\sigma }\Theta _{\mu \nu ,\sigma }$ by
constructing $N_{\sigma }$ identical three-loop systems, and then applying
the "vertical" fusions, by which the $N_{\sigma }$ type-$\sigma $ base loops
all together vanish. And by (\ref{P1}) and (\ref{P3}), (\ref{c8}) and (\ref{c10}) can be proved
similarly.

\section{Solving the Constraints}

\subsection{Category (A)}

The constraints that are related to 2D constraints by dimension
reduction are:
\begin{equation}
\Theta _{000,0}=m\Theta _{000,0}
\end{equation}
\begin{equation}
N_{0}\Theta _{00,0}=\mathcal{F}(N_{0})\Theta _{000,0}=m\Theta _{000,0}
\end{equation}
\begin{equation}
\Theta _{00,0}=\left\{\begin{array}{l}
2\Theta _{0,0}\text{, \ \ if }m\text{ is even%
}\\ 
{4\Theta _{0,0}+\Theta _{000,0}\text{, \ \ if }m\text{ is odd}}
\end{array}\right. 
\end{equation}
\begin{equation}
\left\{\begin{array}{l}
\frac{m}{2}\Theta _{0,0}=0\text{, \ \ if }m\text{ is even}\\ 
m\Theta _{0,0}+\frac{m^{2}-1}{8}\Theta _{000,0}=0\text{, \ \ if }m\text{ is odd}
\end{array}\right. 
\end{equation}

The newly involved 3D constraints are:
\begin{equation}
2\Theta _{000,0}=0\Rightarrow \Theta _{000,0}=0\text{ or }\pi
\end{equation}
\begin{equation}
\Theta _{0,0}=0
\end{equation}

We solve the constraints in two cases:

\subsubsection{$m$ is odd}

By $\Theta _{0,0}=0$ and the constraint $\Theta _{00,0}=4\Theta
_{0,0}+\Theta _{000,0}$, we have $\Theta _{00,0}=\Theta _{000,0}$. Combining 
$\Theta _{00,0}=\Theta _{000,0}$ and $N_{0}\Theta _{00,0}=m\Theta _{000,0}$,
we find $\Theta _{00,0}$ and $\Theta _{000,0}$ can only both be $0$. Hence
\begin{equation}
(\Theta _{0,0},\Theta _{00,0},\Theta _{000,0})=(0,0,0)
\end{equation}

The classification is trivial.

\subsubsection{$m$ is even}

By $\Theta _{0,0}=0$ and the constraint $\Theta _{00,0}=2\Theta _{0,0}$, we
have $\Theta _{00,0}=0$. And the constraint $\Theta _{000,0}=m\Theta
_{000,0} $ ensures that $\Theta _{000,0}=0$. Hence
\begin{equation}
(\Theta _{0,0},\Theta _{00,0},\Theta _{000,0})=(0,0,0)
\end{equation}

The classification is trivial.

\subsection{Category (B)}

For simplicity, we only consider symmetry groups with order being power of 2.

The newly involved 3D constraints are:
\begin{equation}
\Theta _{i,i}=0
\end{equation}
\begin{equation}
N_{0}\Theta _{i,0}=0\text{, \ \ }N_{i}\Theta _{0,i}=0
\end{equation}
\begin{equation}
N_{0}\Theta _{0i,0}=0\text{, \ \ }N_{i}\Theta _{0i,i}=0
\end{equation}
\begin{equation}
N_{0}\Theta _{00i,0}=0\text{, \ \ }N_{i}\Theta _{00i,i}=0
\end{equation}
\begin{equation}
\Theta _{0i,i}=-\frac{N^{0i}}{\widetilde{N}_{i}}\Theta _{i,0}\text{, \ \ }%
\Theta _{0i,0}=-\frac{N^{0i}}{\widetilde{N}_{0}}\Theta _{0,i}
\end{equation}
\begin{equation}
\Theta _{00i,i}=\Theta _{0ii,0}=\Theta _{00i,0}=\Theta _{000,i}
\end{equation}

\subsubsection{ $m$ is odd}

For $m$ is odd, we set $m=1$ for simplicity, i.e. we only consider the
symmetry group $%
\mathbb{Z}
_{2}\times 
\mathbb{Z}
_{N_{i}}$. We can do  this as $\mathbb{Z}_{2m}^f$ is isomorphic to $\mathbb{Z}_{2}^f \times \mathbb{Z}_{m}$, and $\mathbb{Z}_{m}$ can be absorbed into $G_b=\prod_{i}\mathbb{Z}_{N_i}$ part of $G_{f}$, making $G_{f}$ always the form $G_{f}=\mathbb{Z}_{2}^f \times G_b$.

(1) If $\frac{N_{i}}{2}$ is odd (i.e. $%
\mathbb{Z}
_{2}^{f}\times 
\mathbb{Z}
_{2}$), invoking the known 2D results and combining with the 3D constraints $%
N_{0}\Theta _{i,0}=0$, $N_{i}\Theta _{0,i}=0$, $N_{0}\Theta _{0i,0}=0$, $%
N_{i}\Theta _{0i,i}=0$, $N_{0}\Theta _{00i,0}=0$, $N_{i}\Theta _{00i,i}=0$,
the generating phases for the sets (B1), (B2), and (B3) are:
\begin{equation}
\Theta _{0,i}=\frac{2\pi }{m}=0
\end{equation}
\begin{eqnarray}
(\Theta _{i,0},\Theta _{0i,0},\Theta _{00i,0})&=&(\frac{\pi }{2N_{i}},-\frac{%
\pi }{N_{0i}},\pi )\times 2N_{i}a+(0,\frac{4\pi }{N_{0i}},0)
\nonumber\\&=&(\pi ,0,0)a
\end{eqnarray}
\begin{eqnarray}
(\Theta _{i,i},\Theta _{0i,i},\Theta _{00i,i})&=&(\frac{\pi }{2N_{i}},\mp 
\frac{\pi }{N_{0i}},\pi )\times 0+(0,\frac{4\pi }{N_{0i}},0)
\nonumber\\&=&(0,0,0)
\end{eqnarray}
where $a$ is an integer.

By the constraint \ $\Theta _{0i,i}=-\Theta _{i,0}$, $a=0$ (mod $2$). Hence
in this case the classification is trivial.

(2) If $N_{i}=4$ $($mod $8)$, similarly, the generating phases for the sets
(B1), (B2), and (B3) are:
\begin{equation}
\Theta _{0,i}=\frac{2\pi }{m}=0
\end{equation}
\begin{eqnarray}
(\Theta _{i,0},\Theta _{0i,0},\Theta _{00i,0})&=&(\frac{\pi }{N_{i}},\frac{%
2\pi }{N_{0i}},\pi )\times N_{i}a+(0,\frac{2\pi }{N_{0i}},\pi )\times b
\nonumber\\&=&(\pi
,0,0)a+(0,\pi ,\pi )b
\end{eqnarray}
\begin{eqnarray}
(\Theta _{i,i},\Theta _{0i,i},\Theta _{00i,i})&=&(\frac{\pi }{N_{i}},\frac{%
2\pi }{N_{0i}},\pi )\times 0+(0,\frac{2\pi }{N_{0i}},\pi )\times c
\nonumber\\&=&(0,\pi
,\pi )c
\end{eqnarray}
where $a,b,c$ are all integers.

By the constraint \ $\Theta _{0i,0}=-N_{i}\Theta _{0,i}$, $b=0$ (mod $2$).
By the constraint $\Theta _{00i,i}=\Theta _{00i,0}$, $c=0$ (mod $2$). By the
constraint \ $\Theta _{0i,i}=-\Theta _{i,0}$, $a=0$ (mod $2$). Hence in this
case the classification is trivial.

(3) If $N_{i}=0$ $($mod $8)$, the generating phases for the sets (B1), (B2),
and (B3) are:
\begin{equation}
\Theta _{0,i}=\frac{2\pi }{m}=0
\end{equation}
\begin{eqnarray}
(\Theta _{i,0},\Theta _{0i,0},\Theta _{00i,0})&=&(\frac{\pi }{N_{i}},\frac{%
2\pi }{N_{0i}},\pi )\times N_{i}a,(0,\frac{4\pi }{N_{0i}},\pi )\times b
\nonumber\\&=&(\pi
,0,0)a+(0,0,\pi )b
\end{eqnarray}
\begin{eqnarray}
(\Theta _{i,i},\Theta _{0i,i},\Theta _{00i,i})&=&(\frac{\pi }{N_{i}},\frac{%
2\pi }{N_{0i}},\pi )\times 0+(0,\frac{4\pi }{N_{0i}},\pi )\times c
\nonumber\\&=&(0,0,\pi
)c
\end{eqnarray}

By $\Theta _{0,i}=0$ and the constraint $\Theta _{00,i}=4\Theta
_{0,i}+\Theta _{000,i}$, we have $\Theta _{00,i}=\Theta _{000,i}$. Combining 
$\Theta _{00,i}=\Theta _{000,i}$ and $N_{0}\Theta _{00,i}=m\Theta _{000,i}$,
we find $\Theta _{00,0}=\Theta _{000,0}=0$. By $\Theta _{000,i}=0$ and the
constraint $\Theta _{00i,i}=\Theta _{00i,0}=\Theta _{000,i}$, $b=c=0$ (mod $%
2 $). By the constraint \ $\Theta _{0i,i}=-\Theta _{i,0}$, $a=0$ (mod $2$).
Hence in this case the classification is trivial.

\subsubsection{$m$ is even}

(1) If $N_{i}<N_{0}\leq 4$ (i.e. $%
\mathbb{Z}
_{4}^{f}\times 
\mathbb{Z}
_{2}$), the generating phases for the sets (B1), (B2), and (B3) are:
\begin{equation}
\Theta _{0,i}=0
\end{equation}
\begin{eqnarray}
(\Theta _{i,0},\Theta _{0i,0},\Theta _{00i,0})&=&(\frac{\pi }{N_{i}},\frac{%
2\pi }{N_{0i}},0)\times a+(0,\frac{4\pi }{N_{0i}},0)
\nonumber\\&=&(\frac{\pi }{2},\pi ,0)a
\end{eqnarray}
\begin{equation}
(\Theta _{i,i},\Theta _{0i,i},\Theta _{00i,i})=(\frac{\pi }{N_{i}},\frac{%
2\pi }{N_{0i}},0)\times 0+(0,\frac{4\pi }{N_{0i}},0)=(0,0,0)
\end{equation}

By the constraint \ $\Theta _{0i,0}=-\Theta _{0,i}$, $a=0,2$ (mod $4$). And
the remaining generating phase is determined by integer $a$. Hence in this
case the classification is $%
\mathbb{Z}
_{2}$, which belongs to BSPT phases.

(2) If $N_{i}<4<N_{0}$ (i.e. $%
\mathbb{Z}
_{N_{0}}^{f}\times 
\mathbb{Z}
_{2}$), the generating phases for the sets (B1), (B2), and (B3) are:
\begin{equation}
\Theta _{0,i}=\frac{4\pi }{m}\times \frac{m}{4}a=\pi a
\end{equation}
\begin{eqnarray}
(\Theta _{i,0},\Theta _{0i,0},\Theta _{00i,0})&=&(\frac{\pi }{N_{i}},\frac{%
2\pi }{N_{0i}},0)\times b+(0,\frac{4\pi }{N_{0i}},0)\times c
\nonumber\\&=&(\frac{\pi }{2}%
,\pi ,0)b
\end{eqnarray}
\begin{equation}
(\Theta _{i,i},\Theta _{0i,i},\Theta _{00i,i})=(\frac{\pi }{N_{i}},\frac{%
2\pi }{N_{0i}},0)\times 0+(0,\frac{4\pi }{N_{0i}},0)=(0,0,0)
\end{equation}

By the constraint \ $\Theta _{0i,0}=-\Theta _{0,i}$, $b=-a$ (mod $4$). By
the constraint \ $\Theta _{0i,i}=-m\Theta _{i,0}$, $0=-m\frac{\pi }{2}b$,
which is always satisfied as in this case the smallest $m$ is $4$. The
generating phases are generated by integer $b$. Hence in this case the
classification is $%
\mathbb{Z}
_{4}$, which is a $%
\mathbb{Z}
_{2}$ complex fermion layer absorbed into a $%
\mathbb{Z}
_{2}$ BSPT layer as the complex fermion layer indicator is $\Theta _{fi,j}=\Theta _{0i,j}=\pi b$.

(3) If $4\leq N_{i}<N_{0}$, the generating phases for the sets (B1), (B2),
and (B3) are:
\begin{equation}
\Theta_{0,i}=\left\{\begin{array}{l}\frac{4\pi }{m}a\text{, \ \ if }N_{i}>N_{0}/4 \\
{\frac{8\pi }{N_{0}}\times \frac{N_{0}}{4N_{i}}a=\frac{2\pi }{N_{i}}a\text{%
, \ \ if }N_{i}\leq N_{0}/4}
\end{array}\right.
\end{equation}

\begin{eqnarray}
(\Theta _{i,0},\Theta _{0i,0},\Theta _{00i,0})&=&(\frac{\pi }{N_{i}},\frac{%
2\pi }{N_{0i}},0)\times b+(0,\frac{4\pi }{N_{0i}},0)\times c
\nonumber\\&=&(\frac{\pi }{%
N_{i}}b,\frac{2\pi }{N_{i}}b+\frac{4\pi }{N_{i}}c,0)
\end{eqnarray}
\begin{eqnarray}
(\Theta _{i,i},\Theta _{0i,i},\Theta _{00i,i})&=&(\frac{\pi }{N_{i}},\frac{%
2\pi }{N_{0i}},0)\times 0+(0,\frac{4\pi }{N_{0i}},0)\times d
\nonumber\\&=&(0,\frac{4\pi }{%
N_{i}}d,0)
\end{eqnarray}

By the constraint $\Theta _{0i,0}=-\Theta _{0,i}$,
\begin{equation}
\left\{\begin{array}{l}\frac{2\pi}{N_{i}}b+\frac{4\pi}{N_{i}}c=-\frac{8\pi }{%
N_{0}}a\text{, \ \ if }N_{i}>N_{0}/4
\\{\frac{2\pi }{N_{i}}(b+2c)=-\frac{2\pi 
}{N_{i}}a\text{, \ \ if }N_{i}\leq N_{0}/4}
\end{array}\right.
\end{equation}

By the constraint\ $\Theta _{0i,i}=-\frac{N_{0}}{N_{i}}\Theta _{i,0}$, $%
\frac{4\pi }{N_{i}}d=-\frac{N_{0}}{N_{i}}\frac{\pi }{N_{i}}b$, we have:
\begin{equation}
\left\{\begin{array}{l}
b=-\frac{4N_{i}}{N_{0}}d\text{, \ \ if }N_{i}>N_{0}/4 \text{ and } 
N_{0}<2N_{i}^{2} \\
d=-\frac{N_{0}}{4N_{i}}b\text{, \ \ if }N_{i}\leq N_{0}/4
\end{array}\right.
\end{equation}

Combining all the constraints:
\begin{equation}
\left\{\begin{array}{l}
(b+2c)=-2a\text{, }b=-2d\text{, \ \ if }N_{i}=N_{0}/2 \\
(b+2c)=-a\text{, }d=-\frac{N_{0}}{4N_{i}}b\text{, \ \ if }N_{i}\leq N_{0}/4
\end{array}\right.
\end{equation}

\begin{widetext}

The generating phases are determined by the integers:
\begin{equation}
\left\{\begin{array}{l}
a\text{ (mod }N_{0}/4\text{), }b\text{ (mod }\frac{1}{2}\times 2N_{i}=N_{i}%
\text{), \ \ if }N_{i}=N_{0}/2 \\ 
c\text{ (mod }N_{i}/2\text{), }b\text{ (mod }2N_{i}\text{), \ \ if }%
N_{i}\leq N_{0}/4
\end{array}\right.
\end{equation}

Hence in this case the classification is:
\begin{equation}
\left\{\begin{array}{l}

\mathbb{Z}

_{N_{i}}\times 

\mathbb{Z}

_{N_{0}/4}=%

\mathbb{Z}

_{N_{i}}\times 

\mathbb{Z}

_{N_{i}/2}$ (BSPT)$\text{, \ \ if }N_{i}=N_{0}/2 \\

\mathbb{Z}

_{2N_{i}}\times 

\mathbb{Z}

_{N_{i}/2}\text{ (a }

\mathbb{Z}

_{2}\text{ complex fermion layer absorbed into a }

\mathbb{Z}

_{N_{i}}\times 

\mathbb{Z}

_{N_{i}/2}\text{ BSPT layer), \ \ if }N_{i}\leq N_{0}/4
\end{array}\right.
\end{equation}
where the indicators of the complex fermion layer are:
\begin{equation}
\left\{\begin{array}{l}
\Theta _{fi,0}=\frac{N_{i}}{2}\Theta _{0i,0}=-2\pi a\text{, }\Theta _{fi,i}=
\frac{N_{i}}{2}\Theta _{0i,i}=0\text{, \ \ if }N_{i}=N_{0}/2 \\ 
\Theta _{fi,0}=\frac{N_{i}}{2}\Theta _{0i,0}=-\pi a\text{, }\Theta _{fi,i}=%
\frac{N_{i}}{2}\Theta _{0i,i}=0\text{, \ \ if }N_{i}\leq N_{0}/4
\end{array}\right.
\end{equation}

(4) If $4\leq N_{0}\leq N_{i}$, the generating phases for the sets (B1),
(B2), and (B3) are:
\begin{equation}
\Theta _{0,i}=
\left\{\begin{array}{l}
0\text{, \ \ if }N_{0}=4\\
{\frac{4\pi }{m}a
\text{, \ \ if }N_{0}>4}
\end{array}\right.
\end{equation}
\begin{equation}
(\Theta _{i,0},\Theta _{0i,0},\Theta _{00i,0})=(\frac{\pi }{N_{i}},\frac{%
2\pi }{N_{0i}},0)\times \frac{2N_{i}}{N_{0}}b+(0,\frac{4\pi }{N_{0i}}%
,0)\times c
=(\frac{2\pi }{N_{0}}b,\frac{4\pi }{N_{0}}(\frac{N_{i}}{N_{0}}%
b+c),0)
\end{equation}
\begin{equation}
(\Theta _{i,i},\Theta _{0i,i},\Theta _{00i,i})=(\frac{\pi }{N_{i}},\frac{%
2\pi }{N_{0i}},0)\times 0+(0,\frac{4\pi }{N_{0i}},0)\times d
=(0,\frac{4\pi }{%
N_{0}}d,0)
\end{equation}

By the constraint \ $\Theta _{0i,0}=-\frac{N_{i}}{N_{0}}\Theta _{0,i}$, $%
\frac{4\pi }{N_{0}}(\frac{N_{i}}{N_{0}}b+c)=-\frac{N_{i}}{N_{0}}\frac{8\pi }{%
N_{0}}a$, where when $N_{0}=4$ or $N_{i}\geq N_{0}^{2}/4$, the right-hand side becomes 0 (mod $2\pi$). Then we have:
\begin{equation}
\left\{\begin{array}{l}
\frac{4\pi }{N_{0}}(\frac{N_{i}}{N_{0}}b+c)=0\text{, \ \ if }N_{0}=4 \\ 
\frac{4\pi }{N_{0}}(\frac{N_{i}}{N_{0}}b+c)=0\text{, \ \ if }N_{0}>4\text{
and }N_{i}\geq N_{0}^{2}/4 \\ 
(\frac{N_{i}}{N_{0}}b+c)=-\frac{2N_{i}}{N_{0}}a\text{, \ \ if 
}N_{0}>4\text{ and }N_{i}<N_{0}^{2}/4
\end{array}\right.
\end{equation}

By the constraint \ $\Theta _{0i,i}=-\Theta _{i,0}$, $\frac{4\pi }{N_{0}}d=-%
\frac{2\pi }{N_{0}}b$, $b=-2d$.

The generating phases are determined by the integers:
\begin{equation}
\left\{\begin{array}{l}
d\text{ (mod }N_{0}/2\text{), \ \ if }N_{0}=4\\

a\text{ (mod 
}N_{0}/4\text{), }d\text{ (mod }N_{0}/2\text{), \ \ if }N_{0}>4
\end{array}\right.
\end{equation}

Hence in this case the classification is
\begin{equation}
\left\{\begin{array}{l}
\mathbb{Z}
_{N_{0}/2}=%
\mathbb{Z}
_{2\text{ }}\text{ (BSPT), \ \ if }N_{0}=4\\
{%
\mathbb{Z}
_{N_{0}/4}\times 
\mathbb{Z}
_{N_{0}/2}\text{ (BSPT), \ \ if }N_{0}>4}
\end{array}\right.
\end{equation}
where the indicators of the complex fermion layer are:
\begin{equation}
\left\{\begin{array}{l}
\Theta _{fi,0}=\frac{N_{0}}{2}\Theta _{0i,0}=0\text{, }\Theta _{fi,i}=\frac{%
N_{0}}{2}\frac{4\pi }{N_{0}}d=0\text{, \ \ if }N_{0}=4 \\ 
\Theta _{fi,0}=\frac{N_{0}}{2}\Theta _{0i,0}=0
\text{, }\Theta _{fi,i}=\frac{N_{0}}{2}\Theta _{0i,i}=0\text{, \ \ if }
N_{0}>4
\end{array}\right.
\end{equation}

\bigskip

For $m$ is even, combining the cases (1)(2) into (3)(4), in conclusion the
classification is:
\begin{equation}
\left\{\begin{array}{l}
\mathbb{Z}
_{N_{i}}\times 
\mathbb{Z}
_{N_{i}/2}\text{, \ \ if }N_{i}=N_{0}/2 \\ 

\mathbb{Z}
_{2N_{i}}\times 
\mathbb{Z}
_{N_{i}/2}\text{, \ \ if }N_{i}<N_{0}/2 \\ 

\mathbb{Z}
_{N_{0}/2}\times 
\mathbb{Z}
_{N_{0}/4}\text{, \ \ if }N_{0}\leq N_{i}
\end{array}\right.
\end{equation}
which means that the BSPT\ classification is $%
\mathbb{Z}
_{\min \{N_{i},N_{0}/2\}}\times 
\mathbb{Z}
_{\min \{N_{i},N_{0}/2\}/2}$, and a $%
\mathbb{Z}
_{2}$ complex fermion layer will be absorbed in the BSPT layer when $%
N_{i}<N_{0}/2$.

\subsection{Category (C)}

For simplicity, we only consider symmetry groups with order being power of 2. The newly involved constraints in 3D are:
\begin{equation}
N_{\sigma }\Theta _{\mu ,\sigma }=0
\end{equation}
\begin{equation}
N_{\sigma }\Theta _{\mu \nu ,\sigma }=0
\end{equation}
\begin{equation}
N_{\sigma }\Theta _{\mu \nu \lambda ,\sigma }=0
\end{equation}
\begin{equation}
\frac{N^{0ij}}{N^{ij}}\Theta _{ij,0}+\frac{N^{0ij}}{N^{0j}}\Theta _{0j,i}+%
\frac{N^{0ij}}{N^{0i}}\Theta _{0i,j}=0
\end{equation}
\begin{equation}
\Theta _{ij,i}=-\frac{N^{ij}}{\widetilde{N}_{i}}\Theta _{i,j}\text{, \ \ }%
\Theta _{ij,j}=-\frac{N^{ij}}{\widetilde{N}_{j}}\Theta _{j,i}
\end{equation}
\begin{equation}
\Theta _{0ij,0}=\Theta _{00i,j}=\Theta _{0ii,j}=-\Theta _{0ij,i}=-\Theta
_{00j,i}=-\Theta _{0jj,i}=-\Theta _{0ij,j}
\end{equation}

\subsubsection{$m$ is odd}

Similarly we also set $m=1$, so that we only need to consider the symmetry
group $%
\mathbb{Z}
_{2}\times 
\mathbb{Z}
_{N_{i}}\times 
\mathbb{Z}
_{N_{j}}$, and we assume $N_{i}\leq N_{j}$ without loss of generality.

(1) If $\frac{N_{i}}{2},\frac{N_{j}}{2}$ are odd (i.e. $%
\mathbb{Z}
_{2}^{f}\times 
\mathbb{Z}
_{2}\times 
\mathbb{Z}
_{2}$), invoking the known 2D results and combining with the 3D constraints $%
N_{\sigma }\Theta _{\mu ,\sigma }=0$, $N_{\sigma }\Theta _{\mu \nu ,\sigma
}=0$, $N_{\sigma }\Theta _{\mu \nu \lambda ,\sigma }=0$, the generating
phases for the sets (C1), (C2), (C3), (C4) and (C5) are:
\begin{equation}
(\Theta _{ij,0},\Theta _{0ij,0})=(\frac{\pi }{N_{ij}},\frac{2\pi }{N_{0ij}}%
)\times 2a+(0,\frac{4\pi }{N_{0ij}})=(\pi a,0)
\end{equation}
\begin{equation}
(\Theta _{ij,i},\Theta _{0ij,i})=(\frac{\pi }{N_{ij}},\frac{2\pi }{N_{0ij}}%
)\times 2b+(0,\frac{4\pi }{N_{0ij}})=(\pi b,0)
\end{equation}
\begin{equation}
(\Theta _{ij,j},\Theta _{0ij,j})=(\frac{\pi }{N_{ij}},\frac{2\pi }{N_{0ij}}%
)\times 2c+(0,\frac{4\pi }{N_{0ij}})=(\pi c,0)
\end{equation}
\begin{equation}
(\Theta _{i,j},\Theta _{0i,j},\Theta _{00i,j})=(\frac{\pi }{2N_{i}},-\frac{%
\pi }{N_{0i}},\pi )\times 4d+(0,\frac{4\pi }{N_{0i}},0)
=(\pi ,0,0)d
\end{equation}
\begin{equation}
(\Theta _{j,i},\Theta _{0j,i},\Theta _{00j,i})=(\frac{\pi }{2N_{j}},-\frac{%
\pi }{N_{0j}},\pi )\times 4e+(0,\frac{4\pi }{N_{0j}},0)
=(\pi ,0,0)e
\end{equation}
where $a,b,c,d,e$ are integers.

By the constraint $\Theta _{ij,0}+\Theta _{oj,i}+\Theta _{0i,j}=0$, $a=0$
(mod 2).

By the constraint $\Theta _{ij,i}=-\Theta _{i,j}$, $b=-d$ (mod 2).

By the constraint $\Theta _{ij,j}=-\Theta _{j,i}$, $c=-e$ (mod 2).

Hence in this case the classification is $%
\mathbb{Z}
_{2}\times 
\mathbb{Z}
_{2}$, which belongs to BSPT.

(2) If $\frac{N_{i}}{2}$ is odd and $\frac{N_{j}}{2}$ is even (i.e. $%
\mathbb{Z}
_{2}^{f}\times 
\mathbb{Z}
_{2}\times 
\mathbb{Z}
_{N_{j}}$), the generating phases for the sets (C1), (C2), (C3), (C4) and
(C5) are:
\begin{equation}
(\Theta _{ij,0},\Theta _{0ij,0})=(\frac{2\pi }{N_{ij}},0)\times a+(0,\frac{%
2\pi }{N_{0ij}})\times b=(\pi a,\pi b)
\end{equation}
\begin{equation}
(\Theta _{ij,i},\Theta _{0ij,i})=(\frac{2\pi }{N_{ij}},0)\times c+(0,\frac{%
2\pi }{N_{0ij}})\times d=(\pi c,\pi d)
\end{equation}
\begin{equation}
(\Theta _{ij,j},\Theta _{0ij,j})=(\frac{2\pi }{N_{ij}},0)\times e+(0,\frac{%
2\pi }{N_{0ij}})\times f=(\pi e,\pi f)
\end{equation}
\begin{equation}
(\Theta _{i,j},\Theta _{0i,j},\Theta _{00i,j})=
\left\{\begin{array}{l}
(\frac{\pi }{%
2N_{i}},-\frac{\pi }{N_{0i}},\pi )\times 2g+(0,\frac{4\pi }{N_{0i}},0)=(%
\frac{\pi }{2},-\pi ,0)g\text{, \ \ if }N_{j}=4\\
(\frac{\pi }{2N_{i}},-\frac{%
\pi }{N_{0i}},\pi )\times g+(0,\frac{4\pi }{N_{0i}},0)=(\frac{\pi }{4},-%
\frac{\pi }{2},\pi )g\text{, \ \ if }N_{j}>4
\end{array}\right.
\end{equation}
\begin{equation}
(\Theta _{j,i},\Theta _{0j,i},\Theta _{00j,i})=
\left\{\begin{array}{l}
(\frac{\pi }{%
N_{j}},\frac{2\pi }{N_{0j}},0)\times 4h+(0,\frac{2\pi }{N_{0j}},\pi )\times
i=(\pi h,\pi i,\pi i)\text{, \ \ if }N_{j}=4\\
(\frac{\pi }{N_{j}},\frac{2\pi 
}{N_{0j}},0)\times N_{j}h+(0,\frac{4\pi }{N_{0j}},\pi )\times i=(\pi h,0,\pi
i)\text{, \ \ if }N_{j}>4
\end{array}\right.
\end{equation}

By the constraint $\Theta _{ij,0}+\Theta _{oj,i}+\frac{N_{j}}{2}\Theta
_{0i,j}=0$,
\begin{equation}
\left\{\begin{array}{l}
\pi a+\pi i=0\text{, }a=i\text{ (mod 2), \ \ if }N_{j}=4\\
\pi a+\frac{N_{j}}{2}(-\frac{\pi }{2}g)=0\text{, }a=0\text{ (mod 2), \ \ if }%
N_{j}>4
\end{array}\right.
\end{equation}

By the constraint $\Theta _{ij,i}=-\frac{N_{j}}{2}\Theta _{i,j}$,
\begin{equation}
\left\{\begin{array}{l}
\pi c=-\pi g\text{, }c=-g\text{ (mod 4), \ \ if }N_{j}=4 \\ 
\pi c=-\frac{N_{j}}{2}(\frac{\pi }{4}g)\text{, }c=-g\text{ (mod 8), \ \ if }
N_{j}=8 \\ 
\pi c=-\frac{N_{j}}{2}(\frac{\pi }{4}g)\text{, }c=0\text{ (mod 2), \ \ if }%
N_{j}>8
\end{array}\right.
\end{equation}

By the constraint $\Theta _{ij,j}=-\Theta _{j,i}$, $\pi e=-\pi h$, $e=-h$
(mod 2).

By the constraint $\Theta _{0ij,0}=\Theta _{00i,j}=-\Theta _{0ij,i}=-\Theta
_{00j,i}=-\Theta _{0ij,j}$,
\begin{equation}
\left\{\begin{array}{l}
b=d=f=i=0\text{ (mod 2), \ \ if }N_{j}=4\\
b=d=f=g=i\text{
(mod 2), \ \ if }N_{j}>4
\end{array}\right.
\end{equation}

Combining all the constraints:
\begin{equation}
\left\{\begin{array}{l}
a=b=d=f=i=0\text{ (mod 2), }c=-g\text{ (mod 4), }e=-h\text{ (mod 2), \ \ if 
}N_{j}=4 \\ 
a=0\text{ (mod 2)}$, $b=d=f=g=i=-c\text{ (mod 8), }e=-h\text{ (mod 2), \ \
if }N_{j}=8 \\ 
a=0\text{ (mod 2)}, b=d=f=g=i\text{ (mod 8), }c=0\text{ (mod 2), }e=-h%
\text{ (mod 2), \ \ if }N_{j}>8
\end{array}\right.
\end{equation}

Hence in this case the classification is

\begin{equation}
\left\{\begin{array}{l}
\mathbb{Z}
_{4}\times 
\mathbb{Z}
_{2}\text{ (a }%
\mathbb{Z}
_{2}\text{ complex fermion layer aborbed into a }%
\mathbb{Z}
_{2}\times 
\mathbb{Z}
_{2}\text{\ BSPT layer), \ \ if }N_{i}=2\text{, }N_{j}=4\\

\mathbb{Z}
_{8}\times 
\mathbb{Z}
_{2}\text{ (a }%
\mathbb{Z}
_{2}\text{ Kitaev-chain layer further aborbed into the }%
\mathbb{Z}
_{4}\times 
\mathbb{Z}
_{2}\text{ above), \ \ if }N_{i}=2\text{, }N_{j}>4
\end{array}\right.
\end{equation}
where the indicator of the complex fermion layer is:
\begin{equation}
\left\{\begin{array}{l}
\Theta _{fi,j}=\Theta _{0i,j}=-\pi g\text{, \ \ if }N_{i}=2%
\text{, }N_{j}=4\\
\Theta _{fi,j}=\Theta _{0i,j}=-\frac{\pi }{2}g\text{, \ \
if }N_{i}=2\text{, }N_{j}>4
\end{array}\right.
\end{equation}

(3) If $\frac{N_{i}}{2},\frac{N_{j}}{2}$ are even (i.e. $%
\mathbb{Z}
_{2}^{f}\times 
\mathbb{Z}
_{N_{i}}\times 
\mathbb{Z}
_{N_{j}}$) and let $N_{i}\leq N_{j}$ without loss of generality, the
generating phases for the sets (C1), (C2), (C3), (C4) and (C5) are:
\begin{equation}
(\Theta _{ij,0},\Theta _{0ij,0})=(\frac{2\pi }{N_{ij}},0)\times \frac{N_{ij}%
}{2}a+(0,\frac{2\pi }{N_{0ij}})\times \frac{N_{0ij}}{2}b=(\pi a,\pi b)
\end{equation}
\begin{equation}
(\Theta _{ij,i},\Theta _{0ij,i})=(\frac{2\pi }{N_{ij}},0)\times c+(0,\frac{%
2\pi }{N_{0ij}})\times d=(\frac{2\pi }{N_{i}}c,\pi d)
\end{equation}
\begin{equation}
(\Theta _{ij,j},\Theta _{0ij,j})=(\frac{2\pi }{N_{ij}},0)\times e+(0,\frac{%
2\pi }{N_{0ij}})\times f=(\frac{2\pi }{N_{i}}e,\pi f)
\end{equation}
\begin{equation}
(\Theta _{i,j},\Theta _{0i,j},\Theta _{00i,j})=
\left\{\begin{array}{l}
(\frac{\pi }{N_{i}},\frac{2\pi }{N_{0i}},0)\times 2g+(0,\frac{2\pi }{N_{0i}}%
,\pi )\times h=(\frac{\pi }{2}g,\pi h,\pi h)\text{, \ \ if }N_{i}=N_{j}=4
\\ 
(\frac{\pi }{N_{i}},\frac{2\pi }{N_{0i}},0)\times g+(0,\frac{2\pi }{N_{0i}}%
,\pi )\times h=(\frac{\pi }{4}g,\pi (g+h),\pi h)\text{, \ \ if }N_{i}=4, 
N_{j}=8 \\ 
(\frac{\pi }{N_{i}},\frac{2\pi }{N_{0i}},0)\times 2g+(0,\frac{4\pi }{N_{0i}}%
,\pi )\times h=(\frac{2\pi }{N_{i}}g,0,\pi h)\text{, \ \ if }8\leq
N_{i}=N_{j} \\ 
(\frac{\pi }{N_{i}},\frac{2\pi }{N_{0i}},0)\times g+(0,\frac{4\pi }{N_{0i}}%
,\pi )\times h=(\frac{\pi }{N_{i}}g,\pi g,\pi h), \ \ \text{if }8\leq
N_{i}<N_{j}
\end{array}\right.
\end{equation}
\begin{equation}
(\Theta _{j,i},\Theta _{0j,i},\Theta _{00j,i})=
\left\{\begin{array}{l}
(\frac{\pi }{N_{j}},\frac{2\pi }{N_{0j}},0)\times 2l+(0,\frac{2\pi }{N_{0j}}%
,\pi )\times m=(\frac{\pi }{2}l,\pi m,\pi m)\text{, \ \ if }N_{i}=N_{j}=4
\\ 
(\frac{\pi }{N_{j}},\frac{2\pi }{N_{0j}},0)\times 4l+(0,\frac{4\pi }{N_{0j}}%
,\pi )\times m=(\frac{\pi }{2}l,0,\pi m)\text{, \ \ if }N_{i}=4$, $N_{j}=8
\\ 
(\frac{\pi }{N_{j}},\frac{2\pi }{N_{0j}},0)\times 2l+(0,\frac{4\pi }{N_{0j}}%
,\pi )\times m=(\frac{2\pi }{Nj}l,0,\pi m)\text{, \ \ if }8\leq N_{i}=N_{j}
\\ 
(\frac{\pi }{N_{j}},\frac{2\pi }{N_{0j}},0)\times \frac{2N_{j}}{N_{i}}l+(0,%
\frac{4\pi }{N_{0j}},\pi )\times m=(\frac{2\pi }{N_{i}}l,0,\pi m)$, \ \ $%
\text{if }8\leq N_{i}<N_{j}
\end{array}\right.
\end{equation}

By the constraint $\Theta _{ij,0}+\Theta _{oj,i}+\frac{Nj}{N_{i}}\Theta
_{0i,j}=0$,
\begin{equation}
\left\{\begin{array}{l}
\pi a+\pi m+\pi h=0\text{, \ \ if }N_{i}=N_{j}=4 \\ 
\pi a+\pi (g+h)=0$,$\text{ \ \ if }N_{i}=4, N_{j}=8 \\ 
\pi a=0\text{, \ \ if }8\leq N_{i}=N_{j} \\ 
\pi a+\pi g=0, \ \ \text{if }8\leq N_{i}<N_{j}
\end{array}\right.
\end{equation}

By the constraint $\Theta _{ij,i}=-\frac{Nj}{N_{i}}\Theta _{i,j}$,
\begin{equation}
\left\{\begin{array}{l}
c=-g\text{, \ \ if }N_{i}=N_{j}=4 \\ 
c=-g,\text{ \ \ if }N_{i}=4, N_{j}=8 \\ 
c=-g\text{, \ \ if }8\leq N_{i}=N_{j} \\ 
\frac{2\pi }{%
N_{i}}c=0, \ \ \text{if }8\leq N_{i}<N_{j} \text{ and } N_{j}\geq 2N_{i}^{2} \\ 
c=-\frac{Nj}{%
2N_{i}}g, \ \ \text{if }8\leq N_{i}<N_{j} \text{ and } N_{j}<2N_{i}^{2}
\end{array}\right.
\end{equation}

By the constraint $\Theta _{ij,j}=-\Theta _{j,i}$, $e=-l$.

By the constraint $\Theta _{0ij,0}=\Theta _{00i,j}=-\Theta _{0ij,i}=-\Theta
_{00j,i}=-\Theta _{0ij,j}$, $b=d=f=h=m$.

Combine all the constraints:
\begin{equation}
\left\{\begin{array}{l}
a=0\text{, }b=d=f=h=m\text{ (mod 2), }c=-g\text{ (mod 4), }e=-l\text{ (mod 4), \ \ if }%
N_{i}=N_{j}=4  \\ 
a=g+h\text{, }b=d=f=h=m\text{ (mod 2), }c=-g\text{ (mod 8), }e=-l\text{ (mod 4), \ \ if }N_{i}=4, 
N_{j}=8 \\ 
a=0\text{, }b=d=f=h=m\text{ (mod 2), }c=-g\text{ (mod }N_{i}\text{), }e=-l\text{ (mod }N_{i}\text{), \ \ if }8\leq
N_{i}=N_{j} \\ 
a=g\text{ (mod }2N_{i}\text{), }b=d=f=h=m\text{ (mod 2), }c=0\text{, }e=-l\text{ (mod }N_{i}\text{), } \ \ \text{if }
8\leq N_{i}<N_{j} \text{ and } N_{j}\geq 2N_{i}^{2} \\ 
a=g\text{ (mod }2N_{i}\text{), }b=d=f=h=m\text{ (mod 2), }c=-\frac{Nj}{2N_{i}}g\text{, }e=-l\text{ (mod }N_{i}\text{), } \
\ \text{if }8\leq N_{i}<N_{j} \text{ and } N_{j}<2N_{i}^{2}
\end{array}\right.
\end{equation}

Hence in this case the classification is
\begin{equation}
\left\{\begin{array}{l}
\mathbb{Z}
_{4}\times 
\mathbb{Z}
_{4}\times 
\mathbb{Z}
_{2}\text{ (a }%
\mathbb{Z}
_{4}\times 
\mathbb{Z}
_{4}\text{\ BSPT, stacking with a }%
\mathbb{Z}
_{2}\text{\ "Kitaev-chain layer absorbed in complex fermion layer"), \ \ }
\\ 
\text{if }N_{i}=N_{j}=4 \\ 

\mathbb{Z}
_{8}\times 
\mathbb{Z}
_{4}\times 
\mathbb{Z}
_{2}\text{ (a }%
\mathbb{Z}
_{2}\text{\ "complex fermion layer abrobed in a }%
\mathbb{Z}
_{4}\times 
\mathbb{Z}
_{4}\text{\ BSPT", }  \\
 \text{stacking with a }%

\mathbb{Z}
_{2}\text{\ "Kitaev-chain layer absorbed in complex fermion layer")}$,$\text{
\ \ if }N_{i}=4$, $N_{j}=8 \\ 

\mathbb{Z}
_{N_{i}}\times 
\mathbb{Z}
_{N_{i}}\times 
\mathbb{Z}
_{2}\text{ (a }%
\mathbb{Z}
_{N_{i}}\times 
\mathbb{Z}
_{N_{i}}\text{\ BSPT, stacking with a }%
\mathbb{Z}
_{2}\text{\ Kitaev-chain layer), \ \ if }8\leq N_{i}=N_{j} \\ 

\mathbb{Z}
_{2N_{i}}\times 
\mathbb{Z}
_{N_{i}}\times 
\mathbb{Z}
_{2}$ $\text{(a }%
\mathbb{Z}
_{2}\text{\ complex fermion layer abrobed in }%
\mathbb{Z}
_{N_{i}}\times 
\mathbb{Z}
_{N_{i}}\text{\ BSPT, stacking with a }%
\mathbb{Z}
_{2}\text{\ Kitaev-chain layer)}, \\ 
\text{if }8\leq N_{i}<N_{j}
\end{array}\right.
\end{equation}
where the complex fermion layer indicators are $\Theta _{fi,j}=\Theta _{0i,j}$ and $\Theta _{fj,i}=\Theta _{0j,i}$. And the classification can be simplified as:
\begin{equation}
\left\{\begin{array}{l}
\mathbb{Z}
_{N_{i}}\times 
\mathbb{Z}
_{Ni}\times 
\mathbb{Z}
_{2}\text{, \ \ if }N_{i}=N_{j} \\ 

\mathbb{Z}
_{2N_{i}}\times 
\mathbb{Z}
_{Ni}\times 
\mathbb{Z}
_{2}, \ \ \text{if }N_{i}\neq N_{j}
\end{array}\right.
\end{equation}

\subsubsection{$m$ is even}

By the constraint $m\Theta _{00i,j}=\Theta _{00i,j}$, $m\Theta
_{00j,i}=\Theta _{00j,i}$, we have $\Theta _{00i,j}=0$, $\Theta _{00j,i}=0$.

By the constraint $\Theta _{0ij,0}=\Theta _{00i,j}=-\Theta _{0ij,i}=-\Theta
_{00j,i}=-\Theta _{0ij,j}$, we have $\Theta _{0ij,0}=\Theta _{0ij,i}=\Theta
_{0ij,j}=0$, which means that there is no non-Abelian statistics in this
case.

(1) If $\frac{N_{i}}{2},\frac{N_{j}}{2}$ are odd (i.e. $%
\mathbb{Z}
_{N_{0}}^{f}\times 
\mathbb{Z}
_{2}\times 
\mathbb{Z}
_{2}$), the generating phases for the sets (C1), (C2), (C3), (C4) and (C5)
are:
\begin{equation}
(\Theta _{ij,0},\Theta _{0ij,0})=(\frac{2\pi }{N_{ij}},0)\times a+(0,\frac{%
2\pi }{N_{0ij}})\times b=(\pi a,\pi b)
\end{equation}
\begin{equation}
(\Theta _{ij,i},\Theta _{0ij,i})=(\frac{2\pi }{N_{ij}},0)\times c+(0,\frac{%
2\pi }{N_{0ij}})\times d=(\pi c,\pi d)
\end{equation}
\begin{equation}
(\Theta _{ij,j},\Theta _{0ij,j})=(\frac{2\pi }{N_{ij}},0)\times e+(0,\frac{%
2\pi }{N_{0ij}})\times f=(\pi e,\pi f)
\end{equation}
\begin{equation}
(\Theta _{i,j},\Theta _{0i,j},\Theta _{00i,j})=(\frac{\pi }{N_{i}},\frac{%
2\pi }{N_{0i}},0)\times 2g+(0,\frac{4\pi }{N_{0i}},0)=(\pi ,0,0)g
\end{equation}
\begin{equation}
(\Theta _{j,i},\Theta _{0j,i},\Theta _{00j,i})=(\frac{\pi }{N_{j}},\frac{%
2\pi }{N_{0j}},0)\times 2h+(0,\frac{4\pi }{N_{0j}},0)=(\pi ,0,0)h
\end{equation}

By the constraint $\Theta _{0ij,0}=\Theta _{0ij,i}=\Theta _{0ij,j}=\Theta
_{00i,j}=\Theta _{00j,i}=0$, $b=d=f=0$ (mod 2).

By the constraint $\frac{N_{0}}{2}\Theta _{ij,0}+\Theta _{oj,i}+\Theta
_{0i,j}=0$, $\frac{N_{0}}{2}\pi a=0$, which is always satisfied.

By the constraint $\Theta _{ij,i}=-\Theta _{i,j}$, $c=-g$ (mod 2).

By the constraint $\Theta _{ij,j}=-\Theta _{j,i}$, $e=-h$ (mod 2).

The generating phases are determined by integers $a$ (mod 2), $g$ (mod 2), $%
h $ (mod 2). And the classification is $%
\mathbb{Z}
_{2}\times 
\mathbb{Z}
_{2}\times 
\mathbb{Z}
_{2}$ (BSPT).

(2) If $\frac{N_{i}}{2}$ is odd and $\frac{N_{j}}{2}$ is even (i.e. $%
\mathbb{Z}
_{N_{0}}^{f}\times 
\mathbb{Z}
_{2}\times 
\mathbb{Z}
_{N_{j}}$), the generating phases for the sets (C1), (C2), (C3), (C4) and
(C5) are:
\begin{equation}
(\Theta _{ij,0},\Theta _{0ij,0})=(\frac{2\pi }{N_{ij}},0)\times a+(0,\frac{%
2\pi }{N_{0ij}})\times 2b=(\pi a,0)
\end{equation}
\begin{equation}
(\Theta _{ij,i},\Theta _{0ij,i})=(\frac{2\pi }{N_{ij}},0)\times c+(0,\frac{%
2\pi }{N_{0ij}})\times 2d=(\pi c,0)
\end{equation}
\begin{equation}
(\Theta _{ij,j},\Theta _{0ij,j})=(\frac{2\pi }{N_{ij}},0)\times e+(0,\frac{%
2\pi }{N_{0ij}})\times 2f=(\pi e,0)
\end{equation}
\begin{equation}
(\Theta _{i,j},\Theta _{0i,j},\Theta _{00i,j})=(\frac{\pi }{N_{i}},\frac{%
2\pi }{N_{0i}},0)\times g+(0,\frac{4\pi }{N_{0i}},0)\times h=(\frac{\pi }{2}%
,\pi ,0)g
\end{equation}
\begin{equation}
(\Theta _{j,i},\Theta _{0j,i},\Theta _{00j,i})=(\frac{\pi }{N_{j}},\frac{%
2\pi }{N_{0j}},0)\times N_{j}l+(0,\frac{4\pi }{N_{0j}},0)\times \frac{N_{0j}%
}{4}m=(\pi l,\pi m,0)
\end{equation}

By the constraint
\begin{equation}
\left\{\begin{array}{l}
\Theta _{ij,0}+\Theta _{oj,i}+\frac{N_{j}}{N_{0}}\Theta _{0i,j}=0\text{, }%
\pi a+\pi m+\frac{N_{j}}{N_{0}}(\pi g)=0\text{, }a=m\text{ (mod 2), \ \ if }%
N_{0}<N_{j} \\ 
\Theta _{ij,0}+\Theta _{oj,i}+\Theta _{0i,j}=0\text{, }\pi a+\pi m+\pi g=0%
\text{, \ \ if }N_{0}=N_{j} \\ 
\frac{N_{0}}{N_{j}}\Theta _{ij,0}+\Theta _{oj,i}+\Theta _{0i,j}=0\text{, }%
\frac{N_{0}}{N_{j}}\pi a+\pi m+\pi g=0\text{, }m=g\text{ (mod 4) \ \ if }%
N_{j}<N_{0}
\end{array}\right.
\end{equation}
where $g$ is chosen as a generating phase, only $a,m$ need to be considered
here.

By the constraint $\Theta _{ij,i}=-\frac{N_{j}}{2}\Theta _{i,j}$, $\pi c=-%
\frac{N_{j}}{2}(\frac{\pi }{2}g)$,
\begin{equation}
\left\{\begin{array}{l}
\pi c=-\pi g\text{, }c=g\text{ (mod 4), \ \ if }N_{j}=4 \\ 
\pi c=0\text{, }c=0\text{ (mod 2), \ \ if }N_{j}>4
\end{array}\right.
\end{equation}

By the constraint $\Theta _{ij,j}=-\Theta _{j,i}$, $\pi e=-\pi l$, $e=l$
(mod 2).

Combine all the constraints:
\begin{equation}
\left\{\begin{array}{l}
a=m\text{ (mod 2), }g\text{ (mod 4), }c=0\text{ (mod 2)}, e=l \text{(mod 2)}
\text{, \ \ if }N_{0}<N_{j} \\ 
a\text{ (mod 2), }c=g\text{ (mod 4), }e=l\text{ (mod 2), \ \ if }%
4=N_{j}=N_{0} \\ 
a\text{ (mod 2)}, g=m=c \text{(mod 4)},\text{ }e=l\text{ (mod 2), \ \ if }
4=N_{j}<N_{0} \\ 
a\text{ (mod 2), }g\text{ (mod 4), }c=0\text{ (mod 2), }e=l\text{ (mod 2),,
\ \ if }4<N_{j}=N_{0} \\ 
a\text{ (mod 2)}$, $g=m\text{ (mod 4), }c=0\text{ (mod 2), }e=l\text{ (mod
2), \ \ if }4<N_{j}<N_{0}%
\end{array}\right.
\end{equation}
where the generating phases are determined by:
\begin{equation}
a\text{ (mod 2), }g\text{ (mod 4), }l\text{ (mod 2), \ \ for all cases}
\end{equation}

Hence in this case the classification is $%
\mathbb{Z}
_{4}\times 
\mathbb{Z}
_{2}\times 
\mathbb{Z}
_{2}$ (a $\mathbb{Z}_{2}$ complex fermion layer absorbed into a $\mathbb{Z}_{2}\times\mathbb{Z}_{2}\times\mathbb{Z}_{2}$ BSPT layer), and the complex layer indicators are:
\begin{equation}
\Theta _{fi,j}=\Theta _{0i,j}=-\pi g\text{, }
\Theta _{fj,i}=\frac{N_{0}}{2}\Theta _{0j,i}=0\text{, \ \ for all cases }
\end{equation}

(3) If $\frac{N_{i}}{2},\frac{N_{j}}{2}$ are even (i.e. $%
\mathbb{Z}
_{N_{0}}^{f}\times 
\mathbb{Z}
_{N_{i}}\times 
\mathbb{Z}
_{N_{j}}$) and $N_{i}=N_{j}$, the generating phases for the sets (C1), (C2),
(C3), (C4) and (C5) are:
\begin{equation}
(\Theta _{ij,0},\Theta _{0ij,0})=
\left\{\begin{array}{l}
(\frac{2\pi }{N_{ij}},0)\times a,(0,\frac{2\pi }{N_{0ij}})\times N_{0ij}b=(%
\frac{2\pi }{N_{i}}a,0)\text{, \ \ if }N_{0}\geq N_{i} \\ 
(\frac{2\pi }{N_{ij}},0)\times \frac{N_{ij}}{N_{0}}a,(0,\frac{2\pi }{N_{0ij}%
})\times N_{0ij}b=(\frac{2\pi }{N_{0}}a,0)\text{, \ \ if }N_{0}\leq N_{i}
\end{array}\right.
\end{equation}
which can be simplified as (as $N_{0},N_{i}$ are powers of 2, $N_{0i}=\min
\{N_{0},N_{i}\}$):
\begin{equation}
(\Theta _{ij,0},\Theta _{0ij,0})=(\frac{2\pi }{N_{0i}}a,0)
\end{equation}
\begin{equation}
(\Theta _{ij,i},\Theta _{0ij,i})=(\frac{2\pi }{N_{ij}},0)\times c+(0,\frac{%
2\pi }{N_{0ij}})\times N_{0ij}d=(\frac{2\pi }{N_{i}}c,0)
\end{equation}
\begin{equation}
(\Theta _{ij,j},\Theta _{0ij,j})=(\frac{2\pi }{N_{ij}},0)\times e+(0,\frac{%
2\pi }{N_{0ij}})\times N_{0ij}f=(\frac{2\pi }{N_{i}}e,0)
\end{equation}
\begin{equation}
\Theta _{i,j},\Theta _{0i,j},\Theta _{00i,j})=(\frac{\pi }{N_{i}},\frac{2\pi 
}{N_{0i}},0)\times 2g+(0,\frac{4\pi }{N_{0i}},0)\times h=(\frac{2\pi }{N_{i}}%
g,\frac{4\pi }{N_{0i}}g+\frac{4\pi }{N_{0i}}h,0)
\end{equation}
\begin{equation}
(\Theta _{j,i},\Theta _{0j,i},\Theta _{00j,i})=(\frac{\pi }{N_{i}},\frac{%
2\pi }{N_{0i}},0)\times 2l+(0,\frac{4\pi }{N_{0i}},0)\times m=(\frac{2\pi }{%
N_{i}}l,\frac{4\pi }{N_{0i}}l+\frac{4\pi }{N_{0i}}m,0)
\end{equation}

By the constraint:
\begin{equation}
\left\{\begin{array}{l}
\Theta _{ij,0}+\Theta _{oj,i}+\Theta _{0i,j}=0$, $\frac{2\pi }{N_{0i}}a+%
\frac{4\pi }{N_{0i}}l+\frac{4\pi }{N_{0i}}m+\frac{4\pi }{N_{0i}}g+\frac{4\pi 
}{N_{0i}}h=0\text{, \ \ if }N_{0}\leq N_{i} \text{ and } N_{0}/2<N_{i} \\ 
\frac{N_{0}}{N_{i}}\Theta _{ij,0}+\Theta _{oj,i}+\Theta _{0i,j}=0, \frac{%
4\pi }{N_{i}}a+\frac{4\pi }{N_{0i}}l+\frac{4\pi }{N_{0i}}m+\frac{4\pi }{%
N_{0i}}g+\frac{4\pi }{N_{0i}}h=0\text{, \ \ if }N_{0}>N_{i} \text{ and }
N_{0}/2=N_{i} \\ 
\frac{N_{0}}{N_{i}}\Theta _{ij,0}+\Theta _{oj,i}+\Theta _{0i,j}=0, \frac{%
N_{0}}{N_{i}}(\frac{2\pi }{N_{i}}a)+\frac{4\pi }{N_{0i}}l+\frac{4\pi }{N_{0i}%
}m+\frac{4\pi }{N_{0i}}g+\frac{4\pi }{N_{0i}}h=0\text{, \ \ if }N_{0}>N_{i}
\text{ and } N_{0}/2>N_{i}
\end{array}\right.
\end{equation}
where $g,l$ are chosen as generating phases, only $a,h,m$ need to be
considered here.

By the constraint $\Theta _{ij,i}=-\frac{N_{j}}{2}\Theta _{i,j}$, $\frac{2\pi }{N_{i}}c=\pi g$, where the solution is:
\begin{equation}
c=0,\frac{N_{i}}{2}\text{ (mod }N_{i}%
\text{) depending on }g\text{ (mod }N_{i}\text{)}
\end{equation}

By the constraint $\Theta _{ij,j}=-\Theta _{j,i}$, $\frac{2\pi }{N_{i}}e=-%
\frac{2\pi }{N_{i}}l$, $e=-l$ (mod $N_{i}$).

Combine all the constraints, the generating phases are:
\begin{equation}
\left\{\begin{array}{l}
g\text{ (mod }N_{i}\text{), }l\text{ (mod }N_{i}\text{), }a\text{ or }m%
\text{ (mod }N_{0}/2\text{),}\ h\text{ (mod }N_{0i}/2\text{), \ \ if \ if }%
N_{0}\leq N_{i} \text{ and } N_{0}/2<N_{i} \\ 
g\text{ (mod }N_{i}\text{), }l\text{ (mod }N_{i}\text{), }a\text{ (mod }%
N_{i}\text{),}\ h\text{ (mod }N_{0i}/2\text{), \ \ if }N_{0}>N_{i} \text{ and }
N_{0}/2=N_{i} \\ 
g\text{ (mod }N_{i}\text{), }l\text{ (mod }N_{i}\text{), }a\text{ (mod }%
N_{i}\text{),}\ h\text{ (mod }N_{0i}/2\text{), \ \ if }N_{0}>N_{i} \text{ and } 
N_{0}/2>N_{i}
\end{array}\right.
\end{equation}

Hence the classification is $%
\mathbb{Z}
_{N_{i}}\times 
\mathbb{Z}
_{N_{i}}\times 
\mathbb{Z}
_{\min \{N_{0}/2,N_{i}\}}\times 
\mathbb{Z}
_{N_{0i}/2}$ (BSPT).

(4) If $\frac{N_{i}}{2},\frac{N_{j}}{2}$ are even (i.e. $%
\mathbb{Z}
_{N_{0}}^{f}\times 
\mathbb{Z}
_{N_{i}}\times 
\mathbb{Z}
_{N_{j}}$) and $N_{i}<N_{j}$, the generating phases for the sets (C1), (C2),
(C3), (C4) and (C5) are:
\begin{equation}
(\Theta _{ij,0},\Theta _{0ij,0})=(\frac{2\pi }{N_{0i}}a,0)
\end{equation}
\begin{equation}
(\Theta _{ij,i},\Theta _{0ij,i})=(\frac{2\pi }{N_{ij}},0)\times c+(0,\frac{%
2\pi }{N_{0ij}})\times N_{0ij}d=(\frac{2\pi }{N_{i}}c,0)
\end{equation}
\begin{equation}
(\Theta _{ij,j},\Theta _{0ij,j})=(\frac{2\pi }{N_{ij}},0)\times e+(0,\frac{%
2\pi }{N_{0ij}})\times N_{0ij}f=(\frac{2\pi }{N_{i}}e,0)
\end{equation}
\begin{equation}
(\Theta _{i,j},\Theta _{0i,j},\Theta _{00i,j})=(\frac{\pi }{N_{i}},\frac{2\pi 
}{N_{0i}},0)\times g+(0,\frac{4\pi }{N_{0i}},0)\times h=(\frac{\pi }{N_{i}}g,%
\frac{2\pi }{N_{0i}}g+\frac{4\pi }{N_{0i}}h,0)
\end{equation}
\begin{equation}
(\Theta _{j,i},\Theta _{0j,i},\Theta _{00j,i})
=\left\{\begin{array}{l}
(\frac{\pi }{N_{j}},\frac{2\pi }{N_{0j}},0)\times \frac{2N_{j}}{N_{i}}l+(0,%
\frac{4\pi }{N_{0j}},0)\times m=(\frac{2\pi }{N_{i}}l,\frac{4\pi N_{j}}{%
N_{0}N_{i}}l+\frac{4\pi }{N_{0}}m,0),\text{ \ \ if }N_{0}\leq N_{j} \text{ and } 
N_{0}/2\leq N_{i} \\ 
(\frac{\pi }{N_{j}},\frac{2\pi }{N_{0j}},0)\times \frac{2N_{j}}{N_{i}}l+(0,%
\frac{4\pi }{N_{0j}},0)\times \frac{N_{0}}{2N_{i}}m=(\frac{2\pi }{N_{i}}l,%
\frac{4\pi N_{j}}{N_{0}N_{i}}l+\frac{2\pi }{N_{i}}m,0),\text{ \ \ if }%
N_{0}\leq N_{j} \text{ and }  N_{0}/2>N_{i} \\ 
(\frac{\pi }{N_{j}},\frac{2\pi }{N_{0j}},0)\times \frac{2N_{j}}{N_{i}}l+(0,%
\frac{4\pi }{N_{0j}},0)\times m=(\frac{2\pi }{N_{i}}l,\frac{4\pi }{N_{i}}l+\frac{%
2\pi }{N_{i}}m,0),\text{ \ \ if }N_{j}\leq N_{0} \text{ and }  N_{j}/2=N_{i} \\ 
(\frac{\pi }{N_{j}},\frac{2\pi }{N_{0j}},0)\times \frac{2N_{j}}{N_{i}}l+(0,%
\frac{4\pi }{N_{0j}},0)\times \frac{N_{j}}{2N_{i}}m=(\frac{2\pi }{N_{i}}l,%
\frac{4\pi }{N_{i}}l+\frac{2\pi }{N_{i}}m,0),\text{ \ \ if }N_{j}\leq
N_{0} \text{ and }  N_{j}/2>N_{i}
\end{array}\right.
\end{equation}

By the constraint
\begin{equation}
\left\{\begin{array}{l}
\Theta _{ij,0}+\Theta _{oj,i}+\frac{N_{j}}{N^{oi}}\Theta _{0i,j}=0, \frac{%
2\pi }{N_{0i}}a+\frac{4\pi N_{j}}{N_{0}N_{i}}l+\frac{4\pi }{N_{0}}m+\frac{%
N_{j}}{N^{0i}}(\frac{2\pi }{N_{0i}}g+\frac{4\pi }{N_{0i}}h)=0\text{, \ \ if }%
N_{0}\leq N_{j} \text{ and }  N_{0}/2\leq N_{i} \\ 
\Theta _{ij,0}+\Theta _{oj,i}+\frac{N_{j}}{N^{oi}}\Theta _{0i,j}=0, \frac{%
2\pi }{N_{0i}}a+\frac{4\pi N_{j}}{N_{0}N_{i}}l+\frac{2\pi }{N_{i}}m+\frac{%
N_{j}}{N^{0i}}(\frac{2\pi }{N_{0i}}g+\frac{4\pi }{N_{0i}}h)=0\text{, \ \ if }%
N_{0}\leq N_{j} \text{ and }  N_{0}/2>N_{i} \\ 
\frac{N_{0}}{N_{j}}\Theta _{ij,0}+\Theta _{oj,i}+\Theta _{0i,j}=0, \frac{%
N_{0}}{N_{j}}(\frac{2\pi }{N_{i}}a)+\frac{4\pi }{N_{i}}l+\frac{2\pi }{N_{i}}%
m+\frac{2\pi }{N_{i}}g+\frac{4\pi }{N_{i}}h=0\text{, \ \ if }N_{j}\leq N_{0}
\text{ and }  N_{j}/2\geq N_{i}
\end{array}\right.
\end{equation}
where $g,l$ are chosen as generating phases, only $a,m,h$ need to be
considered here.

By the constraint $\Theta _{ij,i}=-\frac{N_{j}}{2}\Theta _{i,j}$, $\frac{%
2\pi }{N_{i}}c=-\frac{N_{j}}{2}(\frac{\pi }{N_{i}}g)$, written further as:
\begin{equation}
\left\{\begin{array}{l}
\frac{2\pi }{N_{i}}c=\pi g\text{, }c=0,\frac{N_{i}}{2}\text{ (mod }N_{i}%
\text{) depending on }g\text{ (mod }2N_{i}\text{), \ \ if }N_{j}=2N_{i} \\ 
\frac{2\pi }{N_{i}}c=-\frac{N_{j}}{2}(\frac{\pi }{N_{i}}g)\text{, }c=0\text{
(mod }N_{i}\text{) and }g\text{ (mod }2N_{i}\text{), \ \ if }N_{j}>2N_{i}
\end{array}\right.
\end{equation}

By the constraint $\Theta _{ij,j}=-\Theta _{j,i}$, $\frac{2\pi }{N_{i}}e=-%
\frac{2\pi }{N_{i}}l$, $e=-l$ (mod $N_{i}$).

Combine all the constraints, the generating phases are:
\begin{equation}
g\text{ (mod }2N_{i}\text{), }l\text{ (mod }N_{i}\text{), }h\text{ (mod }%
N_{0i}/2\text{), }m\text{ (mod }\min \{N_{0}/2,N_{i}\}\text{)}
\end{equation}

Hence the classification is $
\mathbb{Z}_{2N_{i}}\times 
\mathbb{Z}_{N_{i}}\times 
\mathbb{Z}_{\min \{N_{0}/2,N_{i}\}}\times 
\mathbb{Z}_{N_{0i}/2}$ (a $\mathbb{Z}_{2}$ complex fermion layer absorbed into a $\mathbb{Z}_{N_{i}}\times 
\mathbb{Z}_{N_{i}}\times 
\mathbb{Z}_{\min \{N_{0}/2,N_{i}\}}\times 
\mathbb{Z}_{N_{0i}/2}$ BSPT layer), where the complex fermion layer indicators are:
\begin{equation}
\Theta _{fi,j}=\frac{N_{0i}}{2}\Theta _{0i,j}=\pi g\text{, }
\Theta _{fj,i}=\frac{N_{0j}}{2}\Theta _{0j,i}=0
\end{equation}

\end{widetext}

\subsection{Category (D)}

The newly involved constraints in 3D are:
\begin{equation}
N_{k}\Theta _{ij,k}=0\text{, \ \ }N_{i}\Theta _{jk,i}=0\text{, \ \ }%
N_{j}\Theta _{ki,j}=0
\end{equation}
\begin{equation}
N_{k}\Theta _{0ij,k}=0\text{, \ \ }N_{i}\Theta _{0jk,i}=0\text{, \ \ }%
N_{j}\Theta _{0ki,j}=0
\end{equation}
\begin{equation}
\frac{N^{ijk}}{N^{ij}}\Theta _{ij,k}+\frac{N^{ijk}}{N^{jk}}\Theta _{jk,i}+%
\frac{N^{ijk}}{N^{ki}}\Theta _{ki,j}=0
\end{equation}
\begin{equation}
\Theta _{0ij,k}=\Theta _{ijk,0}=\Theta _{0jk,i}=-\Theta _{0ki,j}
\end{equation}
\begin{equation}
\Theta _{ijk,i}=\Theta _{iij,k}=\Theta _{ijj,k}=\Theta _{jjk,i}=\Theta
_{jkk,i}=\Theta _{kki,j}
\end{equation}
where subset (D4) can be totally absorbed into (D1), (D2) and (D3).

\subsubsection{$m$ is odd}

Set $m=1$ (i.e. $G_{f}=%
\mathbb{Z}
_{2}^{f}\times 
\mathbb{Z}
_{N_{i}}\times 
\mathbb{Z}
_{N_{j}}\times 
\mathbb{Z}
_{N_{k}}$) and assume $N_{i}\leq N_{j}\leq N_{k}$ without loss of generality.

(1) If $\frac{N_{i}}{2},\frac{N_{j}}{2},\frac{N_{k}}{2}$ are all odd (i.e. $%
\mathbb{Z}
_{2}^{f}\times 
\mathbb{Z}
_{2}\times 
\mathbb{Z}
_{2}\times 
\mathbb{Z}
_{2}$), invoking the known 2D results and combining with the 3D constraints $%
N_{\sigma }\Theta _{\mu ,\sigma }=0$, $N_{\sigma }\Theta _{\mu \nu ,\sigma
}=0$, $N_{\sigma }\Theta _{\mu \nu \lambda ,\sigma }=0$, the generating
phases for the sets (D1), (D2) and (D3) are:
\begin{equation}
(\Theta _{ij,k},\Theta _{0ij,k})=(\frac{\pi }{N_{ij}},\frac{2\pi }{N_{0ij}}%
)\times 2a+(0,\frac{4\pi }{N_{0ij}})=(\pi a,0)
\end{equation}
\begin{equation}
(\Theta _{jk,i},\Theta _{0jk,i})=(\frac{\pi }{N_{jk}},\frac{2\pi }{N_{0jk}}%
)\times 2b+(0,\frac{4\pi }{N_{0jk}})=(\pi b,0)
\end{equation}
\begin{equation}
(\Theta _{ki,j},\Theta _{0ki,j})=(\frac{\pi }{N_{ki}},\frac{2\pi }{N_{0ki}}%
)\times 2c+(0,\frac{4\pi }{N_{0ki}})=(\pi c,0)
\end{equation}
where $a,b,c$ are integers.

By the constraint $\Theta _{ij,k}+\Theta _{jk,i}+\Theta _{ki,j}=0$, $\pi
a+\pi b+\pi c=0$.

Hence in this case the classification is $%
\mathbb{Z}
_{2}\times 
\mathbb{Z}
_{2}$, which belongs to BSPT.

(2) If $\frac{N_{i}}{2},\frac{N_{j}}{2}$ are odd and $\frac{N_{k}}{2}$ is
even (i.e. $%
\mathbb{Z}
_{2}^{f}\times 
\mathbb{Z}
_{2}\times 
\mathbb{Z}
_{2}\times 
\mathbb{Z}
_{N_{k}}$), the generating phases for the sets (D1), (D2) and (D3) are:
\begin{equation}
(\Theta _{ij,k},\Theta _{0ij,k})=(\frac{\pi }{N_{ij}},\frac{2\pi }{N_{0ij}}%
)\times a+(0,\frac{4\pi }{N_{0ij}})=(\frac{\pi }{2}a,\pi a)
\end{equation}
\begin{equation}
(\Theta _{jk,i},\Theta _{0jk,i})=(\frac{2\pi }{N_{jk}},0)\times b+(0,\frac{%
2\pi }{N_{0jk}})\times c=(\pi b,\pi c)
\end{equation}
\begin{equation}
(\Theta _{ki,j},\Theta _{0ki,j})=(\frac{2\pi }{N_{ki}},0)\times d+(0,\frac{%
2\pi }{N_{0ki}})\times e=(\pi d,\pi e)
\end{equation}

By the constraint $\frac{N_{k}}{2}\Theta _{ij,k}+\Theta _{jk,i}+\Theta
_{ki,j}=0$, $\frac{N_{k}}{2}\frac{\pi }{2}a+\pi b+\pi d=0$.

By the constraint $\Theta _{0ij,k}=\Theta _{0jk,i}=-\Theta _{0ki,j}$, $a=c=e$
(mod $4$).

Combine the two constraints: $a=c=e$ (mod $4$), $b=d$ or $b=d+1$ (mod $2$).

Hence in this case the classification is $%
\mathbb{Z}
_{4}\times 
\mathbb{Z}
_{2}$, which is a $%
\mathbb{Z}
_{2}$\ non-Abelian complex fermion layer absorbed into a $%
\mathbb{Z}
_{2}\times 
\mathbb{Z}
_{2}$ BSPT layer.

(3) If $\frac{N_{i}}{2}$ is odd and $\frac{N_{j}}{2},\frac{N_{k}}{2}$ are
even, or $\frac{N_{i}}{2},\frac{N_{j}}{2},\frac{N_{k}}{2}$ are all even
(i.e. $%
\mathbb{Z}
_{2}^{f}\times 
\mathbb{Z}
_{2}\times 
\mathbb{Z}
_{N_{j}}\times 
\mathbb{Z}
_{N_{k}}$ or $%
\mathbb{Z}
_{2}^{f}\times 
\mathbb{Z}
_{N_{i}}\times 
\mathbb{Z}
_{N_{j}}\times 
\mathbb{Z}
_{N_{k}}$), the generating phases for the sets (D1), (D2) and (D3) are:
\begin{equation}
(\Theta _{ij,k},\Theta _{0ij,k})=(\frac{2\pi }{N_{ij}},0)\times a+(0,\frac{%
2\pi }{N_{0ij}})\times b=(\frac{2\pi }{N_{i}}a,\pi b)
\end{equation}
\begin{equation}
(\Theta _{jk,i},\Theta _{0jk,i})=(\frac{2\pi }{N_{jk}},0)\times \frac{N_{j}}{%
N_{i}}c+(0,\frac{2\pi }{N_{0jk}})\times d=(\frac{2\pi }{N_{i}}c,\pi d)
\end{equation}
\begin{equation}
(\Theta _{ki,j},\Theta _{0ki,j})=(\frac{2\pi }{N_{ki}},0)\times e+(0,\frac{%
2\pi }{N_{0ki}})\times f=(\frac{2\pi }{N_{i}}e,\pi f)
\end{equation}

By the constraint $\frac{N_{k}}{N_{j}}\Theta _{ij,k}+\Theta _{jk,i}+\Theta
_{ki,j}=0$, $\frac{N_{k}}{N_{j}}\frac{2\pi }{N_{i}}a+\frac{2\pi }{N_{i}}c+%
\frac{2\pi }{N_{i}}e=0$.

By the constraint $\Theta _{0ij,k}=\Theta _{0jk,i}=-\Theta _{0ki,j}$, $b=d=f$
(mod 2).

Combine the two constraints, the remaining three generating phases are: $a$
(mod $N_{i}$), $c$ (mod $N_{i}$), $b=d=f$ (mod 2).

Hence in this case the classification is $%
\mathbb{Z}
_{N_{i}}\times 
\mathbb{Z}
_{N_{i}}\times 
\mathbb{Z}
_{2}$ (or $%
\mathbb{Z}
_{N_{ijk}}\times 
\mathbb{Z}
_{N_{ijk}}\times 
\mathbb{Z}
_{2}$), which is a simple stacking of a $%
\mathbb{Z}
_{N_{i}}\times 
\mathbb{Z}
_{N_{i}}$\ BSPT layer and a\textbf{\ }$%
\mathbb{Z}
_{2}$ non-Abelian complex fermion layer.

\subsubsection{$m$ is even}

For symmetry group $%
\mathbb{Z}
_{N_{0}}^{f}\times 
\mathbb{Z}
_{N_{i}}\times 
\mathbb{Z}
_{N_{j}}\times 
\mathbb{Z}
_{N_{k}}$, the generating phases for the sets (D1), (D2) and (D3) are:
\begin{eqnarray}
(\Theta _{ij,k},\Theta _{0ij,k})&=&(\frac{2\pi }{N_{ij}},0)\times a+(0,\frac{%
2\pi }{N_{0ij}})\times \frac{N_{0ij}}{N_{0ijk}}b
\nonumber\\&=&(\frac{2\pi }{N_{i}}a,\frac{%
2\pi }{N_{0ijk}}b)
\end{eqnarray}
\begin{eqnarray}
(\Theta _{jk,i},\Theta _{0jk,i})&=&(\frac{2\pi }{N_{jk}},0)\times \frac{N_{j}}{%
N_{i}}c+(0,\frac{2\pi }{N_{0jk}})\times \frac{N_{0jk}}{N_{0ijk}}d
\nonumber\\&=&(\frac{%
2\pi }{N_{i}}c,\frac{2\pi }{N_{0ijk}}d)
\end{eqnarray}
\begin{eqnarray}
(\Theta _{ki,j},\Theta _{0ki,j})&=&(\frac{2\pi }{N_{ki}},0)\times e+(0,\frac{%
2\pi }{N_{0ki}})\times \frac{N_{0ki}}{N_{0ijk}}f
\nonumber\\&=&(\frac{2\pi }{N_{i}}e,\frac{%
2\pi }{N_{0ijk}}f)
\end{eqnarray}

By the constraint $\frac{N_{k}}{N_{j}}\Theta _{ij,k}+\Theta _{jk,i}+\Theta
_{ki,j}=0$, $\frac{N_{k}}{N_{j}}\frac{2\pi }{N_{i}}a+\frac{2\pi }{N_{i}}c+%
\frac{2\pi }{N_{i}}e=0$.

By the constraint $\Theta _{0ij,k}=\Theta _{0jk,i}=-\Theta _{0ki,j}$, $%
b=d=-f $ (mod $N_{0ijk}$).

Combine the two constraints, the remaining three generating phases are: $a$
(mod $N_{i}$), $c$ (mod $N_{i}$), $b=d=f$ (mod $N_{0ijk}$).

Hence in this case the classification is $%
\mathbb{Z}
_{N_{i}}\times 
\mathbb{Z}
_{N_{i}}\times 
\mathbb{Z}
_{N_{0ijk}}$, which is a $\mathbb{Z}_{2}$ complex fermion layer absorbed into a $%
\mathbb{Z}
_{N_{i}}\times 
\mathbb{Z}
_{N_{i}}\times\mathbb{Z}_{N_{0ijk}/2}$ BSPT layer if the non-Abelian complex fermion layer indicator is $\Theta _{fij,k}=m\Theta _{0ij,k}=\pi b\text{ (mod 2)}$, while it is simply a $
\mathbb{Z}
_{N_{i}}\times 
\mathbb{Z}
_{N_{i}}\times\mathbb{Z}_{N_{0ijk}}$ BSPT layer if the non-Abelian complex fermion layer indicator is $\Theta _{fij,k}=m\Theta _{0ij,k}=0$.

\begin{widetext}

\section{Classification of 3D FSPT phases with unitary finite Abelian $G_f$ using general group super-cohomology theory}
\label{app:group super-cohomology}

In this section, we will derive the classification of 3D FSPT with unitary finite Abelian $G_f$, using the general group super-cohomology theory \cite{supercohomology,WangGu2017,WangGu2018}. We first give a short review of general group super-cohomology theory of FSPT phases in section~\ref{app:subsec:rev}. Some useful group cohomology calculations and relations for cocycles are given in section~\ref{app:subsec:cocycles}. After that, the detailed calculations are given in sections~\ref{app:subsec:nonext} and \ref{app:subsec:ext} for non-extended and extended unitary finite Abelian $G_f$ FSPT, respectively.

\subsection{Review of general group super-cohomology theory}
\label{app:subsec:rev}

The general group super-cohomology theory for FSPT phases is developed in \cite{supercohomology,WangGu2017,WangGu2018}. The classification data for 3D FSPT with \emph{unitary} symmetry group $G_f=\Zf\timesw G_b$ is a triple of cochains $(n_2,n_3,\nu_4)$. The data $n_2\in H^2(G_b,\Z_2)/\Gamma^2$ specifies the Majorana chain decorations to the intersection lines of $G_b$-domain walls. The $n_3$ data is a cochain in $C^3(G_b,\Z_2)/B^3(G_b,\Z_2)/\Gamma^3$, specifying the complex fermion decorations to the intersection points of $G_b$-domain walls. And the last $\nu_4\in C^4(G_b,U(1))/B^3(G_b,U(1))/\Gamma^4$ is the usual bosonic SPT classification data. These classification data satisfy the twisted cocycle equations:
\begin{align}\label{app:dn2}
\dd n_2 &= 0,\\\label{app:dn3}
\dd n_3 &= \om_2\smile n_2+n_2\smile n_2,\\\label{app:dnu4}
\dd \nu_4 &= \mathcal O_5[n_3],
\end{align}
where the most general expression of the last obstruction function is
\begin{align}\label{app:O_5}
\mathcal O_5[n_3](012345) &= (-1)^{(\om_2\smile n_3 + n_3\smile_1 n_3 + n_3\smile_2 \dd n_3)(012345) + \om_2(013)\dd n_3(12345)}\\\nonumber
&\quad\times (-1)^{\dd n_3(02345)\dd n_3(01235) + \om_2(023) \left[\dd n_3(01245) + \dd n_3(01235) + \dd n_3(01234)\right]} \\\nonumber
&\quad\times i^{\dd n_3(01245) \dd n_3(01234) \text{ (mod 2)}}
\times (-i)^{\left[\dd n_3(12345)+\dd n_3(02345)+\dd n_3(01345)\right] \dd n_3(01235) \text{ (mod 2)}}.
\end{align}

All the classification data is defined modulo trivialization subgroup $\Gamma^i$. For state labelled by these data, we can construct a symmetric gapped state without topological order on their boundary. Therefore, they are in fact trivial FSPT states \cite{Anomalous}. For unitary Abelian $G_f$, the trivialization groups $\Gamma^i$ can be calculated from
\begin{align}\label{app:Gamma2}
\Gamma^2 &= \{\omega_2\smile n_0 \in H^2(G_b,\Z_2) | n_0\in H^0(G_b,\Z) \},\\\label{app:Gamma3}
\Gamma^3 &= \{\omega_2\smile n_1 + (\om_2\smile_1 \om_2) \floor{n_0/2} \in H^3(G_b,\Z_2) | n_1\in H^1(G_b,\Z_2), n_0\in H^0(G_b,\Z_T) \},\\\label{app:Gamma4}
\Gamma^4 &= \{ (-1)^{\om_2\smile n_2 + n_2\smile n_2} \in H^4(G_b,U(1)) | n_2 \in H^2(G_b,\Z_2) \}.
\end{align}

In the rest of this section, we will use the above equations to derive a complete classification for unitary finite Abelian $G_f$ FSPT phases.

\subsection{Cohomology groups, explicit cocycles and Bockstein homomorphism}
\label{app:subsec:cocycles}

There are many useful relations for cocycles of the cyclic group $\Z_N$. They can tremendously simplify the FSPT calculations. In the following, we denote the cyclic group as $\Z_N=\{0,1,...,N-1\}$, where the group multiplication is given by the addition of integers mod $N$. We will use the notation $\overset{n}{=}$ to mean equality up to mod $n$. Similarly, $\overset{\Z}{=}$ emphasizes an equality in the ring of integers. And $\overset{n,\dd}{=}$ means an equality up to $\Z_n$-valued coboundaries (a.k.a. they belong to the same $\Z_n$-valued cohomology class). The symbol $\floor{x}$ is the floor function as the largest integer smaller than or equal to $x$. And $[x]_N$ is defined to be the mod $N$ value of an integer $x$.

\subsubsection{$\Z_2$-coefficient cohomology}

The cohomology ring for $\Z_N$ ($N$ even) with $\Z_2$ coefficient is
\begin{align}\label{app:HZNZ2}
H^\ast (\Z_N,\Z_2) =
\begin{cases}
\Z_2[n_1^{\Z_2}],& \text{if } N\overset{4}{=}2,\\
\Z_2[n_1^{\Z_2},n_2^{\Z_2}]/\{ (n_1^{\Z_2})^2 \},& \text{if } N\overset{4}{=}0.
\end{cases}
\end{align}
So we can use the cup product of $n_1^{\Z_2}\in H^1(\Z_N,\Z_2)$ and $n_2^{\Z_2}\in H^2(\Z_N,\Z_2)$ to obtain all cocycles. The superscript $\Z_2$ of $n_i^{\Z_2}$ emphasizes that they are $\Z_2$-valued cocycles. As will shown later, the cocycle $(n_1^{\Z_2})^2:=n_1^{\Z_2}\smile n_1^{\Z_2}$ has different result for different $N$: $(n_1^{\Z_2})^2\overset{2,\dd}{=}n_2^{\Z_2}$ if $N\overset{4}{=}2$, and $(n_1^{\Z_2})^2\overset{2,\dd}{=}0$ if $N\overset{4}{=}0$.

The explicit cocycles in $H^\ast (\Z_N,\Z_2)$ are also very useful in the calculations. The expressions of $n_1^{\Z_2}$ and $n_2^{\Z_2}$ are ($a,b\in\Z_N$):
\begin{align}
n_1^{\Z_2}(a) &\overset{2}{=} [a]_2,\\\label{app:n2Z2}
n_2^{\Z_2}(a,b) &\overset{2}{=} \floor{\frac{a+b}{N}} \overset{2}{=} \frac{a+b-[a+b]_N}{N}.
\end{align}
Other cocycles can be obtained from the cup products of $n_1^{\Z_2}$ and $n_2^{\Z_2}$.

\subsubsection{$\Z$-coefficient cohomology}

The $\Z$-coefficient group cohomology for $\Z_N$ is
\begin{align}
H^\ast(\Z_N,\Z) = \Z[n_2^\Z]/\{N n_2^\Z\} = \Z_N[n_2^\Z].
\end{align}
Again, the superscript $\Z$ of the generator 2-cocycle $n_2^\Z$ is to emphasize that it is $\Z$-valued. In fact, the $\Z_2$-valued cocycle $n_2^{\Z_2}$ in \eq{app:n2Z2} is the same as $\Z$-valued cocycle $n_2^{\Z}$:
\begin{align}\label{app:n2Z}
n_2^\Z(a,b) &\overset{\Z}{=} \floor{\frac{a+b}{N}} \overset{\Z}{=} \frac{a+b-[a+b]_N}{N} \overset{\Z}{=} \left(\frac{1}{N} \dd n_1^{\Z_N}\right)(a,b),
\end{align}
where
\begin{align}\label{app:n1ZN}
n_1^{\Z_N}(a) \overset{N}{=} [a]_N
\end{align}
is the generator of $H^1(\Z_N,\Z_N)=\Z_N$. So it is easy to see that
\begin{align}
\dd n_2^\Z \overset{\Z}{=} \dd \left(\frac{1}{N} \dd n_1^{\Z_N}\right) \overset{\Z}{=} \frac{1}{N} \dd^2 n_1^{\Z_N} \overset{\Z}{=} 0,
\end{align}
from the fact $\dd^2=0$. All other cocycles in $H^\ast(\Z_N,\Z)$ can be obtained from the addition and cup product of several $n_2^\Z$.

Using the relations of $\Z_2$ and $\Z$-valued cocycles, we can show that
\begin{align}
(n_1^{\Z_2})^2
:= n_1^{\Z_2}\smile n_1^{\Z_2}
\overset{\Z}{=} \frac{1}{2} \dd n_1^{\Z_2}
\overset{\Z}{=} \frac{1}{2} \dd n_1^{\Z_N} + \dd \mu_1
\overset{\Z}{=} \frac{N}{2} \frac{\dd n_1^{\Z_N}}{N} + \dd \mu_1
\overset{\Z}{=} \frac{N}{2} n_2^\Z + \dd \mu_1
\overset{\Z,\dd}{=} \frac{N}{2} n_2^\Z.
\end{align}
where we have defined a $\Z$-valued 1-cochain ($a\in \Z_N$):
\begin{align}\label{app:mu1}
\mu_1(a) :=\frac{1}{2}(n_1^{\Z_2}-n_1^{\Z_N})(a) \overset{\Z}{=} \frac{[a]_2-[a]_N}{2}.
\end{align}
Since $N$ is even, the right-hand-side of the above equation is indeed an integer.
So we have
\begin{align}\label{app:n1n1}
(n_1^{\Z_2})^2 \overset{\Z,\dd}{=}\frac{N}{2} n_2^\Z \overset{2,\dd}{=}
\begin{cases}
0,& \text{if } N\overset{4}{=}0,\\
n_2^{\Z_2},& \text{if } N\overset{4}{=}2.
\end{cases}
\end{align}
This is exactly the result claimed below \eq{app:HZNZ2}. When $N\overset{4}{=}0$, we also know the explicit coboundary as
\begin{align}
(n_1^{\Z_2})^2
\overset{\Z}{=} \frac{N}{2} n_2^\Z + \dd \mu_1
\overset{2}{=} \dd \mu_1.
\end{align}

\subsubsection{Bockstein homomorphism}

The notion of Bockstein homomorphism is very useful in checking whether a cocycle $(-1)^{f_k}$ [$f_k\in H^k(G_b,\Z_2)$] is a $U(1)$-valued coboundary or not. It is defined as a mapping from $H^k(G_b,\Z_2)$ to $H^{k+1}(G_b,\Z)$:
\begin{align}\label{app:be}
\be(f_k) \overset{\Z}{=} \frac{\dd f_k}{2},
\end{align}
where $f_k$ is a cocycle in $H^k(G_b,\Z_2)$. The coboundary operator is defined with appropriate plus and minus signs in integers. Because of $\dd f_k \overset{2}{=}0$, the right-hand-side of \eq{app:be} is always an integer. The Bockstein homomorphism is the connecting isomorphism between $H^k(G_b,\Z_2)$ and $H^{k+1}(G_b,\Z)$. So we have the useful relation
\begin{align}
(-1)^{f_k} \in B^k(G_b,U(1)) \Longleftrightarrow \beta(f_k) \in B^{k+1}(G_b,\Z).
\end{align}
We can use it to check whether $(-1)^{f_k}$ is a U(1)-valued coboundary or not.

When acting on the cup product of two cocycles, the Bockstein homomorphism reads
\begin{align}
\be(x\smile y) \overset{\Z}{=} \be(x)\smile y +(-1)^{\operatorname{deg}(x)} x \smile \be(y),
\end{align}
which is essentially the Leibniz's rule for coboundary operators. When modulo two, the Bockstein homomorphism is related to Steenrod square operation and higher cup product as
\begin{align}
\be(f_k) \overset{2}{=} Sq^1(f_k) \overset{2}{=} f_k\smile_{k-1}f_k.
\end{align}

There are also some useful equations for Steenrod squares:
\begin{align}\label{app:Sq0}
Sq^0(x) &\overset{2,\dd}{=} x,\\
Sq^i(x) &\overset{2,\dd}{=} 0,\quad \text{if } i>\operatorname{deg}(x),\\
Sq^i(x) &\overset{2,\dd}{=} x\smile x,\quad \text{if } i=\operatorname{deg}(x),\\
\label{app:Cartan}
Sq^n(x\smile y) &\overset{2,\dd}{=} \sum_{i+j=n} Sq^i(x) \smile Sq^j(y).
\end{align}
The last equation is called Cartan formula.

For the special case of cohomology for group $\Z_N$, the Bockstein homomorphism of the generator $n_2^{\Z_2}\in H^2(\Z_N,\Z_2)=\Z_2$ can be shown to be zero:
\begin{align}\label{app:ben2}
Sq^1(n_2^{\Z_2}) \overset{2}{=} \be(n_2^{\Z_2}) \overset{\Z}{=} \frac{1}{2}\dd n_2^{\Z_2} \overset{\Z}{=} \frac{1}{2}\dd^2 n_1^{\Z} \overset{\Z}{=} 0,
\end{align}
where we used $\dd^2=0$ in the last step.

\subsection{Classification of FSPT with $G_f=\Z_2^f \times \prod_{i=1}^K \Z_{N_i}$}
\label{app:subsec:nonext}

The fermionic symmetry group $G_f=\Z_2^f \times \prod_{i=1}^K \Z_{N_i}$ is associated with bosonic symmetry group
\begin{align}\label{app:App_Gb}
G_b=\prod_{i=1}^K \Z_{N_i},
\end{align}
and trivial central extension $\om_2=0$. It is known that, for a given positive integer $N$, we have a unique factorization $N=\prod_p p^{n_p}$ ($p$ is a prime number and $n_p$ is a positive integer in $\Z_+$) and a group isomorphism $\Z_N\cong \prod_p \Z_{p^{n_p}}$. For prime number $p>2$, the cohomology group $H^\ast (\Z_{p^{n_p}},\Z_2)$ is trivial. Therefore, the symmetry group $\prod_{p>2} \Z_{p^{n_p}}$ can only protect bosonic 
SPT phases, and can only affect the FSPT classifications though adding some BSPT phases. To understand genuine FSPT, we can assume
\begin{align}
N_i=2^{k_i}\quad (1\le i\le K,\ k_i\in\Z_+)
\end{align}
in the bosonic symmetry group $G_b$ \eq{app:App_Gb}. Without loss of generality, we can also reorder the Abelian groups such that
\begin{align}
N_i\le N_{i+1} \quad (1\le i\le K-1).
\end{align}

Using K\"{u}nneth formula and universal coefficient theorem, the relevant cohomology groups with $\Z_2$ and $U(1)$ coefficients for \eq{app:App_Gb} ($N_i=2^{k_i}\ge 2$, $N_i\le N_{i+1}$) are given by
\begin{align}\label{app:H2}
H^2(G_b,\Z_2) &= \Z_2^{C_K^1+C_K^2} = \prod_{1\le i\le K} \langle n_2^{(i)} \rangle
\prod_{1\le i<j\le K} \langle n_1^{(i)}n_1^{(j)} \rangle,\\\label{app:H3}
H^3(G_b,\Z_2) &= \Z_2^{C_K^1+2C_K^2+C_K^3} = \prod_{1\le i\le K} \langle n_1^{(i)}n_2^{(i)} \rangle
\prod_{1\le i<j\le K} \langle n_1^{(i)}n_2^{(j)}, n_2^{(i)}n_1^{(j)} \rangle 
\prod_{1\le i<j<k\le K} \langle n_1^{(i)}n_1^{(j)}n_1^{(k)} \rangle,\\\label{app:H4}
H^4(G_b,U(1)) &= \prod_{1\le i<j\le K} \Z_{N_{ij}}^2 \prod_{1\le i<j<k\le K} \Z_{N_{ijk}}^2 \prod_{1\le i<j<k<l\le K} \Z_{N_{ijkl}},\\\nonumber\label{app:H5}
H^5(G_b,U(1)) &= H^6(G_b,\Z) = \prod_{1\le i\le K} \Z_{N_{i}} \prod_{1\le i<j\le K} \Z_{N_{ij}}^2 \prod_{1\le i<j<k\le K} \Z_{N_{ijk}}^4 \prod_{1\le i<j<k<l\le K} \Z_{N_{ijkl}}^3 \prod_{1\le i<j<k<l<m\le K} \Z_{N_{ijklm}}\\
&= \prod_{1\le i\le K} \langle (n_2^{(i)})^3 \rangle \prod_{1\le i<j\le K} \langle (n_2^{(i)})^2n_2^{(j)},n_2^{(i)}(n_2^{(j)})^2 \rangle \prod_{1\le i<j<k\le K} \langle n_2^{(i)}n_2^{(j)}n_2^{(k)} \rangle \times ...
\end{align}
Here, $C_K^i=\frac{K!(K-i)!}{i!}$ is the binomial coefficient. We have listed the generators for the $\Z_2$ coefficient cohomology groups, as well as the generators of some some relevant subgroups of $\Z$ coefficient cohomology groups. They are expressed as cup products of $n_1^{(i)}$ and $n_2^{(i)}$ ($1\le i\le K$), which are generating cocycles for the $i$-th Abelian group $\Z_{N_i}$ in $G_b$. In the following, all $n_p^{(i)}$ are $\Z_2$-valued $p$-cocycles with the superscript $\Z_2$ omitted.

\subsubsection{Trivialization}

Since $\om_2=0$, we have the trivialization groups $\Gamma^2=0$ and $\Gamma^3=0$ according to \eqs{app:Gamma2}{app:Gamma3}. So all nontrivial obstruction-free $n_2$ and $n_3$ correspond to nontrivial FSPT states.

The trivialization group $\Gamma^4$ is generated $(-1)^{n_2\smile n_2}=(-1)^{Sq^2(n_2)}$ according to \eq{app:Gamma4}. We have two choices of root 2-cocycle $n_2\in H^2(G_b,\Z_2)$ [see \eq{app:H2}]: $n_2=n_2^{(i)}$ and $n_2=n_1^{(i)}n_1^{(j)}$ for some $1\le i<j\le K$. For both of them, one can show that $\be(Sq^2(n_2))\overset{\Z,\dd}{=}0$:
\begin{align}
\be(Sq^2(n_2^{(i)})) &\overset{\Z,\dd}{=} \be(n_2^{(i)}n_2^{(i)}) \overset{\Z,\dd}{=} \be(n_2^{(i)}) n_2^{(i)} + n_2^{(i)} \be(n_2^{(i)}) \overset{\Z}{=} 0,\\
\be(Sq^2(n_1^{(i)}n_1^{(j)})) &\overset{\Z,\dd}{=} \be( Sq^1(n_1^{(i)}) Sq^1(n_1^{(j)}) ) \overset{\Z,\dd}{=} \be( (n_1^{(i)})^2 (n_1^{(j)})^2 ) \overset{\Z,\dd}{=} 0,
\end{align}
where we used \eq{app:ben2} and $\be( (n_1^{(i)})^2) \overset{\Z,\dd}{=} \be(n_1^{(i)})n_1^{(i)}-n_1^{(i)}\be(n_1^{(i)})\overset{\Z,\dd}{=}(n_1^{(i)})^3-(n_1^{(i)})^3\overset{\Z}{=}0$. Therefore, we have $(-1)^{n_2\smile n_2} \overset{\dd}{=} 1$ for all $n_2\in H^2(G_b,\Z_2)$. The trivialization group $\Gamma^4$ is also trivial.

\subsubsection{Obstruction}

To solve the equations \eqs{app:dn3}{app:dnu4}, we have to check that the right-hand-side of the equations are coboundaries, otherwise there are no solutions. We need to check these obstructions layer by layer: we first solve \eq{app:dn3} for $n_3$ with a given 2-cocycle $n_2$; after obtaining $n_3$, we can solve \eq{app:dnu4} for $\nu_4$ with this $n_3$.

\emph{(1) Obstruction for $n_2$.}

The equation for $n_3$ is \eq{app:dn3}, i.e., $\dd n_3 \overset{2}{=} n_2\smile n_2$ (recall that $\om_2=0$). So the obstruction function for $n_2$ is
\begin{align}
\mathcal O_4[n_2] \overset{2}{=} n_2\smile n_2.
\end{align}
Below we will check the obstructions for all possible $n_2\in H^2(G_b,\Z_2)$. In fact, we only need to check the obstructions for generators of $n_2$. All others can be obtained from the cohomology operation property: $\mathcal O_4[n_2+n_2'] \overset{2,\dd}{=} \mathcal O_4[n_2] + \mathcal O_4[n_2']$.

\emph{(1.1) $n_2 \overset{2}{=} n_2^{(i)}$ ($1\le i\le K$) [obstructed].}

According to \eq{app:HZNZ2}, $\mathcal O_4[n_2^{(i)}]=n_2^{(i)}\smile n_2^{(i)}$ is always a nontrivial 4-cocycle in $H^4(G_b,\Z_2)$. So $n_2=n_2^{(i)}$ is obstructed for all $1\le i\le K$.

\emph{(1.2) $n_2 \overset{2}{=} n_1^{(i)}n_1^{(j)}$ ($1\le i<j\le K$) [obstruction-free iff $N_j\ge 4$].}

For $n_2 \overset{2}{=} n_1^{(i)}n_1^{(j)}$ ($1\le i<j\le K$), one can show that
\begin{align}
\mathcal O_4[n_2]
&\overset{2}{=} n_1^{(i)}n_1^{(j)}n_1^{(i)}n_1^{(j)}\\
&\overset{2}{=} n_1^{(i)} \left[n_1^{(i)}n_1^{(j)} + \dd \left(n_1^{(i)}\smile_1n_1^{(j)}\right) \right] n_1^{(j)}\\
&\overset{2}{=} (n_1^{(i)})^2 (n_1^{(j)})^2 + \dd \left[ n_1^{(i)} \left(n_1^{(i)}\smile_1n_1^{(j)}\right) n_1^{(j)} \right],
\end{align}
where we have used Steenrod's higher cup product $\smile_i$ \cite{Steenrod1947}. It can be used to switch the cup product of two cocycles as
\begin{align}
\dd( n_p \smile_i n_q ) \overset{2}{=} \dd n_p \smile_i n_q + n_p \smile_i \dd n_q + n_p \smile_{i-1} n_q + n_q \smile_{i-1} n_p.
\end{align}
From \eq{app:n1n1}, we can further simplify $\mathcal O_4[n_2]$ as $\mathcal O_4[n_2] \overset{2,\dd}{=} n_2^{(i)}n_2^{(j)}$ if $N_i=N_j=2$, and $\mathcal O_4[n_2] \overset{2,\dd}{=} 0$, if $N_j\ge 4$ (note that we have assumed $N_i\le N_j$). For the latter case ($N_j\ge 4$), we also have
\begin{align}
\mathcal O_4[n_2]
&\overset{2}{=} (n_1^{(i)})^2 (n_1^{(j)})^2 + \dd \left[ n_1^{(i)} \left(n_1^{(i)}\smile_1n_1^{(j)}\right) n_1^{(j)} \right]\\
&\overset{2}{=} (n_1^{(i)})^2 \dd \mu_1^{(j)} + \dd \left[ n_1^{(i)} \left(n_1^{(i)}\smile_1n_1^{(j)}\right) n_1^{(j)} \right]\\
&\overset{2}{=} \dd \left[ (n_1^{(i)})^2 \mu_1^{(j)} + n_1^{(i)} \left(n_1^{(i)}\smile_1n_1^{(j)}\right) n_1^{(j)} \right]
\end{align}
where $\mu_1^{(j)}$ is the cochain defined in \eq{app:mu1} for the subgroup $\Z_{N_j}$.
So we can get a special solution of $n_3$ as
\begin{align}\label{app:n_3}
\bar n_3 \overset{2}{=} (n_1^{(i)})^2 \mu_1^{(j)} + n_1^{(i)} \left(n_1^{(i)}\smile_1n_1^{(j)}\right) n_1^{(j)}.
\end{align}

In summary, $n_2 \overset{2}{=} n_1^{(i)}n_1^{(j)}$ ($1\le i<j\le K$) is obstructed by $\mathcal O_4[n_2]$ iff $N_i=N_j=2$. For other cases (i.e., $N_j\ge 4$), $\dd n_3 \overset{2}{=} n_2\smile n_2$ has a special $n_3$ solution \eq{app:n_3}.

\emph{(2) Obstruction for $n_3$.}

After checked the obstruction function $\mathcal O_4[n_2]$, we now can check the obstruction function $\mathcal O_5[n_3]$. There are two cases we need to calculate. The first case (2.0) below is that $n_2=n_1^{(i)}n_1^{(j)}$ for some $1\le i<j\le K$ and $N_j\ge 4$, and $n_3$ has a special solution $\bar n_3$ shown in \eq{app:n_3}. We then need to calculate the full complicated obstruction function $\mathcal O_5[\bar n_3]$ in \eq{app:O_5}. The second case (from 2.1 to 2.4) is associated with $n_2 \overset{2}{=}0$, and a 3-cocycle $n_3$ in $H^3(G_b,\Z_2)$. And the obstruction function in this case is merely
\begin{align}\label{app:O_5_}
\mathcal O_5[n_3]\big |_{\om_2\overset{2}{=}0,n_2 \overset{2}{=} 0} &= (-1)^{Sq^2(n_3)} = (-1)^{n_3\smile_1 n_3},
\end{align}
for $\dd n_3 \overset{2}{=} n_2\smile n_2 \overset{2}{=} 0$.

\emph{(2.0) $n_2 \overset{2}{=} n_1^{(i)}n_1^{(j)}$ ($1\le i<j\le K$, $N_j\ge 4$) and $n_3 \overset{2}{=} \bar n_3$ [obstruction-free iff $N_iN_j\ge 16$].}

In this case, we should use the obstruction function $\mathcal O_5[\bar n_3]$ in \eq{app:O_5} which involves some non-higher-cup-product terms. Therefore, we can use the complete U(1) cocyle invariants for unitary finite Abelian groups to check whether $\mathcal O_5[\bar n_3]$ is trivial or not. After some tedious calculations, the possibly nontrivial invariants associated with $\mathcal O_5[\bar n_3]$ are
\begin{align}
e^{i\Omega_{ij}} &= i^{-N^{ij}(N^{ij}-1) N_j(N_j-1) /4},\\
e^{i\Omega_{ji}} &= i^{N^{ij}(N^{ij}-1) N_i(N_i-1) /4}.
\end{align}
For all the invariants to be trivial and $\bar n_3$ to be obstruction-free, we need $N^{ij}N_i/4 \overset{4}{=} 0$. Recall that $N_i\le N_j$, so the obstruction-free condition reduces to
\begin{align}
N_iN_j\ge 16.
\end{align}

In summary, the classification data $(n_2\overset{2}{=}n_1^{(i)}n_1^{(j)}, n_3\overset{2}{=}\bar n_3)$ is obstruction-free ($1\le i<j\le K$), iff the parameters of the symmetry group satisfy $N_iN_j\ge 16$. For $G_b=\Z_2\times\Z_2$, it is obstructed by $\mathcal O_4[n_2]$. For $G_b=\Z_2\times\Z_4$, $\mathcal O_5[n_3]$ is nontrivial although $\mathcal O_4[n_2]$ is trivial.

For the following cases, we have $n_2 \overset{2}{=}0$. So we can use the simpler obstruction function \eq{app:O_5_}.

\emph{(2.1) $n_3 \overset{2}{=} n_1^{(i)}n_2^{(i)}$ ($1\le i\le K$) [obstructed].}

Using \eq{app:ben2} and the formulas \eqs{app:Sq0}{app:Cartan} related to Steenrod square, we have $Sq^2(n_3) \overset{2}{=} Sq^2(n_1^{(i)}n_2^{(i)}) \overset{2}{=} n_1^{(i)}(n_2^{(i)})^2$. We can use the Bockstein homomorphism of $Sq^2(n_3)$ is 
\begin{align}
\be[Sq^2(n_3)] \overset{\Z}{=} \be[n_1^{(i)}(n_2^{(i)})^2] \overset{\Z}{=} (n_1^{(i)})^2(n_2^{(i)})^2 \overset{\Z,\dd}{=} \frac{N_i}{2} (n_2^{(i)})^3,
\end{align}
which is the $\frac{N_i}{2}$-th nontrivial cocycle in $H^6(\Z_{N_i},\Z)=\Z_{N_i}=\langle (n_2^{(i)})^3\rangle$. So $n_3 \overset{2}{=} n_1^{(i)}n_2^{(i)}$ is obstructed.

\emph{(2.2) $n_3 \overset{2}{=} n_1^{(i)}n_2^{(j)}$ ($1\le i<j\le K$) [obstructed].}

Similar to the previous case, we have $Sq^2(n_3) \overset{2}{=} Sq^2(n_1^{(i)}n_2^{(j)}) \overset{2}{=} n_1^{(i)}(n_2^{(j)})^2$. The Bockstein homomorphism is then
\begin{align}
\be[Sq^2(n_3)] \overset{\Z}{=} \be[n_1^{(i)}(n_2^{(j)})^2] \overset{\Z}{=} (n_1^{(i)})^2(n_2^{(j)})^2 \overset{\Z,\dd}{=} \frac{N_i}{2} n_2^{(i)}(n_2^{(j)})^2.
\end{align}
The generator $n_2^{(i)}(n_2^{(j)})^2$ of $\Z_{N_{ij}} \subset H^6(\Z_{N_i}\times\Z_{N_j},\Z)$ [see \eq{app:H5}] has order $N_{ij}=N_i$ (note that $N_i\le N_j$). Therefore, $\be[Sq^2(n_3)]$ is a nontrivial cocycle in $H^6(\Z_{N_i}\times\Z_{N_j},\Z)$, and $n_3 \overset{2}{=} n_1^{(i)}n_2^{(j)}$ is obstructed.

\emph{(2.3) $n_3 \overset{2}{=} n_2^{(i)}n_1^{(j)}$ ($1\le i<j\le K$) [obstruction-free iff $N_i<N_j$].}

In this case, we have $Sq^2(n_3) \overset{2}{=} Sq^2(n_2^{(i)}n_1^{(j)}) \overset{2}{=} (n_2^{(i)})^2n_1^{(j)}$. Its Bockstein homomorphism is
\begin{align}\label{app:n3=n2n1}
\be[Sq^2(n_3)] \overset{\Z}{=} \be[(n_2^{(i)})^2n_1^{(j)}] \overset{\Z}{=} (n_2^{(i)})^2(n_1^{(j)})^2 \overset{\Z,\dd}{=} \frac{N_j}{2} (n_2^{(i)})^2n_2^{(j)}.
\end{align}
The generator $(n_2^{(i)})^2n_2^{(j)}$ of $\Z_{N_{ij}} \subset H^6(\Z_{N_i}\times\Z_{N_j},\Z)$ [see \eq{app:H5}] also has order $N_{ij}=N_i$. If $N_i=N_j$, \eq{app:n3=n2n1} is a nontrivial cocycle. If $N_i<N_j$, then \eq{app:n3=n2n1} is a coboundary, since the coefficient $N_j/2$ is a multiple of the order $N_i$.

In summary, $n_3 \overset{2}{=} n_2^{(i)}n_1^{(j)}$ is obstruction-free iff $N_i<N_j$. We can also derive the explicit $\nu_4$.

\emph{(2.4) $n_3 \overset{2}{=} n_1^{(i)}n_1^{(j)}n_1^{(k)}$ ($1\le i<j<k\le K$) [obstruction-free iff $N_k\ge 4$].}

In this case, we can show that
\begin{align}
Sq^2(n_3) &\overset{2}{=} Sq^2[n_1^{(i)}n_1^{(j)}n_1^{(k)}]
\overset{2}{=} (n_1^{(i)})^2(n_1^{(j)})^2n_1^{(k)} + (n_1^{(i)})^2n_1^{(j)}(n_1^{(k)})^2 + n_1^{(i)}(n_1^{(j)})^2(n_1^{(k)})^2,\\
\be[Sq^2(n_3)] &\overset{\Z}{=} 3(n_1^{(i)})^2(n_1^{(j)})^2(n_1^{(k)})^2
\overset{\Z,\dd}{=} \frac{3N_iN_jN_k}{8}n_2^{(i)}n_2^{(j)}n_2^{(k)}.
\end{align}
The generator $n_2^{(i)}n_2^{(j)}n_2^{(k)}$ in the subgroup $\Z_{N_{ijk}} \subset H^6(G_b,\Z)$ has order $N_{ijk}$ [see \eq{app:H5}]. So $n_3 \overset{2}{=} n_1^{(i)}n_1^{(j)}n_1^{(k)}$ is obstruction-free iff $\frac{3N_iN_jN_k}{8}\in\Z$. Using the fact $2\le N_i \le N_j \le N_k$, the obstruction-free condition is reduced to $N_k\ge 4$.

\subsubsection{Summary}

Note that all the obstruction functions are different for the above generators $n_3$ that are obstructed. Therefore, the summation of several generators, as a generic 3-cocycle $n_3\in H^3(G_b,\Z_2)$, is obstructed if one of the generator is obstructed. To summarize, the obstruction-free classification data $(n_2,n_3,\nu_4)$ belongs to the groups:
\begin{align}
n_2&\in
\prod_{1\le i<j\le K}
\begin{cases}
\Z_2,&(N_iN_j\ge 16)\\
0,&(N_iN_j\le 8)
\end{cases}
,\\
n_3&\in
\prod_{1\le i<j\le K}
\begin{cases}
\Z_2,&(N_i<N_j)\\
0,&(N_i=N_j)
\end{cases}
\times
\prod_{1\le i<j<k\le K}
\begin{cases}
\Z_2,&(N_k\ge 4)\\
0,&(N_i=N_j=N_k=2)
\end{cases}
,\\
\nu_4&\in
\prod_{1\le i<j\le K} \Z_{N_{ij}}^2 \prod_{1\le i<j<k\le K} \Z_{N_{ijk}}^2 \prod_{1\le i<j<k<l\le K} \Z_{N_{ijkl}}.
\end{align}


%

\subsection{Classification of FSPT with $G_f=\Z_{2m}^f \times \prod_{i=1}^K \Z_{N_i}$}
\label{app:subsec:ext}

With the definition
\begin{align}
N_0=2m,
\end{align}
the fermionic symmetry group $G_f=\Z_{2m}^f \times \prod_{i=1}^K \Z_{N_i}$ can be also written as $G_f=\prod_{i=0}^K \Z_{N_i}$. It is associated with bosonic symmetry group
\begin{align}\label{app:App_Gb_}
G_b=\Z_m\times\prod_{i=1}^K \Z_{N_i}=\Z_{N_0/2}\times\prod_{i=1}^K \Z_{N_i}.
\end{align}
and extension specified by
\begin{align}
\om_2(a,b) \overset{2}{=} n_2^{(0)}(a,b) \overset{2}{=} \floor{\frac{a_0+b_0}{m}}.
\end{align}
Without loss of generality, we can assume $m=2^{k}$ to be the $k$-th power of 2 ($k\ge 1$). Otherwise $\Z_{2m}^f$ is isomorphic to $\Z_{2^{k+1}}^f\times\Z_{m/2^k}$, and the latter subgroup can be absorbed to $\Z_{N_i}$ with $i>0$. We note that $n_2^{(0)}$ is the nontrivial $\Z_2$-valued 2-cocycle of $\Z_m$, rather than that of $\Z_{N_0}=\Z_{2m}$. This is different from $n_2^{(i)}$ for $1\le i\le K$.

Similar to the previous section, we assume
\begin{align}
N_0&=2^{k_0}\quad (k_0\ge 2),\\
N_i&=2^{k_i}\quad (0\le i\le K,\ k_i\ge 1).
\end{align}
Without loss of generality, we can also reorder the Abelian groups such that
\begin{align}
N_i\le N_{i+1} \quad (1\le i\le K-1).
\end{align}

Using K\"{u}nneth formula and universal coefficient theorem, the relevant cohomology groups with $\Z_2$ and $U(1)$ coefficients for \eq{app:App_Gb_} are given by
\begin{align}\label{app:H2_}
H^2(G_b,\Z_2) &= \Z_2^{1+2C_K^1+C_K^2} = \langle n_2^{(0)} \rangle \prod_{1\le i\le K} \langle n_2^{(i)},n_1^{(0)}n_1^{(i)} \rangle
\prod_{1\le i<j\le K} \langle n_1^{(i)}n_1^{(j)} \rangle,\\\label{app:H3_}\nonumber
H^3(G_b,\Z_2) &= \Z_2^{1+3C_K^1+3C_K^2+C_K^3} = \langle n_1^{(0)}n_2^{(0)} \rangle 
\prod_{1\le i\le K} \langle n_1^{(i)}n_2^{(i)},n_1^{(0)}n_2^{(i)},n_1^{(i)}n_2^{(0)} \rangle\\
&\quad\quad\quad\quad\quad\quad\quad\quad\quad\quad\times
\prod_{1\le i<j\le K} \langle n_1^{(i)}n_2^{(j)}, n_2^{(i)}n_1^{(j)}, n_1^{(0)}n_1^{(i)}n_1^{(j)} \rangle 
\prod_{1\le i<j<k\le K} \langle n_1^{(i)}n_1^{(j)}n_1^{(k)} \rangle,\\\label{app:H4_}\nonumber
H^4(G_b,U(1)) &= \prod_{1\le i\le K} \Z_{\gcd(m,N_{i})}^2 \prod_{1\le i<j\le K} \Z_{N_{ij}}^2 \Z_{\gcd(m,N_{ij})}^2 \prod_{1\le i<j<k\le K} \Z_{N_{ijk}}^2 \Z_{\gcd(m,N_{ijk})} \prod_{1\le i<j<k<l\le K} \Z_{N_{ijkl}},\\\nonumber
&= \prod_{1\le i\le K} \langle e^{2\pi i \frac{k_{0i}}{{\gcd(m,N_i)}}n_1^{(0)}n_2^{(0)}n_1^{(i)}}, e^{2\pi i \frac{k_{i0}}{{\gcd(m,N_i)}}n_1^{(i)}n_2^{(i)}n_1^{(0)}} \rangle
\prod_{1\le i<j\le K} \langle e^{2\pi i \frac{k_{ij}}{N_{ij}}n_1^{(i)}n_2^{(i)}n_1^{(j)}}, e^{2\pi i \frac{k_{ji}}{N_{ij}}n_1^{(j)}n_2^{(j)}n_1^{(i)}} \rangle\\\nonumber
&\quad\times \prod_{1\le i<j\le K} \langle e^{2\pi i \frac{k_{0ij}}{{\gcd(m,N_{i})}}n_1^{(0)}n_1^{(i)}n_2^{(j)}}, e^{2\pi i \frac{k_{0ji}}{{\gcd(m,N_{j})}}n_1^{(0)}n_1^{(j)}n_2^{(i)}} \rangle
\prod_{1\le i<j<k\le K} \langle e^{2\pi i \frac{k_{ijk}}{N_{ij}}n_1^{(i)}n_1^{(j)}n_2^{(k)}}, e^{2\pi i \frac{k_{ikj}}{N_{ik}}n_1^{(i)}n_1^{(k)}n_2^{(j)}} \rangle\\
&\quad\times \prod_{1\le i<j<k\le K} \langle e^{2\pi i \frac{k_{0ijk}}{{\gcd(m,N_{ijk})}}n_1^{(0)}n_1^{(i)}n_1^{(j)}n_1^{(k)}} \rangle
\prod_{1\le i<j<k<l\le K} \langle e^{2\pi i \frac{k_{ijkl}}{N_{ijkl}}n_1^{(i)}n_1^{(j)}n_1^{(k)}n_1^{(l)}} \rangle.
\end{align}
We have also listed explicitly the ``cononical'' U(1)-valued cocycles, in terms of $n_1^{(i)}\in H^1(\Z_{N_i},\Z_{N_i})$ and $n_2^{(i)}\in H^2(\Z_{N_i},\Z)$ [see \eqs{app:n1ZN}{app:n2Z} for their expressions]. The parameters $k$'s are integers modulo the corresponding subgroup orders:
\begin{align}
&k_{0i},k_{i0}\in\gcd(m,N_i),\quad\quad
k_{ij},k_{ji}\in N_{ij},\quad\quad
k_{0ij},k_{0ji}\in \gcd(m,N_{ij}),\\
&k_{ijk},k_{ikj}\in N_{ijk},\quad\quad
k_{0ijk}\in \gcd(m,N_{ijk}),\quad\quad
k_{ijkl}\in \gcd(m,N_{ijk}).
\end{align}
There cohomology results are essentially the same as the previous \eq{app:H2} to \eq{app:H4}. The only difference is that we have one more subgroup $\Z_{N_0/2}=\Z_m$ in $G_b$. The cohomology group of $H^5(G_b,U(1))$ can be similarly obtained.

\subsubsection{Trivialization}

Since $\om_2=n_2^{(0)}$, we have the trivialization group
\begin{align}\label{app:G2}
\Gamma^2=\Z_2=\langle n_2^{(0)}\rangle,
\end{align}
according to \eq{app:Gamma2}. As $n_2^{(0)}\smile_1n_2^{(0)} \overset{2,\dd}{=} Sq^1(n_2^{(0)}) \overset{2,\dd}{=} 0$,
we have trivialization group
\begin{align}\label{app:G3}
\Gamma^3=\Z_2^{K+1}=\prod_{0\le i\le K} \langle n_2^{(0)}n_1^{(i)}\rangle,
\end{align}
according to \eq{app:Gamma3}.

The trivialization group $\Gamma^4$ in \eq{app:Gamma4} is much more complicated. We have several choices of $n_2$ in \eq{app:H2_}. (1) For $n_2 \overset{2}{=} n_2^{(i)}$ ($0\le i\le K$), the trivialization cocycle $(-1)^{\om_2n_2+n_2n_2}=(-1)^{n_2^{(0)}n_2^{(i)}+n_2^{(i)}n_2^{(i)}}$ is a U(1) coboundary. This is because $\be(n_2^{(0)}n_2^{(i)}+n_2^{(i)}n_2^{(i)}) \overset{\Z}{=}0$, as $\be(n_2^{(i)}) \overset{\Z}{=}0$ from \eq{app:ben2}. (2) For $n_2 \overset{2}{=} n_1^{(0)}n_1^{(i)}$ ($1\le i\le K$), we have
\begin{align}
(-1)^{\om_2n_2+n_2n_2} &= (-1)^{n_2^{(0)}n_1^{(0)}n_1^{(i)}+n_1^{(0)}n_1^{(i)}n_1^{(0)}n_1^{(i)}}
\overset{\dd}{=} (-1)^{n_2^{(0)}n_1^{(0)}n_1^{(i)}+(n_1^{(0)})^2(n_1^{(i)})^2}\\
&\overset{\dd}{=} (-1)^{n_2^{(0)}n_1^{(0)}n_1^{(i)}}
= e^{2\pi i \frac{{\gcd(m,N_i)}/2}{{\gcd(m,N_i)}}n_2^{(0)}n_1^{(0)}n_1^{(i)}}
\end{align}
which is a nontrivial element in $H^5(\Z_m\times\Z_{N_i},U(1))$ compared to the explicit cocycles in \eq{app:H4_}. (3) For $n_2 \overset{2}{=} n_1^{(i)}n_1^{(j)}$, we can do similar calculations and find that
\begin{align}
(-1)^{\om_2n_2+n_2n_2}
\overset{\dd}{=} (-1)^{n_2^{(0)}n_1^{(i)}n_1^{(j)}}
= e^{2\pi i \frac{N_{ij}/2}{N_{ij}}n_2^{(0)}n_1^{(i)}n_1^{(j)}}.
\end{align}
This cocycle is a coboundary in $\Z_{\gcd(m,N_{ij})}\subset H^4(\Z_m\times\Z_{N_i}\times \Z_{N_j},U(1))$, iff $N_{ij}/2\in \gcd(m,N_{ij})\Z$, which is equivalent to $m<N_{ij}$. In summary, the trivialization group for BSPT 4-cocycles is
\begin{align}
\Gamma^4 = \Z_2^K \times \prod_{1\le i<j\le K}
\begin{cases}
\Z_2,&(m\ge N_{ij})\\
\Z_1,&(m<N_{ij})
\end{cases}
= \prod_{1\le i\le K} \langle (-1)^{n_2^{(0)}n_1^{(0)}n_1^{(i)}} \rangle
\prod_{1\le i<j\le K\text{ and }m\ge N_{ij}} \langle (-1)^{n_2^{(0)}n_1^{(i)}n_1^{(j)}} \rangle.
\end{align}
We note that all the nontrivial elements in the trivialization group are of order two.

\subsubsection{Obstruction}

Now we check the obstructions for different choices of $n_2$ and $n_3$ shown in \eqs{app:H2_}{app:H3_}.

\emph{(1) Obstruction for $n_2$.}

From the equation of $n_3$ \eq{app:dn3}, the obstruction function for $n_2$ is
\begin{align}
\mathcal O_4[n_2] \overset{2}{=} \om_2\smile n_2 + n_2\smile n_2.
\end{align}
Again, we need to check the obstructions for generators of $n_2$. All others can be obtained from $\mathcal O_4[n_2+n_2'] \overset{2,\dd}{=} \mathcal O_4[n_2] + \mathcal O_4[n_2']$.

\emph{(1.1) $n_2 \overset{2}{=} n_2^{(0)}$ [trivialized].}

Although $n_2 \overset{2}{=} n_2^{(0)}$ is obstruction-free, it is trivialized by $\Gamma^2$ in \eq{app:G2}.

\emph{(1.2) $n_2 \overset{2}{=} n_2^{(i)}$ ($1\le i\le K$) [obstructed].}

The obstruction function $\mathcal O_4[n_2^{(i)}] \overset{2}{=}n_2^{(0)}\smile n_2^{(i)}+n_2^{(i)}\smile n_2^{(i)}$ is a nontrivial 4-cocycle in $H^4(G_b,\Z_2)$. So $n_2=n_2^{(i)}$ is obstructed for all $1\le i\le K$.

\emph{(1.3) $n_2 \overset{2}{=} n_1^{(0)}n_1^{(i)}$ ($1\le i\le K$) [obstructed].}

In this case, the obstruction function is
\begin{align}
\mathcal O_4[n_2]
\overset{2}{=}n_2^{(0)} n_1^{(0)}n_1^{(i)}+n_1^{(0)}n_1^{(i)}n_1^{(0)}n_1^{(i)}
\overset{2,\dd}{=}n_2^{(0)} n_1^{(0)}n_1^{(i)} + (n_1^{(0)})^2 (n_1^{(i)})^2.
\end{align}
The first part $n_2^{(0)} n_1^{(0)}n_1^{(i)}$ is always a nontrivial cocycle and can not be cancelled by the second. So $n_2 \overset{2}{=} n_1^{(0)}n_1^{(i)}$ are all obstructed for ($1\le i\le K$).

\emph{(1.4) $n_2 \overset{2}{=} n_1^{(i)}n_1^{(j)}$ ($1\le i<j\le K$) [obstructed].}

Similar to the previous case, the obstruction function is
\begin{align}
\mathcal O_4[n_2]
\overset{2}{=} n_2^{(0)} n_1^{(i)}n_1^{(j)} + n_1^{(i)}n_1^{(j)}n_1^{(i)}n_1^{(j)}
\overset{2,\dd}{=} n_2^{(0)} n_1^{(i)}n_1^{(j)} + (n_1^{(i)})^2(n_1^{(j)})^2.
\end{align}
The first part is always nontrivial, and can not be cancelled by the second. So $n_2 \overset{2}{=} n_1^{(i)}n_1^{(j)}$ are all obstructed for ($1\le i<j\le K$).

In summary, all nontrivial $n_2$ (even for the summation of some generators of $n_2$ discussed above) are either trivialized or obstructed. So there is no Majorana chain decoration for FSPT with arbitrary unitary finite Abelian symmetry group $G_f$, if the symmetry is extended by $\om_2\ne 0$. The only possibility is complex fermion decoration layer $n_3$ which will be discussed below.

\emph{(2) Obstruction for $n_3$.}

Since all nontrivial $n_2$ are trivialized or obstructed, we only need to consider $n_2 \overset{2}{=}0$. Then the obstruction function $\mathcal O_5[n_3]$ in \eq{app:O_5} becomes a simpler form
\begin{align}\label{app:O_5__}
\mathcal O_5[n_3]\big |_{n_2 \overset{2}{=} 0} &= (-1)^{\om_2n_3+Sq^2(n_3)} = (-1)^{\om_2n_3+n_3\smile_1 n_3}.
\end{align}

\emph{(2.1) $n_3 \overset{2}{=} n_1^{(0)}n_2^{(0)}$ [trivialized].}

The complex fermion decoration data $n_3 \overset{2}{=} n_1^{(0)}n_2^{(0)}$ is trivialized by \eq{app:G3}.

\emph{(2.2) $n_3 \overset{2}{=} n_1^{(i)}n_2^{(i)}$ ($1\le i\le K$) [obstructed].}

Since $Sq^2(n_3) \overset{2}{=} Sq^2(n_1^{(i)}n_2^{(i)}) \overset{2}{=} n_1^{(i)}(n_2^{(i)})^2$, we have obstruction function \eq{app:O_5__} as
\begin{align}
\mathcal O_5[n_3] = (-1)^{n_2^{(0)}n_1^{(i)}n_2^{(i)}+n_1^{(i)}(n_2^{(i)})^2}=e^{2\pi i \frac{N_i/2}{N_i}n_2^{(0)}n_1^{(i)}n_2^{(i)} + 2\pi i \frac{N_i/2}{N_i}n_1^{(i)}(n_2^{(i)})^2}.
\end{align}
The second term $e^{2\pi i \frac{N_i/2}{N_i}n_1^{(i)}(n_2^{(i)})^2}$ is always a nontrivial element in $H^5(\Z_{N_i},U(1))=\Z_{N_i}=\langle e^{2\pi i \frac{1}{N_i}n_1^{(i)}(n_2^{(i)})^2} \rangle$. And it cannot be cancelled by the first term. Therefore, $n_3 \overset{2}{=} n_1^{(i)}n_2^{(i)}$ is always obstructed.

\emph{(2.3) $n_3 \overset{2}{=} n_1^{(0)}n_2^{(i)}$ ($1\le i\le K$) [obstruction-free iff $m>N_i$].}

In this case, we have $Sq^2(n_3) \overset{2}{=} Sq^2(n_1^{(0)}n_2^{(i)}) \overset{2}{=} n_1^{(0)}(n_2^{(i)})^2$, and
\begin{align}
\mathcal O_5[n_3] = (-1)^{n_2^{(0)}n_1^{(0)}n_2^{(i)}+n_1^{(0)}(n_2^{(i)})^2}=e^{2\pi i \frac{m/2}{m}n_2^{(0)}n_1^{(0)}n_2^{(i)} + 2\pi i \frac{m/2}{m}n_1^{(0)}(n_2^{(i)})^2}.
\end{align}
The cohomology group of $H^5(\Z_m\times\Z_{N_i})$ has subgroups $\Z_{\gcd(m,N_i)}^2 = \langle e^{2\pi i \frac{1}{m}n_2^{(0)}n_1^{(0)}n_2^{(i)}}, e^{2\pi i \frac{1}{m}n_1^{(0)}(n_2^{(i)})^2} \rangle$. So $\mathcal O_5[n_3]$ is a trivial cocycle iff $m/2 \in \gcd(m,N_i)\Z$, which is equivalent to $m>N_i$. Therefore, $n_3 \overset{2}{=} n_1^{(0)}n_2^{(i)}$ ($1\le i\le K$) is obstruction-free iff $m>N_i$.

\emph{(2.4) $n_3 \overset{2}{=} n_2^{(0)}n_1^{(i)}$ ($1\le i\le K$) [trivialized].}

The complex fermion decoration data $n_3 \overset{2}{=} n_2^{(0)}n_1^{(i)}$ is trivialized by \eq{app:G3}.

\emph{(2.5) $n_3 \overset{2}{=} n_1^{(i)}n_2^{(j)}$ ($1\le i<j\le K$) [obstructed].}

We have $Sq^2(n_3) \overset{2}{=} Sq^2(n_1^{(i)}n_2^{(j)}) \overset{2}{=} n_1^{(i)}(n_2^{(j)})^2$, and
\begin{align}
\mathcal O_5[n_3] = (-1)^{n_2^{(0)}n_1^{(i)}n_2^{(j)}+n_1^{(i)}(n_2^{(j)})^2}=e^{2\pi i \frac{N_i/2}{N_i}n_2^{(0)}n_1^{(i)}n_2^{(j)} + 2\pi i \frac{N_i/2}{N_i}n_1^{(i)}(n_2^{(j)})^2}.
\end{align}
The cohomology group $H^5(G_,U(1))$ has subgroups $\Z_{\gcd(m,N_{ij})}=\langle e^{2\pi i \frac{1}{N_i}n_2^{(0)}n_1^{(i)}n_2^{(j)}} \rangle$ and $\Z_{N_{ij}} = \langle e^{2\pi i \frac{1}{N_i}n_1^{(i)}(n_2^{(j)})^2} \rangle$. So $\mathcal O_5[n_3]$ is trivial iff $N_i/2 \in \gcd(m,N_{ij})\Z$ and $N_i/2 \in N_{ij}\Z$. This is impossible since we have assumed $N_i\le N_j$. Therefore, $n_3 \overset{2}{=} n_1^{(i)}n_2^{(j)}$ ($1\le i<j\le K$) is always obstructed.

\emph{(2.6) $n_3 \overset{2}{=} n_1^{(j)}n_2^{(i)}$ ($1\le i<j\le K$) [obstruction-free iff $N_i<N_j$].}

This case is similar to the above one with $i$ and $j$ switched. So the conclusion is that, $\mathcal O_5[n_3]$ is trivial iff $N_j/2 \in \gcd(m,N_{ij})\Z$ and $N_j/2 \in N_{ij}\Z$. These conditions are satisfied iff $N_i<N_j$ (note that we have assumed $N_i\le N_j$). Therefore, $n_3 \overset{2}{=} n_1^{(j)}n_2^{(i)}$ ($1\le i<j\le K$) is obstruction-free iff $N_i<N_j$.

\emph{(2.7) $n_3 \overset{2}{=} n_1^{(0)}n_1^{(i)}n_1^{(j)}$ ($1\le i<j\le K$) [obstructed].}

Using the properties of Steenrod square, we have
\begin{align}
Sq^2(n_3) &\overset{2}{=} Sq^2[n_1^{(0)}n_1^{(i)}n_1^{(j)}]
\overset{2}{=} (n_1^{(0)})^2(n_1^{(i)})^2n_1^{(j)} + (n_1^{(0)})^2n_1^{(i)}(n_1^{(j)})^2 + n_1^{(0)}(n_1^{(i)})^2(n_1^{(j)})^2.
\end{align}
The obstruction function is
\begin{align}
\mathcal O_5[n_3] &= (-1)^{n_2^{(0)}n_1^{(0)}n_1^{(i)}n_1^{(j)} + (n_1^{(0)})^2(n_1^{(i)})^2n_1^{(j)} + (n_1^{(0)})^2n_1^{(i)}(n_1^{(j)})^2 + n_1^{(0)}(n_1^{(i)})^2(n_1^{(j)})^2}\\
&= (-1)^{n_2^{(0)}n_1^{(0)}n_1^{(i)}n_1^{(j)} + (n_1^{(0)})^2(n_1^{(i)})^2n_1^{(j)}}
\times e^{(\pi i/2) \dd[n_1^{(0)}n_1^{(i)}(n_1^{(j)})^2 + n_1^{(0)}n_1^{(i)}(n_1^{(j)})^2]}\\
&\overset{\dd}{=}(-1)^{n_2^{(0)}n_1^{(0)}n_1^{(i)}n_1^{(j)} + (n_1^{(0)})^2(n_1^{(i)})^2n_1^{(j)} }\\
&=e^{2\pi i \frac{\gcd(m,N_{ij})/2}{\gcd(m,N_{ij})} n_2^{(0)}n_1^{(0)}n_1^{(i)}n_1^{(j)} + \pi i (n_1^{(0)})^2(n_1^{(i)})^2n_1^{(j)}}.
\end{align}
The first term $e^{2\pi i \frac{\gcd(m,N_{ij})/2}{\gcd(m,N_{ij})} n_2^{(0)}n_1^{(0)}n_1^{(i)}n_1^{(j)}}$ is always a nontrivial cocycle in the subgroup $\Z_{\gcd(m,N_{ij})}$ [of $H^5(G_b,U(1))$] generated by $e^{2\pi i \frac{1}{\gcd(m,N_{ij})} n_2^{(0)}n_1^{(0)}n_1^{(i)}n_1^{(j)}}$. So $n_3 \overset{2}{=} n_1^{(0)}n_1^{(i)}n_1^{(j)}$ ($1\le i<j\le K$) is always obstructed.

\emph{(2.8) $n_3 \overset{2}{=} n_1^{(i)}n_1^{(j)}n_1^{(k)}$ ($1\le i<j<k\le K$) [obstruction-free iff $m<N_{ijk}$].}

Similar to the previous case, we have
\begin{align}
Sq^2(n_3) &\overset{2}{=} Sq^2[n_1^{(i)}n_1^{(j)}n_1^{(k)}]
\overset{2}{=} (n_1^{(i)})^2(n_1^{(j)})^2n_1^{(k)} + (n_1^{(i)})^2n_1^{(j)}(n_1^{(k)})^2 + n_1^{(i)}(n_1^{(j)})^2(n_1^{(k)})^2.
\end{align}
The obstruction function is
\begin{align}
\mathcal O_5[n_3] &= (-1)^{n_2^{(0)}n_1^{(i)}n_1^{(j)}n_1^{(k)} + (n_1^{(i)})^2(n_1^{(j)})^2n_1^{(k)} + (n_1^{(i)})^2n_1^{(j)}(n_1^{(k)})^2 + n_1^{(i)}(n_1^{(j)})^2(n_1^{(k)})^2}\\
&\overset{\dd}{=}(-1)^{n_2^{(0)}n_1^{(i)}n_1^{(j)}n_1^{(k)} + (n_1^{(i)})^2(n_1^{(j)})^2n_1^{(k)} }\\
&=e^{2\pi i \frac{N_{ijk}/2}{N_{ijk}} n_2^{(0)}n_1^{(i)}n_1^{(j)}n_1^{(k)} + 2\pi i \frac{N_iN_jN_k/8}{N_k}n_2^{(i)}n_2^{(j)}n_1^{(k)}}.
\end{align}
The cohomology group $H^5(G_b,U(1))$ has subgroups $\Z_{\gcd(m,N_{ijk})}$ generated by $e^{2\pi i \frac{1}{N_{ijk}} n_2^{(0)}n_1^{(i)}n_1^{(j)}n_1^{(k)}}$ and $\Z_{N_{ijk}}$ generated by $e^{2\pi i \frac{1}{N_k}n_2^{(i)}n_2^{(j)}n_1^{(k)}}$. So $\mathcal O_5[n_3]$ is a trivial cocycle iff $N_{ijk}/2 \in \gcd(m,N_{ijk})\Z$ and $N_iN_jN_k/8\in N_{ijk}\Z$. Using the fact $2\le N_i\le N_j\le N_k$, these conditions are equivalent to $m<N_{i}$ and $N_jN_k/8\in\Z$. As $m\ge 2$, they can be further simplified to $m<N_i$ only. In summary, $n_3 \overset{2}{=} n_1^{(i)}n_1^{(j)}n_1^{(k)}$ ($1\le i<j<k\le K$) is obstruction-free iff $m<N_{i}$.

\subsubsection{Summary}

We note that all the obstruction functions are different for the above obstructed generating $n_3$'s. So the summation of several generators is obstructed if one of the generator is obstructed. In summary, the trivialization-free and obstruction-free classification data $(n_2=0,n_3,\nu_4)$ belongs to the groups:
\begin{align}
n_3&\in
\prod_{1\le i\le K}
\begin{cases}
\Z_2,&(m>N_i)\\
0,&(m\le N_i)
\end{cases}
\times
\prod_{1\le i<j\le K}
\begin{cases}
\Z_2,&(N_i<N_j)\\
0,&(N_i=N_j)
\end{cases}
\times
\prod_{1\le i<j<k\le K}
\begin{cases}
\Z_2,&(m<N_{ijk})\\
0,&(m\ge N_{ijk})
\end{cases}
,\\\nonumber
\nu_4&\in
\prod_{1\le i\le K} \Z_{\gcd(m,N_{i})} \Z_{\gcd(m,N_{i})/2}
\times\prod_{1\le i<j\le K} \Z_{N_{ij}}^2 \Z_{\gcd(m,N_{ij})}
\times
\begin{cases}
\Z_{\gcd(m,N_{ij})/2},&(m\ge N_{ij})\\
\Z_{\gcd(m,N_{ij})},&(m<N_{ij})
\end{cases}
\\
&\quad\times\prod_{1\le i<j<k\le K} \Z_{N_{ijk}}^2 \Z_{\gcd(m,N_{ijk})}
\times\prod_{1\le i<j<k<l\le K} \Z_{N_{ijkl}}.
\end{align}



\end{widetext}

\section{Stacking additivity of topological invariants}

\subsection{Physical argument}

In this section, we give a justification for the additivity of topological invariants under stacking of FSPT states. An intuitive justification on additivity is given in \Ref{2DFSPTbraiding}. Here, we give a more microscopic justification. If stacking is done after gauging, then it is obvious that the topological invariants, being Abelian Berry phases, must be additive. However, stacking is done before gauging, so additivity cannot be immediately observed. The following discussion applies to both bosonic and fermionic systems.

We recall that the gauging procedure of \Ref{braiding12} (also see Appendix A of \Ref{topoinvar15} for a general description) has a special property: the gauge flux operator on \emph{every} plaquette of the lattice commutes with the gauged Hamiltonian. That is, flux on every plaquette is conserved. Accordingly, eigenstates of the gauged system are all of the form $|\Psi_{\rm SPT}(\phi)\rangle\otimes |\Psi_{\rm gauge}(\phi)\rangle$, where $\phi$ is a fixed flux configuration,  $|\Psi_{\rm SPT}(\phi)\rangle$ describes the state of the SPT degrees of freedom in the presence of $\phi$, and $|\Psi_{\rm gauge}(\phi)\rangle$ describes the state of the gauge field. 

For a given eigenstate, we now adiabatically deform the state in a cyclic fashion and assume that an Abelian Berry phase $\Theta$ results, which may be any of the topological invariants defined in this paper. There are two contributions, $\Theta_{\rm SPT}$ and $ \Theta_{\rm gauge}$, from $|\Psi_{\rm SPT}(\phi)\rangle $ and $|\Psi_{\rm gauge}(\phi)\rangle$ respectively. The total Berry phase is $\Theta=\Theta_{\rm SPT} + \Theta_{\rm gauge}$. The gauging procedure in \Ref{braiding12} is designed in a way such that $\Theta_{\rm gauge} =0$. In general, $\Theta_{\rm gauge}$ does not have to be zero, e.g., if the gauge field has a Chern-Simon type interaction. However, for our purpose, the gauge field is a tool to detect SPT physics, so it is a natural choice not to introduce any topological Berry phase that is purely due to the gauge field. In fact, $\Theta_{\rm gauge}=0$ is implicitly assumed in almost all works that use the method of gauging symmetries in the field of symmetry-protected and symmetry-enriched topological phases. 

With the above understanding, we now consider stacking two SPT systems $a$ and $b$. For individual systems, the Berry phases associated with an adiabatic process is $\Theta^a = \Theta_{\rm SPT}^a$ and $\Theta^b = \Theta_{\rm SPT}^b$, respectively. After stacking, eigenstates of the gauged system are of the form $|\Psi_{\rm SPT}^a(\phi)\rangle\otimes|\Psi_{\rm SPT}^b(\phi)\rangle\otimes |\Psi_{\rm gauge}(\phi)\rangle$. Therefore, the total Berry phase associated with the same adiabatic process is given by $\Theta^{a+b}=\Theta_{\rm SPT}^a + \Theta_{\rm SPT}^b + \Theta_{\rm gauge} =\Theta^a + \Theta^b$, where the fact $\Theta_{\rm gauge}=0$ is used. This is the stacking additivity of topological invariants.

\subsection{Example}

The stacking additivity of the topological invariants is verified by the additivity of solutions to the constraints of topological invariants. Furthermore, solutions to the constraints under addition gives rise to the Abelian classification group  $H_{stack}$ of FSPT phases, with total number of FSPT phases  given by the order of the group, $\lvert H_{stack} \rvert$.  More specifically, solutions to the constraints satisfy the following group properties: (1) ``identity": the trivial solution with all topological invariants being zero corresponds the trivial FSPT phase, i.e., the conventional atomic insulators; (2) ``group multiplication": addition of two solutions gives rise to a new solution, corresponding to the fact that stacking two FSPT phases gives a new phase; (3) ``inverse": the nagative of a solution is also a solution, corresponding to the fact that given an FSPT phase, there exists an inverse phase, such that stacking the two produces the trivial phase. 

To better illustrate the additivity, we show the additive structure by ``adding" two FSPT phases in the example $G^f=
\mathbb{Z}
_{2}^{f}\times 
\mathbb{Z}
_{2}\times 
\mathbb{Z}
_{8}$.
Let the first FSPT phase be associated with the following topological invariants:
\begin{eqnarray}
&&(\Theta _{0ij,0},\Theta _{ij,i},\Theta _{0ij,i},\Theta _{0ij,j},\Theta
_{i,j},\Theta _{0i,j},\Theta _{00i,j},\Theta _{00j,i})\nonumber\\&=&(0 ,0 ,0,0,%
\frac{\pi }{2},\pi,0,0 ),
\end{eqnarray}
\begin{equation}
(\Theta _{ij,j},\Theta _{j,i})=(0 ,0 ).
\end{equation}%
One can check the it is a solution to the constraints of topological invariants.  Let the second FSPT phase be associated with the following topological invariants:
\begin{eqnarray}
&&(\Theta _{0ij,0},\Theta _{ij,i},\Theta _{0ij,i},\Theta _{0ij,j},\Theta
_{i,j},\Theta _{0i,j},\Theta _{00i,j},\Theta _{00j,i})\nonumber\\&=&(\pi ,\pi ,\pi ,\pi ,%
\frac{3\pi }{4},\frac{\pi }{2},\pi ,\pi ),
\end{eqnarray}
\begin{equation}
(\Theta _{ij,j},\Theta _{j,i})=(0,0 ).
\end{equation}%
It is again a solution. Stacking these two FSPT phases, we obtain the total topological invariants to be:
\begin{eqnarray}
&&(\Theta _{0ij,0},\Theta _{ij,i},\Theta _{0ij,i},\Theta _{0ij,j},\Theta
_{i,j},\Theta _{0i,j},\Theta _{00i,j},\Theta _{00j,i})\nonumber\\&=&(\pi ,\pi ,\pi ,\pi ,%
\frac{5\pi }{4},\frac{3\pi }{2},\pi ,\pi ),
\end{eqnarray}
\begin{equation}
(\Theta _{ij,j},\Theta _{j,i})=(0 ,0 ).
\end{equation}%
One can check again that it satisfies all of the 14 constraints. Therefore, the added new phase is again a valid FSPT phase with the symmetry group  $G^f=
\mathbb{Z}
_{2}^{f}\times 
\mathbb{Z}
_{2}\times 
\mathbb{Z}
_{8}$.

 \bibliography{Nonabelian.bib}

\begin{thebibliography}{30}%
\makeatletter
\providecommand \@ifxundefined [1]{%
 \@ifx{#1\undefined}
}%
\providecommand \@ifnum [1]{%
 \ifnum #1\expandafter \@firstoftwo
 \else \expandafter \@secondoftwo
 \fi
}%
\providecommand \@ifx [1]{%
 \ifx #1\expandafter \@firstoftwo
 \else \expandafter \@secondoftwo
 \fi
}%
\providecommand \natexlab [1]{#1}%
\providecommand \enquote  [1]{``#1''}%
\providecommand \bibnamefont  [1]{#1}%
\providecommand \bibfnamefont [1]{#1}%
\providecommand \citenamefont [1]{#1}%
\providecommand \href@noop [0]{\@secondoftwo}%
\providecommand \href [0]{\begingroup \@sanitize@url \@href}%
\providecommand \@href[1]{\@@startlink{#1}\@@href}%
\providecommand \@@href[1]{\endgroup#1\@@endlink}%
\providecommand \@sanitize@url [0]{\catcode `\\12\catcode `\$12\catcode
  `\&12\catcode `\#12\catcode `\^12\catcode `\_12\catcode `\%12\relax}%
\providecommand \@@startlink[1]{}%
\providecommand \@@endlink[0]{}%
\providecommand \url  [0]{\begingroup\@sanitize@url \@url }%
\providecommand \@url [1]{\endgroup\@href {#1}{\urlprefix }}%
\providecommand \urlprefix  [0]{URL }%
\providecommand \Eprint [0]{\href }%
\providecommand \doibase [0]{http://dx.doi.org/}%
\providecommand \selectlanguage [0]{\@gobble}%
\providecommand \bibinfo  [0]{\@secondoftwo}%
\providecommand \bibfield  [0]{\@secondoftwo}%
\providecommand \translation [1]{[#1]}%
\providecommand \BibitemOpen [0]{}%
\providecommand \bibitemStop [0]{}%
\providecommand \bibitemNoStop [0]{.\EOS\space}%
\providecommand \EOS [0]{\spacefactor3000\relax}%
\providecommand \BibitemShut  [1]{\csname bibitem#1\endcsname}%
\let\auto@bib@innerbib\@empty
\bibitem [{\citenamefont {Wang}\ and\ \citenamefont {Levin}(2014)}]{threeloop}%
  \BibitemOpen
  \bibfield  {author} {\bibinfo {author} {\bibfnamefont {C.}~\bibnamefont
  {Wang}}\ and\ \bibinfo {author} {\bibfnamefont {M.}~\bibnamefont {Levin}},\
  }\href {\doibase 10.1103/PhysRevLett.113.080403} {\bibfield  {journal}
  {\bibinfo  {journal} {Phys. Rev. Lett.}\ }\textbf {\bibinfo {volume} {113}},\
  \bibinfo {pages} {080403} (\bibinfo {year} {2014})}\BibitemShut {NoStop}%
\bibitem [{\citenamefont {{Wang}}\ \emph {et~al.}(2018)\citenamefont {{Wang}},
  \citenamefont {{Cheng}}, \citenamefont {{Wang}},\ and\ \citenamefont
  {{Gu}}}]{WCWG2018}%
  \BibitemOpen
  \bibfield  {author} {\bibinfo {author} {\bibfnamefont {Q.-R.}\ \bibnamefont
  {{Wang}}}, \bibinfo {author} {\bibfnamefont {M.}~\bibnamefont {{Cheng}}},
  \bibinfo {author} {\bibfnamefont {C.}~\bibnamefont {{Wang}}}, \ and\ \bibinfo
  {author} {\bibfnamefont {Z.-C.}\ \bibnamefont {{Gu}}},\ }\href@noop {}
  {\bibfield  {journal} {\bibinfo  {journal} {ArXiv e-prints}\ } (\bibinfo
  {year} {2018})},\ \Eprint {http://arxiv.org/abs/1810.13428} {arXiv:1810.13428
  [cond-mat.str-el]} \BibitemShut {NoStop}%
\bibitem [{\citenamefont {Lin}\ and\ \citenamefont {Levin}(2015)}]{lin15}%
  \BibitemOpen
  \bibfield  {author} {\bibinfo {author} {\bibfnamefont {C.-H.}\ \bibnamefont
  {Lin}}\ and\ \bibinfo {author} {\bibfnamefont {M.}~\bibnamefont {Levin}},\
  }\href {\doibase 10.1103/PhysRevB.92.035115} {\bibfield  {journal} {\bibinfo
  {journal} {Phys. Rev. B}\ }\textbf {\bibinfo {volume} {92}},\ \bibinfo
  {pages} {035115} (\bibinfo {year} {2015})}\BibitemShut {NoStop}%
\bibitem [{\citenamefont {Levin}\ and\ \citenamefont {Gu}(2012)}]{braiding12}%
  \BibitemOpen
  \bibfield  {author} {\bibinfo {author} {\bibfnamefont {M.}~\bibnamefont
  {Levin}}\ and\ \bibinfo {author} {\bibfnamefont {Z.-C.}\ \bibnamefont {Gu}},\
  }\href {\doibase 10.1103/PhysRevB.86.115109} {\bibfield  {journal} {\bibinfo
  {journal} {Phys. Rev. B}\ }\textbf {\bibinfo {volume} {86}},\ \bibinfo
  {pages} {115109} (\bibinfo {year} {2012})}\BibitemShut {NoStop}%
\bibitem [{\citenamefont {Gu}\ and\ \citenamefont {Levin}(2014)}]{gu14b}%
  \BibitemOpen
  \bibfield  {author} {\bibinfo {author} {\bibfnamefont {Z.-C.}\ \bibnamefont
  {Gu}}\ and\ \bibinfo {author} {\bibfnamefont {M.}~\bibnamefont {Levin}},\
  }\href {\doibase 10.1103/PhysRevB.89.201113} {\bibfield  {journal} {\bibinfo
  {journal} {Phys. Rev. B}\ }\textbf {\bibinfo {volume} {89}},\ \bibinfo
  {pages} {201113} (\bibinfo {year} {2014})}\BibitemShut {NoStop}%
\bibitem [{\citenamefont {Wang}\ and\ \citenamefont
  {Levin}(2015)}]{topoinvar15}%
  \BibitemOpen
  \bibfield  {author} {\bibinfo {author} {\bibfnamefont {C.}~\bibnamefont
  {Wang}}\ and\ \bibinfo {author} {\bibfnamefont {M.}~\bibnamefont {Levin}},\
  }\href {\doibase 10.1103/PhysRevB.91.165119} {\bibfield  {journal} {\bibinfo
  {journal} {Phys. Rev. B}\ }\textbf {\bibinfo {volume} {91}},\ \bibinfo
  {pages} {165119} (\bibinfo {year} {2015})}\BibitemShut {NoStop}%
\bibitem [{\citenamefont {Cheng}\ \emph {et~al.}(2018)\citenamefont {Cheng},
  \citenamefont {Tantivasadakarn},\ and\ \citenamefont {Wang}}]{threeloop2018}%
  \BibitemOpen
  \bibfield  {author} {\bibinfo {author} {\bibfnamefont {M.}~\bibnamefont
  {Cheng}}, \bibinfo {author} {\bibfnamefont {N.}~\bibnamefont
  {Tantivasadakarn}}, \ and\ \bibinfo {author} {\bibfnamefont {C.}~\bibnamefont
  {Wang}},\ }\href {\doibase 10.1103/PhysRevX.8.011054} {\bibfield  {journal}
  {\bibinfo  {journal} {Phys. Rev. X}\ }\textbf {\bibinfo {volume} {8}},\
  \bibinfo {pages} {011054} (\bibinfo {year} {2018})}\BibitemShut {NoStop}%
\bibitem [{\citenamefont {Chen}\ \emph {et~al.}(2012)\citenamefont {Chen},
  \citenamefont {Gu}, \citenamefont {Liu},\ and\ \citenamefont
  {Wen}}]{BSPT2012}%
  \BibitemOpen
  \bibfield  {author} {\bibinfo {author} {\bibfnamefont {X.}~\bibnamefont
  {Chen}}, \bibinfo {author} {\bibfnamefont {Z.-C.}\ \bibnamefont {Gu}},
  \bibinfo {author} {\bibfnamefont {Z.-X.}\ \bibnamefont {Liu}}, \ and\
  \bibinfo {author} {\bibfnamefont {X.-G.}\ \bibnamefont {Wen}},\ }\href
  {\doibase 10.1126/science.1227224} {\bibfield  {journal} {\bibinfo  {journal}
  {Science}\ }\textbf {\bibinfo {volume} {338}},\ \bibinfo {pages} {1604}
  (\bibinfo {year} {2012})}\BibitemShut {NoStop}%
\bibitem [{\citenamefont {Chen}\ \emph {et~al.}(2013)\citenamefont {Chen},
  \citenamefont {Gu}, \citenamefont {Liu},\ and\ \citenamefont {Wen}}]{BSPT13}%
  \BibitemOpen
  \bibfield  {author} {\bibinfo {author} {\bibfnamefont {X.}~\bibnamefont
  {Chen}}, \bibinfo {author} {\bibfnamefont {Z.-C.}\ \bibnamefont {Gu}},
  \bibinfo {author} {\bibfnamefont {Z.-X.}\ \bibnamefont {Liu}}, \ and\
  \bibinfo {author} {\bibfnamefont {X.-G.}\ \bibnamefont {Wen}},\ }\href
  {\doibase 10.1103/PhysRevB.87.155114} {\bibfield  {journal} {\bibinfo
  {journal} {Phys. Rev. B}\ }\textbf {\bibinfo {volume} {87}},\ \bibinfo
  {pages} {155114} (\bibinfo {year} {2013})}\BibitemShut {NoStop}%
\bibitem [{\citenamefont {Gu}\ and\ \citenamefont {Wen}(2009)}]{gu09}%
  \BibitemOpen
  \bibfield  {author} {\bibinfo {author} {\bibfnamefont {Z.-C.}\ \bibnamefont
  {Gu}}\ and\ \bibinfo {author} {\bibfnamefont {X.-G.}\ \bibnamefont {Wen}},\
  }\href {\doibase 10.1103/PhysRevB.80.155131} {\bibfield  {journal} {\bibinfo
  {journal} {Phys. Rev. B}\ }\textbf {\bibinfo {volume} {80}},\ \bibinfo
  {pages} {155131} (\bibinfo {year} {2009})}\BibitemShut {NoStop}%
\bibitem [{\citenamefont {Gu}\ and\ \citenamefont
  {Wen}(2014)}]{supercohomology}%
  \BibitemOpen
  \bibfield  {author} {\bibinfo {author} {\bibfnamefont {Z.-C.}\ \bibnamefont
  {Gu}}\ and\ \bibinfo {author} {\bibfnamefont {X.-G.}\ \bibnamefont {Wen}},\
  }\href {\doibase 10.1103/PhysRevB.90.115141} {\bibfield  {journal} {\bibinfo
  {journal} {Phys. Rev. B}\ }\textbf {\bibinfo {volume} {90}},\ \bibinfo
  {pages} {115141} (\bibinfo {year} {2014})}\BibitemShut {NoStop}%
\bibitem [{\citenamefont {{Kapustin}}\ \emph {et~al.}(2014)\citenamefont
  {{Kapustin}}, \citenamefont {{Thorngren}}, \citenamefont {{Turzillo}},\ and\
  \citenamefont {{Wang}}}]{kapustin14}%
  \BibitemOpen
  \bibfield  {author} {\bibinfo {author} {\bibfnamefont {A.}~\bibnamefont
  {{Kapustin}}}, \bibinfo {author} {\bibfnamefont {R.}~\bibnamefont
  {{Thorngren}}}, \bibinfo {author} {\bibfnamefont {A.}~\bibnamefont
  {{Turzillo}}}, \ and\ \bibinfo {author} {\bibfnamefont {Z.}~\bibnamefont
  {{Wang}}},\ }\href@noop {} {\bibfield  {journal} {\bibinfo  {journal} {arXiv
  e-prints}\ } (\bibinfo {year} {2014})},\ \Eprint
  {http://arxiv.org/abs/1406.7329} {arXiv:1406.7329} \BibitemShut {NoStop}%
\bibitem [{\citenamefont {{Freed}}(2014)}]{freed14}%
  \BibitemOpen
  \bibfield  {author} {\bibinfo {author} {\bibfnamefont {D.~S.}\ \bibnamefont
  {{Freed}}},\ }\href@noop {} {\bibfield  {journal} {\bibinfo  {journal} {arXiv
  e-prints}\ } (\bibinfo {year} {2014})},\ \Eprint
  {http://arxiv.org/abs/1406.7278} {arXiv:1406.7278} \BibitemShut {NoStop}%
\bibitem [{\citenamefont {{Cheng}}\ \emph {et~al.}(2015)\citenamefont
  {{Cheng}}, \citenamefont {{Bi}}, \citenamefont {{You}},\ and\ \citenamefont
  {{Gu}}}]{cheng15}%
  \BibitemOpen
  \bibfield  {author} {\bibinfo {author} {\bibfnamefont {M.}~\bibnamefont
  {{Cheng}}}, \bibinfo {author} {\bibfnamefont {Z.}~\bibnamefont {{Bi}}},
  \bibinfo {author} {\bibfnamefont {Y.-Z.}\ \bibnamefont {{You}}}, \ and\
  \bibinfo {author} {\bibfnamefont {Z.-C.}\ \bibnamefont {{Gu}}},\ }\href@noop
  {} {\bibfield  {journal} {\bibinfo  {journal} {arXiv e-prints}\ } (\bibinfo
  {year} {2015})},\ \Eprint {http://arxiv.org/abs/1501.01313}
  {arXiv:1501.01313} \BibitemShut {NoStop}%
\bibitem [{\citenamefont {Gaiotto}\ and\ \citenamefont
  {Kapustin}(2016)}]{Gaiotto2016}%
  \BibitemOpen
  \bibfield  {author} {\bibinfo {author} {\bibfnamefont {D.}~\bibnamefont
  {Gaiotto}}\ and\ \bibinfo {author} {\bibfnamefont {A.}~\bibnamefont
  {Kapustin}},\ }\href {\doibase 10.1142/S0217751X16450445} {\bibfield
  {journal} {\bibinfo  {journal} {International Journal of Modern Physics A}\
  }\textbf {\bibinfo {volume} {31}},\ \bibinfo {pages} {1645044} (\bibinfo
  {year} {2016})}\BibitemShut {NoStop}%
\bibitem [{\citenamefont {Freed}\ and\ \citenamefont
  {Hopkins}(2016)}]{freed16}%
  \BibitemOpen
  \bibfield  {author} {\bibinfo {author} {\bibfnamefont {D.~S.}\ \bibnamefont
  {Freed}}\ and\ \bibinfo {author} {\bibfnamefont {M.~J.}\ \bibnamefont
  {Hopkins}},\ }\href@noop {} {\bibfield  {journal} {\bibinfo  {journal} {arXiv
  e-prints}\ } (\bibinfo {year} {2016})},\ \Eprint
  {http://arxiv.org/abs/1604.06527} {arXiv:1604.06527} \BibitemShut {NoStop}%
\bibitem [{\citenamefont {Kapustin}\ and\ \citenamefont
  {Thorngren}(2017)}]{Kapustin2017}%
  \BibitemOpen
  \bibfield  {author} {\bibinfo {author} {\bibfnamefont {A.}~\bibnamefont
  {Kapustin}}\ and\ \bibinfo {author} {\bibfnamefont {R.}~\bibnamefont
  {Thorngren}},\ }\href {\doibase 10.1007/JHEP10(2017)080} {\bibfield
  {journal} {\bibinfo  {journal} {Journal of High Energy Physics}\ }\textbf
  {\bibinfo {volume} {2017}},\ \bibinfo {pages} {80} (\bibinfo {year}
  {2017})}\BibitemShut {NoStop}%
\bibitem [{\citenamefont {Wang}\ and\ \citenamefont
  {Gu}(2018{\natexlab{a}})}]{WangGu2017}%
  \BibitemOpen
  \bibfield  {author} {\bibinfo {author} {\bibfnamefont {Q.-R.}\ \bibnamefont
  {Wang}}\ and\ \bibinfo {author} {\bibfnamefont {Z.-C.}\ \bibnamefont {Gu}},\
  }\href {\doibase 10.1103/PhysRevX.8.011055} {\bibfield  {journal} {\bibinfo
  {journal} {Phys. Rev. X}\ }\textbf {\bibinfo {volume} {8}},\ \bibinfo {pages}
  {011055} (\bibinfo {year} {2018}{\natexlab{a}})}\BibitemShut {NoStop}%
\bibitem [{\citenamefont {Wang}\ and\ \citenamefont
  {Gu}(2018{\natexlab{b}})}]{WangGu2018}%
  \BibitemOpen
  \bibfield  {author} {\bibinfo {author} {\bibfnamefont {Q.-R.}\ \bibnamefont
  {Wang}}\ and\ \bibinfo {author} {\bibfnamefont {Z.-C.}\ \bibnamefont {Gu}},\
  }\href@noop {} {\bibfield  {journal} {\bibinfo  {journal} {arXiv e-prints}\ }
  (\bibinfo {year} {2018}{\natexlab{b}})},\ \Eprint
  {http://arxiv.org/abs/arXiv:1811.00536} {arXiv:arXiv:1811.00536} \BibitemShut
  {NoStop}%
\bibitem [{\citenamefont {Cheng}\ and\ \citenamefont {Wang}(2018)}]{Meng2018}%
  \BibitemOpen
  \bibfield  {author} {\bibinfo {author} {\bibfnamefont {M.}~\bibnamefont
  {Cheng}}\ and\ \bibinfo {author} {\bibfnamefont {C.}~\bibnamefont {Wang}},\
  }\href@noop {} {\bibfield  {journal} {\bibinfo  {journal} {ArXiv e-prints}\ }
  (\bibinfo {year} {2018})},\ \Eprint {http://arxiv.org/abs/1810.12308}
  {arXiv:1810.12308 [cond-mat.str-el]} \BibitemShut {NoStop}%
\bibitem [{\citenamefont {Kitaev}(2001)}]{Kitaev2001}%
  \BibitemOpen
  \bibfield  {author} {\bibinfo {author} {\bibfnamefont {A.~Y.}\ \bibnamefont
  {Kitaev}},\ }\href {http://stacks.iop.org/1063-7869/44/i=10S/a=S29}
  {\bibfield  {journal} {\bibinfo  {journal} {Physics-Uspekhi}\ }\textbf
  {\bibinfo {volume} {44}},\ \bibinfo {pages} {131} (\bibinfo {year}
  {2001})}\BibitemShut {NoStop}%
\bibitem [{\citenamefont {{Tantivasadakarn}}(2017)}]{Tantivasadakarn}%
  \BibitemOpen
  \bibfield  {author} {\bibinfo {author} {\bibfnamefont {N.}~\bibnamefont
  {{Tantivasadakarn}}},\ }\href {\doibase 10.1103/PhysRevB.96.195101}
  {\bibfield  {journal} {\bibinfo  {journal} {\prb}\ }\textbf {\bibinfo
  {volume} {96}},\ \bibinfo {eid} {195101} (\bibinfo {year} {2017})},\ \Eprint
  {http://arxiv.org/abs/1706.09769} {arXiv:1706.09769 [cond-mat.str-el]}
  \BibitemShut {NoStop}%
\bibitem [{\citenamefont {Kitaev}(2006)}]{Kitaev2006}%
  \BibitemOpen
  \bibfield  {author} {\bibinfo {author} {\bibfnamefont {A.}~\bibnamefont
  {Kitaev}},\ }\href@noop {} {\bibfield  {journal} {\bibinfo  {journal} {Annals
  of Physics}\ }\textbf {\bibinfo {volume} {321}},\ \bibinfo {pages} {2}
  (\bibinfo {year} {2006})}\BibitemShut {NoStop}%
\bibitem [{\citenamefont {Wang}\ \emph {et~al.}(2017)\citenamefont {Wang},
  \citenamefont {Lin},\ and\ \citenamefont {Gu}}]{2DFSPTbraiding}%
  \BibitemOpen
  \bibfield  {author} {\bibinfo {author} {\bibfnamefont {C.}~\bibnamefont
  {Wang}}, \bibinfo {author} {\bibfnamefont {C.-H.}\ \bibnamefont {Lin}}, \
  and\ \bibinfo {author} {\bibfnamefont {Z.-C.}\ \bibnamefont {Gu}},\ }\href
  {\doibase 10.1103/PhysRevB.95.195147} {\bibfield  {journal} {\bibinfo
  {journal} {Phys. Rev. B}\ }\textbf {\bibinfo {volume} {95}},\ \bibinfo
  {pages} {195147} (\bibinfo {year} {2017})}\BibitemShut {NoStop}%
\bibitem [{\citenamefont {Preskill}(1999)}]{preskill1999}%
  \BibitemOpen
  \bibfield  {author} {\bibinfo {author} {\bibfnamefont {J.}~\bibnamefont
  {Preskill}},\ }\href@noop {} {\bibfield  {journal} {\bibinfo  {journal}
  {Caltech Lecture Notes}\ } (\bibinfo {year} {1999})}\BibitemShut {NoStop}%
\bibitem [{\citenamefont {Pachos}(2012)}]{pachos2012}%
  \BibitemOpen
  \bibfield  {author} {\bibinfo {author} {\bibfnamefont {J.~K.}\ \bibnamefont
  {Pachos}},\ }\href@noop {} {\emph {\bibinfo {title} {Introduction to
  topological quantum computation}}}\ (\bibinfo  {publisher} {Cambridge
  University Press},\ \bibinfo {year} {2012})\BibitemShut {NoStop}%
\bibitem [{\citenamefont {Bonderson}(2007)}]{anyoncondensation2007}%
  \BibitemOpen
  \bibfield  {author} {\bibinfo {author} {\bibfnamefont {P.~H.}\ \bibnamefont
  {Bonderson}},\ }\emph {\bibinfo {title} {Non-Abelian anyons and
  interferometry}},\ \href@noop {} {Ph.D. thesis},\ \bibinfo  {school}
  {California Institute of Technology} (\bibinfo {year} {2007})\BibitemShut
  {NoStop}%
\bibitem [{\citenamefont {Eli{\"e}ns}(2010)}]{anyoncondensation2010}%
  \BibitemOpen
  \bibfield  {author} {\bibinfo {author} {\bibfnamefont {S.}~\bibnamefont
  {Eli{\"e}ns}},\ }\href@noop {} {\bibfield  {journal} {\bibinfo  {journal}
  {Master's thesis, Universiteit van Amsterdam, Netherlands}\ } (\bibinfo
  {year} {2010})}\BibitemShut {NoStop}%
\bibitem [{\citenamefont {Wang}\ \emph {et~al.}(2018)\citenamefont {Wang},
  \citenamefont {Qi},\ and\ \citenamefont {Gu}}]{Anomalous}%
  \BibitemOpen
  \bibfield  {author} {\bibinfo {author} {\bibfnamefont {Q.-R.}\ \bibnamefont
  {Wang}}, \bibinfo {author} {\bibfnamefont {Y.}~\bibnamefont {Qi}}, \ and\
  \bibinfo {author} {\bibfnamefont {Z.-C.}\ \bibnamefont {Gu}},\ }\href@noop {}
  {\bibfield  {journal} {\bibinfo  {journal} {ArXiv e-prints}\ } (\bibinfo
  {year} {2018})},\ \Eprint {http://arxiv.org/abs/1810.12899} {arXiv:1810.12899
  [cond-mat.str-el]} \BibitemShut {NoStop}%
\bibitem [{\citenamefont {Steenrod}(1947)}]{Steenrod1947}%
  \BibitemOpen
  \bibfield  {author} {\bibinfo {author} {\bibfnamefont {N.~E.}\ \bibnamefont
  {Steenrod}},\ }\href {http://www.jstor.org/stable/1969172} {\bibfield
  {journal} {\bibinfo  {journal} {Annals of Mathematics}\ }\textbf {\bibinfo
  {volume} {48}},\ \bibinfo {pages} {290} (\bibinfo {year} {1947})}\BibitemShut
  {NoStop}%
\end{thebibliography}%

\end{document}